# Multi-Agent Path Planning via Compatible Encounter-Space Allocation in Crowded Spaces Dynamically Occupied by Work Activities

| | | |
|---|---|---|
| Boda Liu | Institute of Intelligent Manufacturing Technology, Hefei University of Technology, Hefei, China | wlbdwww@163.com |
| Xuezhi Yang* | School of Computer Science and Information Engineering, Hefei University of Technology, Hefei, China | xzyang@hfut.edu.cn |
| Bin Yang* | College of Civil Engineering, Tongji University, Shanghai, China | yangbi@tongji.edu.cn |
| Shu Li | Hefei Institute of Public Safety Research, Tsinghua University, Hefei, China | lishu@tsinghua-hf.edu.cn |

* Corresponding authors

## Abstract

Lifelong multi-agent path planning on construction floors must accommodate continually changing goals and work occupancy. Existing planners can update paths and timing online, but work–motion space coupling changes the spatial basis of unfinished encounters, while rerouting affects later encounters. Planning must therefore construct and maintain local encounter passability. We propose Width-based Planning with Compatible Encounter-Space Allocation (WP), which constructs shared passage relationships from actual encounter demands and real-time widths. Shared channels, ordered anchors, and urgency-based updates coordinate successive encounters while retaining individual greedy motion between them. A necessary and sufficient width condition for fixed-point position exchange and a necessary directional condition for composite exchange provide geometric grounds for channel construction and encounter relocation. In crowded-floor computational experiments, WP maintained route continuation after occupancy changes and enabled multi-agent clearing, with infrequent, brief near-stops. Complete-run comparisons with four methods showed 25.2%–52.2% more mean goal arrivals than the best baseline at each configuration of 5–10 agents, with lower mean computation time than the rolling-horizon K-CBS adaptation. The contribution is a planning principle that constructs and continuously maintains local passage relationships from real-time space, unifying shared spatial allocation and conditional local clearing to extend lifelong planning to work–motion space coupling.

## 1. Introduction

Confined activity areas on building floors are linked by narrow entrances. Workers and mobile equipment repeatedly travel between work locations and material storage areas, receiving a new motion goal after reaching the current one. This gives rise to lifelong multi-agent path planning: agents must select routes as goal demands change and repeatedly coordinate passage through shared space.

Multi-agent path planning coordinates individual paths and passage times to resolve motion conflicts and has been applied to warehouse transport and ongoing pickup-and-delivery tasks [1,2]. Continually arriving goals, neighboring motion, and local blockages require online adjustment. Existing approaches use reciprocal collision avoidance to respond to neighbors [3,4], rolling-horizon planning and continuous conflict search to update paths and timing [5,6], and topological guidance, corridor reasoning, safe regions, and behavioral rules to organize passage and recovery from blockage in crowded spaces [7–14].

Construction tasks both generate motion demands and occupy or release surrounding motion space as work starts and ends. This coupling between work and motion spaces changes not only motion goals but also the spatial conditions supporting unfinished encounters. New occupancy may reduce the space available for velocity-based avoidance or invalidate paths, passage positions, and ordering arrangements. Detours then alter encounters with other agents, so a local repair must also connect to subsequent passage arrangements. Previous studies have addressed ongoing tasks, environment configuration, and dynamic navigation [2,5,15–17], but have given less focused attention to how work activation and completion interact with lifelong motion to change encounter conditions. Persistent blockage interrupts movement during construction activities. The resulting planning problem is **the construction and maintenance of local encounter passability in changing confined spaces**: establishing compatible passage arrangements from the available space, retaining valid relationships as conditions change, and reconstructing the affected parts.

We address this problem through **Width-based Planning with Compatible Encounter-Space Allocation (WP)**. Joint decisions focus on predicted body conflicts along current routes. Real-time width, body size, and relative motion determine passage arrangements, which shared channels and ordered anchors connect across multiple agents. New urgent encounters trigger updates to affected arrangements while valid relationships are retained; continuous, greedily guided routes remain available between encounters. A necessary and sufficient width condition for fixed-point exchange and a necessary directional condition for composite exchange establish the geometry underlying channel construction and encounter relocation. Compatible channels organize normal passage and provide space for temporal yielding and unidirectional clearing when entry is blocked. Constructing and maintaining passability thus become parts of one planning process.

Computational experiments on a crowded building floor with complex corridors and narrow entrances showed that WP continued trips toward their original goals after work occupancy changed and enabled passage and clearing in encounters involving two to five agents. Across multiple complete WP runs, near-stops accounted for 0.393% of navigating agent-steps; the longest continuous near-stop within a trip toward the same goal lasted 1.0 s. The encounter sequences and motion records demonstrate local blockage clearing in the crowded scenarios examined. In complete-run comparisons with four representative adapted methods, WP achieved a higher mean number of goal arrivals at every configuration of 5–10 agents. The increase over the best baseline at each configuration was 25.2%–52.2%, with larger relative gains at higher densities. Mean per-step computation time was lower than that of the rolling-horizon K-CBS adaptation. Together, these results show that WP combined sustained passage, efficient goal arrival, and computational economy under the tested occupancy changes and continuous motion demands.

The method combines group coordination with individual freedom to follow greedy routes, complementing representations based on grid positions, spatial partitions, and sampled connections [6,18,19] with joint structures generated from actual encounter demands. Geometric derivations establish the conditions for local construction, encounter records demonstrate continued motion, and complete-run comparisons provide system-level evidence. This architecture extends lifelong planning to work–motion space coupling and offers a methodological basis for engineering settings with both task occupancy and continual travel demands. It may also inform local collision-avoidance dependencies in dynamic configuration spaces, including those of robotic manipulators; the corresponding separation, reachability, and clearing conditions must be established for their kinematics.

## 2. Related work

Multi-agent pathfinding (MAPF) coordinates occupancy of shared positions and passage timing as agents travel between their goals. Spatial representations, conflict definitions, and optimization objectives jointly specify solution requirements [20]. For work–motion space coupling, we review three successive questions: how joint decisions can focus on necessary interactions, how local space supports mutual passage, and how passage relationships persist as motion and occupancy change.

### 2.1 Localizing joint decisions and representing space

Reactive collision avoidance adjusts velocities directly in response to neighboring motion. Velocity obstacles exclude relative velocities that can cause collisions [21]; Reciprocal Velocity Obstacles (RVO) distribute avoidance responsibility between agents, and Optimal Reciprocal Collision Avoidance (ORCA) formulates velocity selection with linear constraints [3,4]. Combined with obstacle-avoiding paths, these methods provide hierarchical navigation, although narrow spaces still require coordination of several bodies passing together.

Search methods coordinate paths and occupancy order in advance. Conflict-Based Search (CBS) adds constraints incrementally to eliminate path conflicts [22], while Priority-Based Search (PBS) compares precedence relationships between agents [23]. Graph resolution affects both geometric freedom and computational scale: coarse graphs may omit lateral room, whereas fine graphs increase search and update costs. Safe Interval Path Planning (SIPP) represents passage opportunities as collision-free time intervals [24], and Kinodynamic Conflict-Based Search (K-CBS) directly coordinates continuous trajectories [6]. These representations expand space–time choices but increase the burden of joint planning.

Local coupling jointly plans only agents or paths that affect one another. M* expands joint search as new interactions arise [25]; Experience-based Multi-Agent Path Finding (EMP) retrieves offline solutions for narrow-corridor conflicts, with reuse restricted by stored coordination patterns [9]; CDR-RRT guides continuous search using a passage skeleton and its capacity [7]. Interaction scope, existing solutions, and spatial structure each provide a way to reduce search.

New construction occupancy can restrict avoidance directions, invalidate connections, and require local coordination to be rebuilt. Online adjustment must therefore determine lateral positions and passage order from current clearance. Existing routes already encode connectivity, obstacle avoidance, and forward direction. Predicting simultaneous body conflicts along them identifies where joint decisions are needed and provides a basis for constructing body-placement choices from real-time geometry.

### 2.2 Conditions for spatial separation and encounter completion

Spatial-separation methods first assign admissible regions and then generate body motion. Buffered Voronoi Cells (BVC) subtract body margins from neighboring agents' partitions and use a right-hand rule to restore blocked motion [18]. This simplifies collision constraints, but local configurations still need adjustment when the partitions restrict progress toward a goal.

Safety constraints can combine with goal-directed motion to produce blocked equilibria [26]. IMPC-DR adjusts spatial separation over a receding horizon, excludes stable deadlocks under specified goal conditions, and analyzes recursive feasibility [11]. DLSC-CDPOP combines time-varying safe domains, neighboring-trajectory exchange, and distributed optimization, then applies rightward yielding and priority-based retreat to address blockage [13]. These studies connect body separation, sustained progress, and recovery; their respective guarantee conditions are distinct.

Passage rules must also fit the available geometry. Social-rule coordination uses door location, width, and entry or exit direction to prioritize departing agents [14]. The narrow-road passing method SM-NR searches for gaps based on width occupancy and selects passing positions and motion plans [27]. Roads have a defined longitudinal direction, whereas general planar encounters require current routes to determine the measurement direction.

When construction clearance changes, existing partitions, preferred directions, and capacities must be rechecked. Independent goals produce different motion directions, so a single capacity value cannot determine passage: opposing bodies require separated positions, same-direction bodies may follow one another, and side changes involve approach and departure. Slowing, turning, and waiting must leave others room to advance and clear; waiting longer cannot clear a channel blocked by stationary bodies. Local passability must therefore relate clearance, body size, relative motion, and temporal yielding to placement spacing, route continuation, and actual clearing.

## 2.3 Maintaining and reconstructing relationships under dynamic changes

Velocities, goals, and occupancy continue to change after passage arrangements have been established. Planning must respond while preserving unfinished coordination. Rolling-Horizon Collision Resolution (RHCR) decomposes continuing goal demands into finite-horizon conflict problems [5]. Each window's final state becomes the next window's initial state, so current arrangements must also account for subsequent release of bottleneck space.

Execution deviations need not require every relationship to be rebuilt. Homotopy-based coordination on fixed paths permits pauses or delays to adjust timing [28]. Switchable Action Dependency Graphs (SADG) reorder actions over a receding horizon and maintain execution under conditions that prevent circular waiting [29]. Distributed coordination can also propagate motion constraints across multi-task journeys [30]. These methods support retention of valid relationships while adjusting motion; for physical encounters, the dependencies must still be realized in shared passage space.

Environment updates concern different objects. Configurable-environment MAPF (C-MAPF) jointly selects layouts and routes [16]; path and action planning (PAPO) coordinates width, load, and directional constraints in construction transport [2]; dynamic construction navigation combines maps, real-time perception, and constrained planning to update motion [17]. Active configuration changes layouts, perception revises passage judgments, and work activation or completion directly changes occupancy.

Construction occupancy may invalidate avoidance positions or reduce parallel passage to sequential passage; new members, delays, and rerouting also change encounter urgency. Retaining dependencies requires their geometry to remain valid, and motion updates must respect other unfinished encounters. Replacing everything can disrupt valid arrangements, whereas unconditional reuse can retain invalid constraints. Planning therefore needs conditions for retaining, rebuilding, and releasing relationships, admitting new conflicts promptly and releasing space after bodies clear it.

## 2.4 Synthesis and research question

Existing research uses dynamic avoidance, local search, spatial separation, and dependency maintenance to support safe motion, blockage recovery, and continuous execution. Work–motion space coupling makes these requirements act simultaneously on unfinished encounters: new occupancy changes geometry, rerouting changes participants, and goals and actual progress change subsequent order. Restoring a single route must therefore be accompanied by connecting the agents' passage relationships; local arrangements also need executable entry and clearing conditions.

The resulting scientific problem is **the construction and maintenance of local encounter passability in changing confined spaces**. First, how do clearance, body size, and relative motion determine compatible arrangements and provide space for clearing when entry is blocked? Second, when space, membership, or motion changes, which relationships should be retained, rebuilt, or released, and how should subsequent encounters be connected? WP constructs compatible spatial allocations from real-time widths and maintains continuation through anchored routes and urgency-based updates. Experiments evaluate passage, sustained arrivals, and computational cost. We first establish spatial compatibility conditions, then use them to organize planning and multi-agent coordination.

## 3. Encounter Primitives and Spatial Compatibility Conditions

Two-body motion about an encounter point provides the basic construction. Connecting these motions relates body size and local space to encounter completion.

### 3.1 Scenario and motion constraints

Building components enclose connected activity regions. Workers operate near the components and travel between work locations and material storage areas. When work starts, its operating space is removed from motion space (Figure 1). Continuing construction and material demands generate motion goals; an agent receives its next goal after reaching the current one, producing lifelong multi-agent path planning. Agents are modeled as omnidirectional disks of equal radius $r > 0$. Agent $i$ has current position $x_i$, goal $g_i$, and nominal speed $v_i > 0$. At time $t$, the obstacle set $O_t$ contains fixed components and active work spaces. Excluding these occupancies and the exterior of the site leaves the closed physical space $\Omega_t$ available to bodies. Positions at least $r$ from obstacles and site boundaries form the center free space $F_t$. For $M$ agents, let $y_i(\tau)$ be an absolutely continuous candidate center motion in the current environment, with speed at most $v_i$. Here $\tau$ is elapsed time from planning time $t$, and $H > 0$ is the prediction horizon. Motion, waiting, and their combinations satisfy

$$\begin{cases} y_i(0) = x_i, \quad y_i(\tau) \in F_t (1 \le i \le M), \\ \| y_i(\tau) - y_j(\tau) \| \ge 2r \qquad (1 \le i < j \le M). \end{cases} \quad 0 \le \tau \le H. \tag{1}$$

Equation (1) specifies the initial position, obstacle clearance, and simultaneous body separation; $\|\cdot\|$ is the Euclidean norm. Waiting corresponds to $y_i(\tau) \equiv x_i$. During prediction, $H$ is determined by the routes and their timing. During execution, it is the motion step $\Delta t$, and the constraints are rechecked after environmental changes.

An **encounter** is a local motion process requiring coordinated passage because the current routes and timing predict a body conflict. The routes used for that prediction are termed the **source routes**.

Passage positions are planned with a clearance margin $b \ge 0$. The corresponding obstacle-clearance radius is $R = r + b$, and the safe two-body center distance is $d = 2r + b$. Within a closed, bounded local region $U$ for encounter $e$, define the admissible center set and the compatible two-body configuration set as

$$\begin{aligned} F_e(t) &= \{x \in U : x + B_R \subseteq \Omega_t\}, \\ Q_e(t) &= \{(x_i, x_j) : x_i, x_j \in F_e(t), \ \| x_i - x_j \| \ge d\}. \end{aligned} \tag{2}$$

$B_R$ is the closed disk of radius $R$ centered at the origin, and $x + B_R$ is its placement at $x$. The set $U$ bounds the analysis region, whereas $\Omega_t$ determines the available physical space. The quantities $R$ and $d$ constrain obstacle clearance and inter-agent separation, respectively. A **compatible passage** consists of bounded-speed, absolutely continuous motions that remain in compatible configurations and complete the prescribed approach, mutual passage, and body clearing. These planning margins govern local space and channel construction; actual body motions are checked using Eq. (1). Figure 1 illustrates the spatial occupancy and both distance constraints.

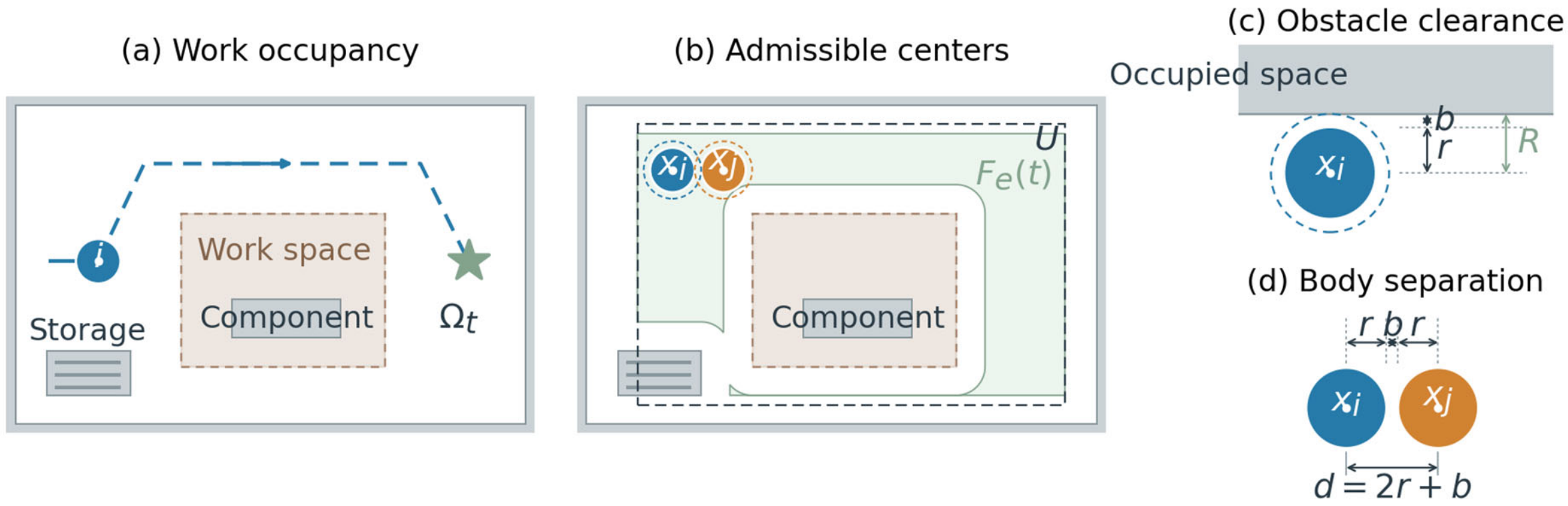

**Figure 1.** Scenario and common motion constraints. Work occupancy changes the physical space $\Omega_t$. Accounting for body size and obstacle clearance within $U$ gives the admissible center domain $F_e(t)$. The body radius is $r$, the planning obstacle-clearance radius is $R = r + b$, and the safe two-body center distance is $d = 2r + b$. Actual bodies remain separated under Eq. (1), while planned placements satisfy the margins in Eq. (2).

Continuous width measurement also requires an unobstructed connection between the bodies. In the current environment, this **mutual-visibility condition** is

$$[y_i(\tau), y_j(\tau)] \subseteq F_e(t), \qquad 0 \leq \tau \leq H. \tag{3}$$

The closed segment $[y_i(\tau), y_j(\tau)]$ joins the simultaneous centers. Equation (3) requires radius-$R$ obstacle clearance at every point on it. In a nonconvex space, the segment between two admissible centers may cross an obstacle. Continuous width constructions therefore use Eq. (3), while general two-body motions use Eq. (2). A strict clearance margin requires the center distance to exceed $d$ and each center's distance from obstacles to exceed $R$.

Environmental occupancy remains fixed within each step and is updated between steps. Newly occupied areas and newly introduced agents must satisfy the initial body-clearance constraints. Connections affected by changes in work occupancy are checked again.

## 3.2 Fixed-point encounter primitives and spatial conditions

In a **complete position exchange**, two agents move from $p, q$ to $q, p$, respectively, while remaining separated. A **fixed-point encounter primitive** is a compatible motion segment whose center-to-center segment always passes through the same fixed reference point. A complete exchange may use one primitive or a sequence of primitives. The environment is fixed below, and the admissible center domain and compatible configuration set are abbreviated as $F_e$ and $Q_e$. The required space remains available during the motion; environmental changes trigger replanning.

### 3.2.1 Fixed-point encounter primitives and their composition

Let $X = (x_i, x_j)$ denote a two-body configuration. Choose a reference point $a$ on the segment joining the centers. Let $\theta$ be the angle from $j$ toward $i$, with unit direction $e(\theta) = (\cos\theta, \sin\theta)$. Then

$$\begin{gathered} x_i = a + \alpha e(\theta), \qquad x_j = a - \beta e(\theta), \\ \alpha, \beta \geq 0, \quad \alpha + \beta \geq d, \quad X \in Q_e. \end{gathered} \tag{4}$$

The radial distances $\alpha$ and $\beta$ locate the centers of $i$ and $j$ relative to $a$; their sum is the center distance. A fixed-point encounter primitive keeps $a$ fixed while allowing $\theta, \alpha, \beta$ to change subject to Eq. (4). In Figure 2, both agents move while their center-to-center segment continues to pass through the same $a$. When one agent waits, its center can serve as the reference point.

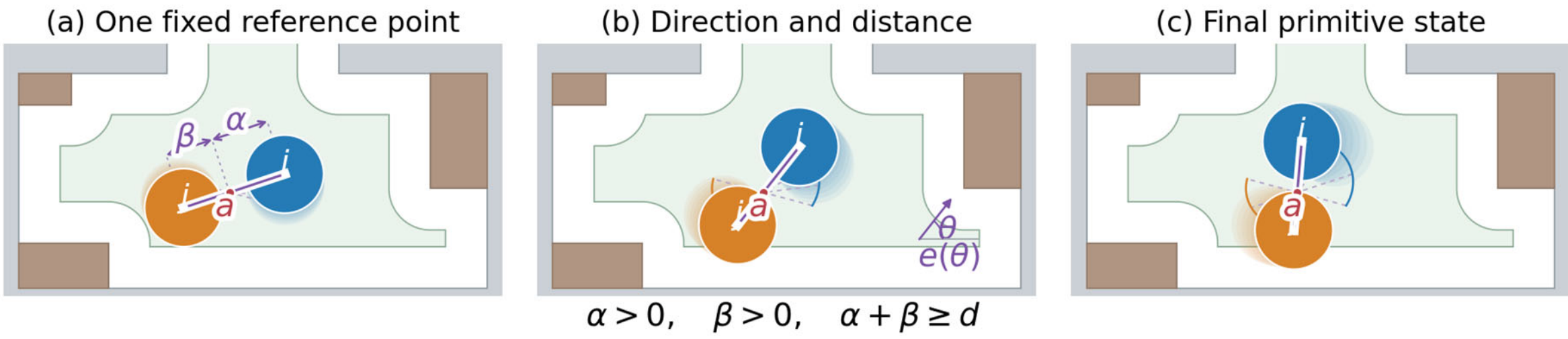

**Figure 2.** Fixed-point encounter primitive. Panels (a–c) show successive motion in the same wide region. The center-to-center line always passes through fixed point $a$, while its direction $\theta$ and the two radial distances $\alpha, \beta$ change. Blue and orange disks are current bodies, pale traces show recent motion, and thin solid lines are center trajectories.

Let $X_k(u)$ be the two-body configuration in primitive $k$, where $u \in [0,1]$ is the segment parameter and $a_k$ is its fixed reference point. Connecting the endpoint configurations of $m$ primitives gives a complete position exchange:

$$\Gamma = X_1 \to \cdots \to X_m, \qquad X_k(1) = X_{k+1}(0),$$
$$X_1(0) = (p, q), \qquad X_m(1) = (q, p). \tag{5}$$

$\Gamma$ is the composite motion, and the arrows denote sequential connection. The intermediate equality holds for $k = 1, \ldots, m-1$, and every segment satisfies Eq. (4). Reference points may change between segments while body positions remain continuous; the endpoint configurations complete the position exchange. Figure 3 illustrates this connection using rotations about three reference points.

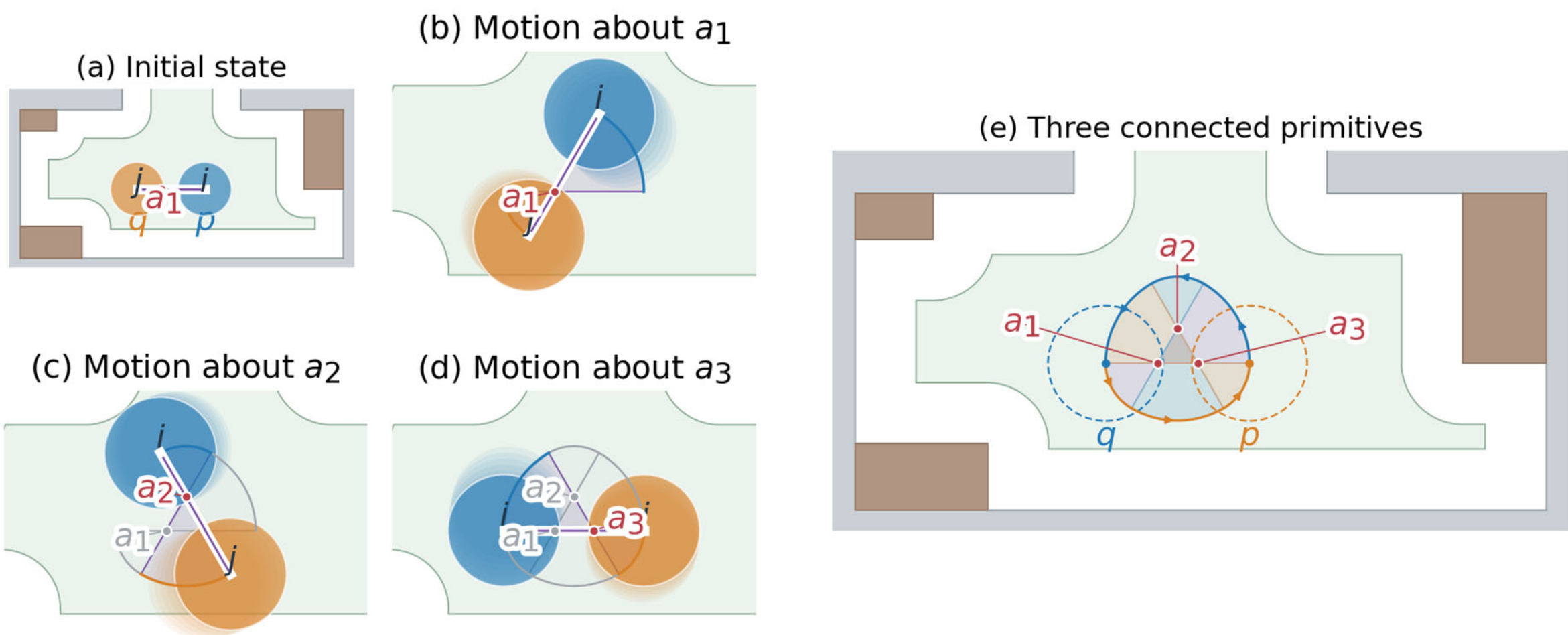

**Figure 3.** A complete exchange composed of fixed-point encounter primitives. Panels (b–d) highlight the active reference points $a_1, a_2, a_3$ in sequence; completed reference points, labels, and trajectories turn gray. Each pale sector extends from the active reference point to the two center trajectories and identifies the associated motion segment. Current bodies are darker than their traces. Panel (e) combines the three primitives; each initial position $p, q$ is finally occupied by the other agent. The sectors associate reference points with trajectories; safety is checked using the actual body sweeps.

**Lemma 1 (Representation by fixed-point encounter primitives).** Any feasible motion over a finite interval, with both centers remaining in the interior of the admissible center domain and strictly more than $d$ apart, admits a finite composition of fixed-point encounter primitives with the same endpoints. This composition approximates the original trajectories in motion order to arbitrarily small spatial error.

**Proof.** Locally, a two-body configuration changes through its center-to-center direction, center distance, and midpoint. Rotation about the midpoint changes the direction; radial adjustments change the distance and the along-line position. Two small opposite rotations generate transverse displacement. Let $c = (x_i + x_j)/2$ be the midpoint and $s = \| x_i - x_j \|/2$ the half-distance. Choose $a = c - \kappa e(\theta)$ inside the center segment, where $\kappa$ is a signed offset satisfying $|\kappa| < s$. First rotate by $\varphi$ with offset $\kappa_1$. Then select $\kappa_2 \neq \kappa_1$ relative to the updated midpoint and direction, and rotate back by $\varphi$. The net midpoint displacement is

$$\Delta c = (\kappa_1 - \kappa_2)[e(\theta + \varphi) - e(\theta)]. \tag{6}$$

Direction and separation are restored. The lateral displacement $(\kappa_1 - \kappa_2)\sin\varphi$ varies continuously with a small rotation angle; radial motion removes the accompanying longitudinal displacement. These primitives can therefore realize any sufficiently small change in the two-body configuration. A finite subdivision of the original path, followed by connection of neighboring configurations, gives a finite composition.

Lemma 1 represents complex two-body motion through connected primitives. Planning can construct locally compatible placements and then connect the motions before and after them, giving encounter relocation and multi-agent extension a common representation.

### 3.2.2 Width condition for a fixed-point encounter

Once the encounter point is fixed, the ability to maintain separation as relative direction changes depends on the center positions available in each direction. For $a \in F_e$, let $\ell$ denote the distance measured along a ray and define

$$\begin{aligned} h_a(\theta) &= \max\{\ell \geq 0 : [a, a + \ell e(\theta)] \subseteq F_e\}, \\ W_a(\theta) &= h_a(\theta) + h_a(\theta + \pi). \end{aligned} \tag{7}$$

The notation $[a, x]$ denotes a closed segment. The radial distance $h_a(\theta)$ measures how far a center can move continuously from $a$ along $e(\theta)$ before leaving the admissible domain. The sum $W_a(\theta)$ is the admissible center-chord width. Opposite directions define the same measurement axis, so $W_a(\theta + \pi) = W_a(\theta)$. The union of all continuous admissible center chords through $a$ is its **width space**. For the centers in Eq. (4) to lie in this space, their radial distances must satisfy $\alpha \leq h_a(\theta)$ and $\beta \leq h_a(\theta + \pi)$. a two-body placement with center distance at least $d$ exists in that direction if and only if

$$W_a(\theta) \geq d. \tag{8}$$

Equation (8), the **directional placement condition**, states that the center chord accommodates two safely separated positions. A complete fixed-point exchange must traverse every unoriented direction over a half-turn, with opposite directions treated as one measurement axis. The shortest center chord determines whether the point can support the exchange.

**Theorem 1 (Width condition for a fixed-point encounter).** Suppose $(p, q) \in \mathcal{Q}_e$, $[p, q] \subseteq F_e$, the reference point $a \in [p, q]$ lies in the interior of $F_e$, and the radial function $h_a$ is Lipschitz continuous in its angle. A single fixed-point encounter primitive can complete $(p, q) \to (q, p)$ within this width space if and only if

$$W_{\min}(a) = \min_{0 \leq \theta \leq \pi} W_a(\theta) \geq d. \tag{9}$$

$W_{\min}(a)$ is the **minimum center-chord width**, and Eq. (9) is the **fixed-point exchange condition**. Lipschitz continuity bounds changes in radial reach by a constant times the angular change. It yields continuous, finite-length constructed trajectories that can satisfy the speed limits by time rescaling. Figure 4 compares sufficient width in every direction with insufficient width in one direction at the same point.

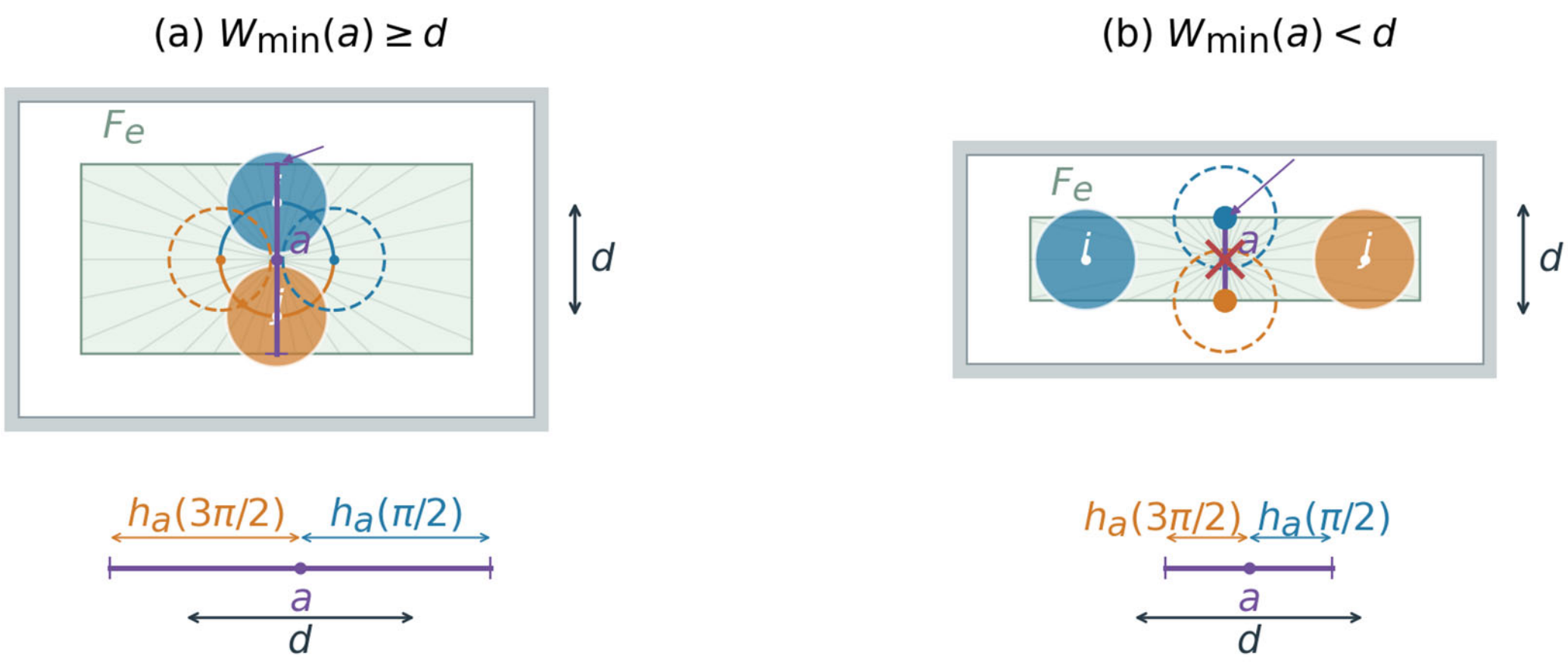

**Figure 4.** Fixed-point exchange condition. (a) Every direction satisfies $W_a(\theta) \geq d$, permitting a continuous half-turn. (b) The narrowest direction has $W_a(\theta) < d$, which the primitive cannot traverse; a complete exchange about that reference point is therefore impossible. Thick lines show the narrowest center chord, pale lines show other directions, and disks show the bodies. Center distance and chord width use the same scale.

**Proof.** For sufficiency, suppose Eq. (9) holds. Distribute the safe center distance $d$ in proportion to the available radial reaches:

$$\alpha(\theta) = \frac{d\, h_a(\theta)}{W_a(\theta)}, \quad \beta(\theta) = \frac{d\, h_a(\theta+\pi)}{W_a(\theta)}, \\ \alpha + \beta = d, \quad \alpha(\theta + \pi) = \beta(\theta). \tag{10}$$

Since $d/W_a \leq 1$, both centers lie on admissible rays and are separated by $d$. First interpolate the original placements along their initial rays to those in Eq. (10), then rotate continuously through a half-turn, and finally connect radially to the exchanged positions $q, p$. On both radial connections, the center distance lies between the original distance and $d$; during rotation, it remains $d$. After a half-turn, the radial distances exchange, completing an admissible motion to the swapped positions. The Lipschitz radial function gives a finite-length route, and extending the execution time satisfies the speed limits.

For necessity, the center distance is at least $d > 0$, so the center-to-center direction is always defined and its angle can be accumulated continuously along the motion. Exchanging the endpoint positions requires

$$e(\theta(T)) = -e(\theta(0)) \quad \Rightarrow \quad |\theta(T) - \theta(0)| \geq \pi. \tag{11}$$

$T$ is the completion time. Continuity forces the angle to cover every unoriented direction in a half-turn. In each direction, $d \leq \alpha + \beta \leq W_a(\theta)$, establishing Eq. (9) and completing the proof.

### 3.2.3 Directional condition for composite encounters

If one location cannot support the exchange, Eq. (5) allows different reference points to accommodate successive direction changes. Matching two-body endpoint configurations connect the segments. To define the spatial range in which these locations can be sought, introduce the single-body reachable set from $p$:

$$\mathcal{A} = \{a \in F_e : \exists\, \eta : [0,1] \to F_e,\ \eta(0) = p,\ \eta(1) = a\}. \tag{12}$$

$\eta$ is an absolutely continuous center route whose radius-$R$ disk maintains obstacle clearance. Any two points in $\mathcal{A}$ can be connected through their routes to $p$, so $\mathcal{A}$ describes the range of continuous single-body passage.

**Theorem 2 (Directional condition for composite encounters).** Any compatible motion completing $(p,q) \to (q,p)$, including a primitive composition under Eq. (5), must satisfy $q \in \mathcal{A}$ and

$$\forall \theta \in [0,\pi), \quad \exists\, a \in \mathcal{A},\ \ell \geq d: \quad a + \ell e(\theta) \in \mathcal{A}. \tag{13}$$

Equation (13) requires two admissible centers separated by at least $d$ in every unoriented direction within the same reachable set. The point $a$ is one endpoint and $\ell$ is the endpoint distance. If Eq. (3) holds throughout the motion, the continuous-width condition also follows:

$$\forall \theta \in [0,\pi), \quad \exists\, a \in \mathcal{A}: \quad W_a(\theta) \geq d. \tag{14}$$

**Proof.** During a safe exchange, the center distance remains nonzero and the center-to-center direction is continuous. By Eq. (11), exchanging the endpoint positions requires coverage of every unoriented direction over a half-turn. Agent $i$'s trajectory connects $p$ to $q$, giving $q \in \mathcal{A}$. Following that trajectory to $q$ and then agent $j$'s trajectory shows that all centers on both trajectories belong to $\mathcal{A}$. When a direction occurs, select one center as $a$ and the other as $a + \ell e(\theta)$, exchanging endpoints if necessary. Body separation gives $\ell \geq d$, proving Eq. (13). Under mutual visibility, the entire center segment is admissible and its length cannot exceed the chord width through $a$, yielding Eq. (14). Primitive compositions preserve endpoint continuity under Eq. (5) and are included among these safe exchanges.

Equation (13) specifies the separation directions that any exchange must cover; when the center segment remains admissible, the same proof yields Eq. (14). Connecting these directional placements into a complete motion additionally requires continuous two-body connections. For example, consider alternating waiting and moving. If, in segment $k$, the waiting center $a_k \in F_e$ is the reference point and the other body follows an absolutely continuous route $\eta_k$, then

$$\eta_k([0,1]) \subseteq F_e, \qquad \min_{0 \leq u \leq 1} \| \eta_k(u) - a_k \| \geq d, \qquad X_k(1) = X_{k+1}(0). \tag{15}$$

The first two conditions enforce the moving body's obstacle clearance and the inter-body distance; the last requires matching two-body endpoint configurations. When both bodies move, Eqs. (4) and (5) provide the connection. Figure 5 uses three narrow channels to illustrate the directional coverage in Eq. (13) and connections satisfying Eq. (15). First, $i$ waits at $a_1 = p$ while $j$ moves from $q$ in channel 2 to $a_2$ in channel 3. Next, $j$ waits while $i$ moves from $p$ in channel 1 to $q$ in channel 2. Finally, $i$ waits at $a_3 = q$ while $j$ moves from $a_2$ to $p$. The two-body configurations are $(p,q)$, $(p,a_2)$, $(q,a_2)$, and $(q,p)$. The relative angle $\theta$ changes successively from $\pi$ to $5\pi/4$, $7\pi/4$, and $2\pi$, so the three primitives complete a continuous half-turn exchange.

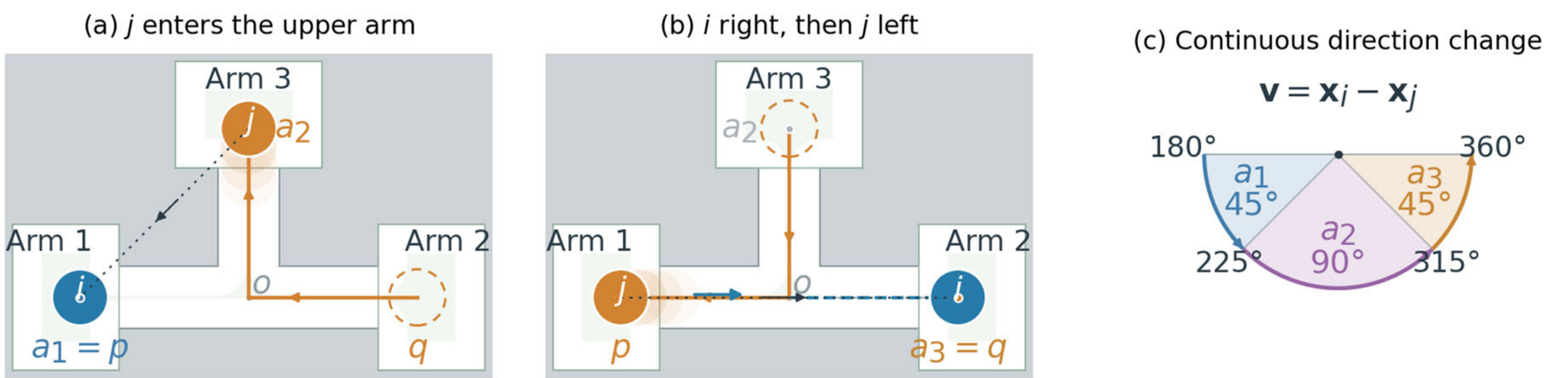

**Figure 5.** Complete position exchange through three reference points. (a) Agent $i$ waits in the left arm while $j$ enters the upper arm. (b) Agent $i$ first moves to the right arm; $j$ then moves from the upper arm to the left. Solid disks show final positions and dashed disks show intermediate states. (c) The relative direction continuously covers a half-turn. The three passages admit one body at a time, and $o$ is the central junction. Points $a_1, a_2, a_3$ contribute 45°, 90°, and 45°, totaling 180°. Dotted lines show relative center direction $v = x_i - x_j$. Each stage satisfies Eq. (15); adjacent stages have identical two-body endpoint configurations. Colors in the final panel identify each primitive's angular range.

Alternating waiting connects the directional changes into a complete exchange, illustrating the relationship between Theorem 2's directional condition and an actual continuous motion. Around corners, center segments may cross obstacles. The example is therefore checked using the general configuration constraints in Eq. (2) and the directional condition in Eq. (13). Equation (14) applies to the next section's width-space constructions, where the center segment remains continuously admissible.

Theorem 1 establishes the width required at a single exchange point, and Theorem 2 identifies the separation directions required across a composite exchange. Together, they guide the selection and relocation of encounter points. WP first constructs compatible channels at predicted conflicts and uses continuous routes to approach, pass, and connect successive encounters. If local space is insufficient, the encounter location is adjusted within the reachable region. Joint planning is thereby concentrated at the separation positions needed for passage, while individual route choices remain available between encounters.

## 4. WP planning with width spaces

Section 3 represents local motion through fixed-point encounter primitives and establishes spatial conditions for compatible passage. WP uses these results to organize planning around encounters: compatible channels are assigned at predicted conflicts, channel anchors connect continuous routes guided by greedy intentions, and the arrangements are updated as actual motion and work occupancy change.

The architecture proceeds from intention to encounter, channel assignment, route construction, and execution, with actual-state feedback closing the cycle (Figure 6). Goals and obstacles determine greedy intentions; the current implemented routes and arrival timing determine encounters. Real-time width measurements and source-route relations jointly determine channel assignments. Anchors specify required passage positions, while curve and velocity adjustments coordinate approach and departure. When approach is blocked, motion is cleared along the established channel directions. Actual motion determines anchor visits and body clearing, and provides the starting state for the next prediction.

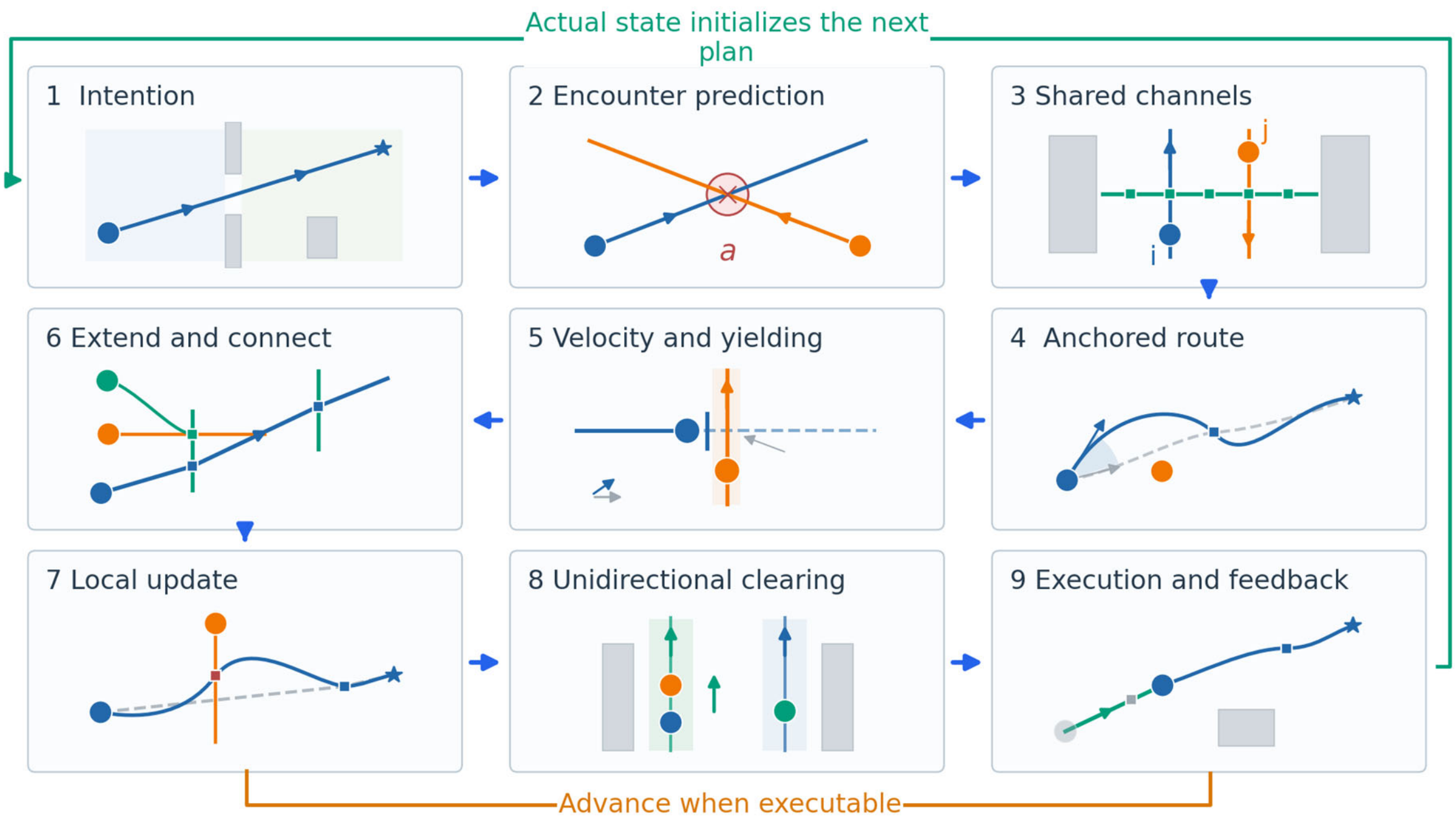

**Figure 6.** WP planning architecture and continuous feedback.

### 4.1 Intention: a greedy motion reference updated with position and space

Each agent first constructs a greedy path from the fixed enclosures, current work occupancy, and goal. Enclosures and opening cross-sections, including doorways and corridor entrances, partition the motion space into internally connected regions. A traversable opening between adjacent regions is a portal; its usable span accounts for body clearance. A topological graph uses regions as vertices and portal connections as edges. Estimated travel between the current position, representative portal positions, and the goal selects a region and portal sequence:

$$R_0 \xrightarrow{P_1} R_1 \xrightarrow{P_2} \cdots \xrightarrow{P_m} R_m. \tag{16}$$

The regions $R_0, R_m$ contain the current position $x_i$ and goal $g_i$, respectively; $P_j$ connects $R_{j-1}$ to $R_j$, and $m$ is the number of remaining portals. Equation (16) determines the portals and their order; within-region geometry determines the crossing positions. Activating or releasing work occupancy prompts checks of the internal connections and portal traversability.

Let $p_j$ be the crossing position on $P_j$, with $p_0 = x_i, p_{m+1} = g_i$. Each $p_j$ is initialized at the portal center and then adjusted using intersections between the portal and its preceding and following connections. If an intersection is obstructed, the nearest admissible portal point is selected. In region $R_j$, a visibility graph is built from $p_j, p_{j+1}$ and obstacle-boundary vertices offset for body clearance. Straight segments that permit body passage form the edges, weighted by length. Let $\mathcal{C}_j$ be the graph paths connecting the endpoints, and let $\ell(c)$ be the length of path $c$. The shortest connection within the region is

$$L_j \in \underset{c \in \mathcal{C}_j}{\operatorname{argmin}} \ell(c). \tag{17}$$

$j = 0, \ldots, m$, and argmin denotes the set of paths attaining the minimum length. A direct connection is used when feasible; otherwise, the path follows visible obstacle boundaries. Segments are connected in the order of Eq. (16), redundant vertices are removed, and turns are smoothed subject to clearance. Obstacle-avoidance adjustments preserve the portals and their order while updating the crossing positions $p_j$. If the straight line from the current position to the goal is feasible and traverses the selected portals in order, its portal intersections are used to preserve the passage direction. A goal in the same region requires only one connection.

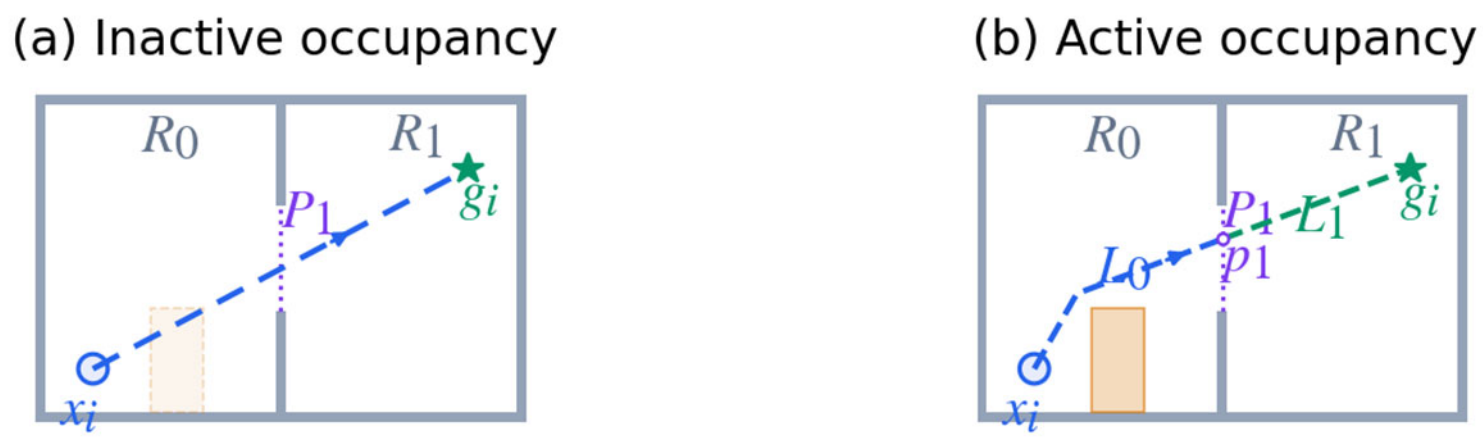

**Figure 7.** Intention under region and portal constraints. A fixed enclosure separates $R_0$ and $R_1$; portal $P_1$ connects them, and $p_1$ is the actual crossing position. (a) Without active work occupancy, the route connects $x_i$ directly to $g_i$. (b) After work occupancy is activated, the within-region connection avoids it and approaches the goal through the same portal. The connections on either side are $L_0, L_1$.

The greedy path formed by smoothing these local shortest connections subject to clearance is the **motion intention**. When a feasible connection exists between start $p$ and goal $g$, denote the intention at time $t$ by $\rho_t(p, g)$. Agent $i$'s current intention is

$$\rho_i = \rho_t(x_i, g_i). \tag{18}$$

The intention is updated with the start, goal, and available motion space. The portal order and crossing positions define the local problems in Eq. (17). Equal-length candidates are resolved in a fixed order, so identical inputs select the same intention.

Let $\gamma_i$ denote the planned route currently executed by agent $i$. Without encounter passage requirements, it follows the intention. When specific passage positions are required, they are connected in order, followed by the intention from the last position to the goal.

## 4.2 Encounter: route-timing prediction and a shared conflict point

When two routes pass through neighboring space, safe body passage also depends on arrival times. WP predicts simultaneous body contact from current routes, route progress, and velocity schedules to identify the local space requiring coordination.

Parameterize $\gamma_i$ by arc length, with total length $L_i$ and current progress $s_i \in [0, L_i]$, so that $x_i = \gamma_i(s_i)$. Arrival timing is calculated from segment distances, speeds, and waiting durations. Let $\tau \geq 0$ be prediction time from the present and $\ell_i(\tau)$ the additional distance traveled along the route. The predicted center is $\hat{x}_i(\tau) = \gamma_i(s_i + \ell_i(\tau))$, where $\ell_i(0) = 0$ and $0 \leq \ell_i(\tau) \leq L_i - s_i$. Progress stops increasing at the goal. Prediction assumes that goals, routes, and the environment remain unchanged over the considered interval. Its finite horizon is bounded by the remaining predictable route and the local prediction duration.

Within the common prediction horizon, the planner tests whether

$$\| \hat{x}_i(\tau) - \hat{x}_j(\tau) \| \leq 2r. \tag{19}$$

Equation (19) detects simultaneous contact or overlap of the predicted disks, triggering encounter planning for the pair. Prediction is partitioned at route-segment boundaries and velocity changes. Segmentwise relative motion gives the interval $I$ from first contact until separation or the end of the horizon. Select the time $\tau_*$ of minimum body distance within $I$ and define the conflict point $a$ as the midpoint of the two predicted centers:

$$\begin{aligned} \tau_* &\in \operatorname*{argmin}_{\tau \in I} \| \hat{x}_i(\tau) - \hat{x}_j(\tau) \|, \\ a &= \frac{\hat{x}_i(\tau_*) + \hat{x}_j(\tau_*)}{2}. \end{aligned} \tag{20}$$

Ties in Eq. (20) select the earliest time. The point $a$ is determined by simultaneous predicted centers and becomes the width-measurement origin. First-contact time, forward route distance, body positions at closest approach, and source routes support encounter ordering, relation classification, and width measurement. Figure 8 distinguishes geometric route intersection from simultaneous body contact.

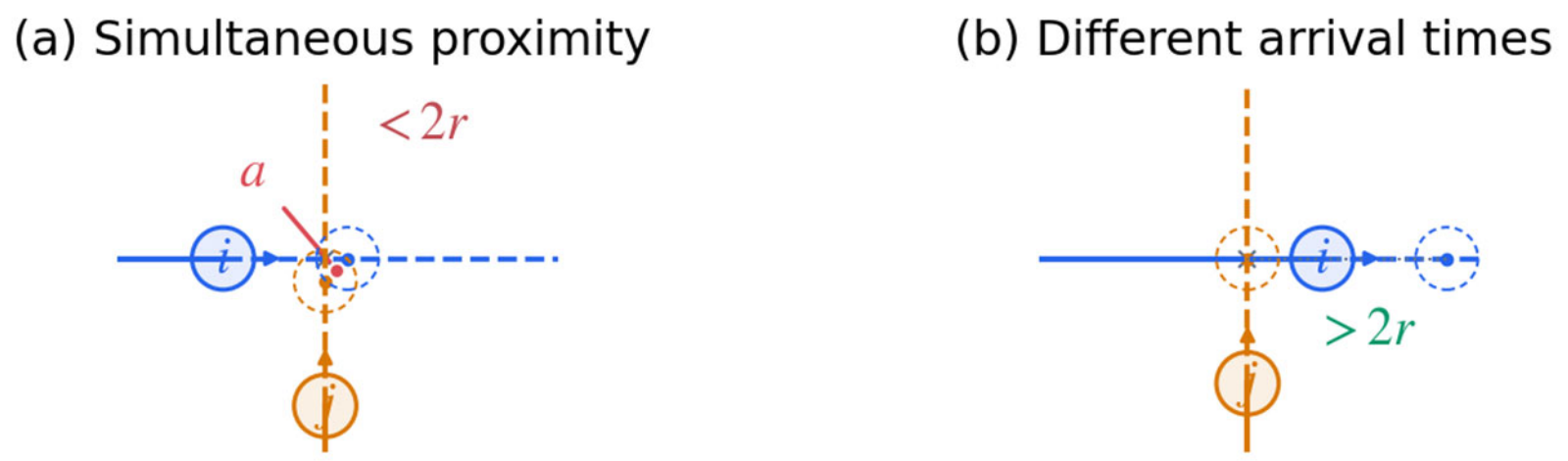

**Figure 8.** Route intersections and body encounters. Solid lines end at the current bodies. Colored dashed lines and dashed disks indicate predicted routes and bodies at the same future time. (a) At the closest approach within the first contact interval, the midpoint $a$ of the predicted centers is the measurement origin. (b) When $j$ reaches the geometric intersection, $i$ has already passed it, and the predicted bodies remain separated.

Prediction using current routes and expected along-route speeds is termed **nominal prediction**. Without existing encounter constraints, nominal prediction is evaluated first; speed perturbations supplement the candidates if no contact is found. For already coordinated routes, current timing is checked before the effects of speed changes are assessed. Each candidate retains mutually consistent contact times, body positions, and source routes, and candidates are filtered and ordered by risk weight.

Speed error affects predicted positions through route progress. Suppose both agents follow their predicted routes over the interval, starting from their known current positions, and their absolute along-route speed errors are bounded by $\varepsilon_i, \varepsilon_j$. Since the distance between two points on a route does not exceed the intervening arc length, after time $\tau$,

$$\begin{aligned} \| x_i(t+\tau) - \hat{x}_i(\tau) \| &\leq \varepsilon_i \tau, \\ \| x_i(t+\tau) - x_j(t+\tau) \| &\geq \| \hat{x}_i(\tau) - \hat{x}_j(\tau) \| - (\varepsilon_i + \varepsilon_j)\tau. \end{aligned} \tag{21}$$

$x_i(t+\tau)$ is the actual future position. Equation (21) first bounds position error by accumulated speed error and then applies the triangle inequality to lower-bound the actual center distance. If this lower bound remains at least $2r$ throughout the interval, the bodies remain separated, linking the valid prediction horizon to speed uncertainty.

Prediction is updated with actual progress. Changes in travel or speed due to collision avoidance trigger arrival-time recalculation. If actual motion leaves the route, a connection is rebuilt from the current position before encounter prediction resumes. Goals, routes, and changes in work occupancy likewise trigger recalculation. An agent blocking another agent's next motion step also defines an encounter through Eqs. (19)–(20). Immediate blockage and distant conflicts thus share a spatial representation, while stepwise body checks constrain execution.

## 4.3 Shared channels: from fixed-point encounters to compatible assignments

Conflict prediction identifies where coordination is needed, and Theorem 1 supplies the width condition for a fixed-point exchange there. WP measures the available space, constructs shared channels, and assigns passage positions according to the source routes' approach and departure relations. When the origin lacks sufficient space, alternative encounter points are probed.

### 4.3.1 Disk-based width measurement and the narrowest direction

A disk moving along a straight segment sweeps a capsule-shaped region. The measurement origin must satisfy $a \in F_e$, permitting placement of a radius-$R$ disk. Along $e(\theta)$, let $d_R(\theta) = h_a(\theta)$ and $d_L(\theta) = h_a(\theta + \pi)$ be the maximum continuous center distances before the first unavailable boundary on each side. The planning clearance envelope is

$$\bigcup_{-d_L(\theta) \leq u \leq d_R(\theta)} (a + ue(\theta) + B_R) \subseteq \Omega_t. \tag{22}$$

The union in Eq. (22) places a radius-$R = r + b$ disk along the center segment, producing a capsule-shaped envelope of thickness $2R$. The actual body sweep has radius $r$. The distances $d_L, d_R$ terminate at the end-cap centers, and their sum is the admissible center-chord width, already accounting for obstacle clearance. The capsule's full length additionally includes the radius $R$ at each end (Figure 9). If the conflict point lacks this clearance or the required admissible center interval, an alternative measurement point is selected under Section 4.3.4.

The fixed-point exchange condition in Eq. (9) tests the narrowest direction. Online computation generates candidates from a nonempty finite direction set $\Theta \subset [0, \pi)$ and selects

$$\theta_* \in \arg\min_{\theta \in \Theta} [d_L(\theta) + d_R(\theta)], \qquad w = e(\theta_*). \tag{23}$$

$w$ is the unit width axis. Direction ties are resolved first by angular relations to source-route tangents and then by a fixed direction order. An end with no obstacle detected within the observation range is marked as open. Equation (23) selects a candidate from finitely many directions, whereas Eq. (9) states the fixed-point exchange condition over all directions. Actual candidates undergo further disk-clearance and continuous-route checks.

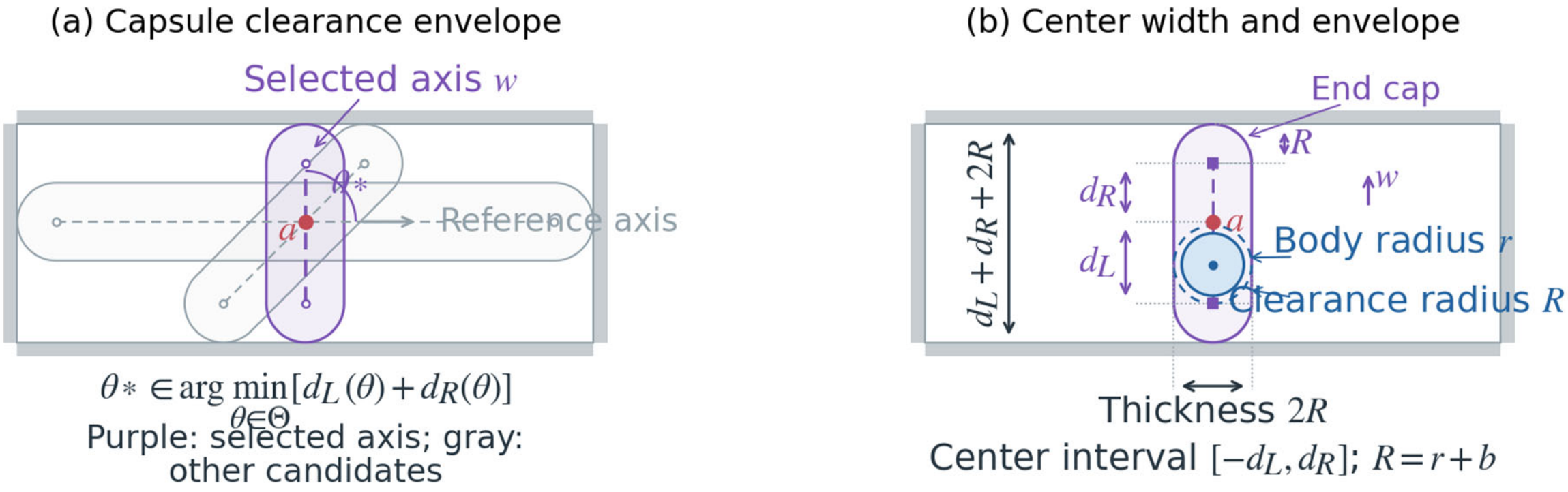

**Figure 9.** Disk-based width measurement. (a) Capsules in finitely many directions are compared at the same origin; $w$ is the selected axis. (b) The distances $d_L, d_R$ terminate at the end-cap centers. Both the end-cap radius and half-thickness are $R$; the body radius is $r$. Candidate centers are assigned directly within the available length $d_L + d_R$.

### 4.3.2 From width space to an encounter grid

Passage positions assigned to two agents may affect a third. WP therefore discretizes the measured admissible center interval into channel points, giving encounters at the same location a shared basis for body placement and lateral ordering.

For a continuous admissible center interval $[u_L, u_R]$, arrange channel points according to the safe center distance $d = 2r + b$:

$$N = 1 + \left\lfloor \frac{u_R - u_L}{2r+b} \right\rfloor, \qquad q_k = a + u_k w,$$
$$u_k = u_L + \frac{k-1}{N-1}(u_R - u_L) \quad (N \geq 2). \tag{24}$$

A full two-sided measurement gives $[u_L, u_R] = [-d_L, d_R]$. The integer $N$ is the maximum number of stationary bodies that fit simultaneously within the interval, $\lfloor\cdot\rfloor$ denotes rounding down, and $k = 1, \ldots, N$. For $N = 1$, the midpoint is used. Any $N$ ordered centers require a total adjacent spacing of at least $(N-1)d$, so Eq. (24) attains the maximum count. Equal spacing also maximizes the minimum adjacent distance.

The measurement origin, axis, and point set form an **encounter grid**:

$$h = (a, w, Q_h), \qquad Q_h = \{q_1, \ldots, q_N\}. \tag{25}$$

Choose separated channel points $q_k, q_l$ and a unit passage direction $n$ perpendicular to $w$. Over the range in which both axial center routes satisfy clearance, use longitudinal displacement $s$ to construct

$$a_{kl} = \frac{q_k + q_l}{2}, \qquad x_i(s) = q_k + sn, \quad x_j(s) = q_l - sn,$$
$$\| x_i - x_j \|^2 = \| q_k - q_l \|^2 + 4s^2 \geq d^2. \tag{26}$$

Because the center segment always passes through $a_{kl}$, Eq. (26) defines a fixed-point encounter primitive segment. Channel spacing maintains transverse separation; approach and departure routes determine this segment's direction changes and its connection to surrounding motion. A channel point assigned to an agent is its **anchor**. Figure 10 shows widths and candidate placements at different measurement points.

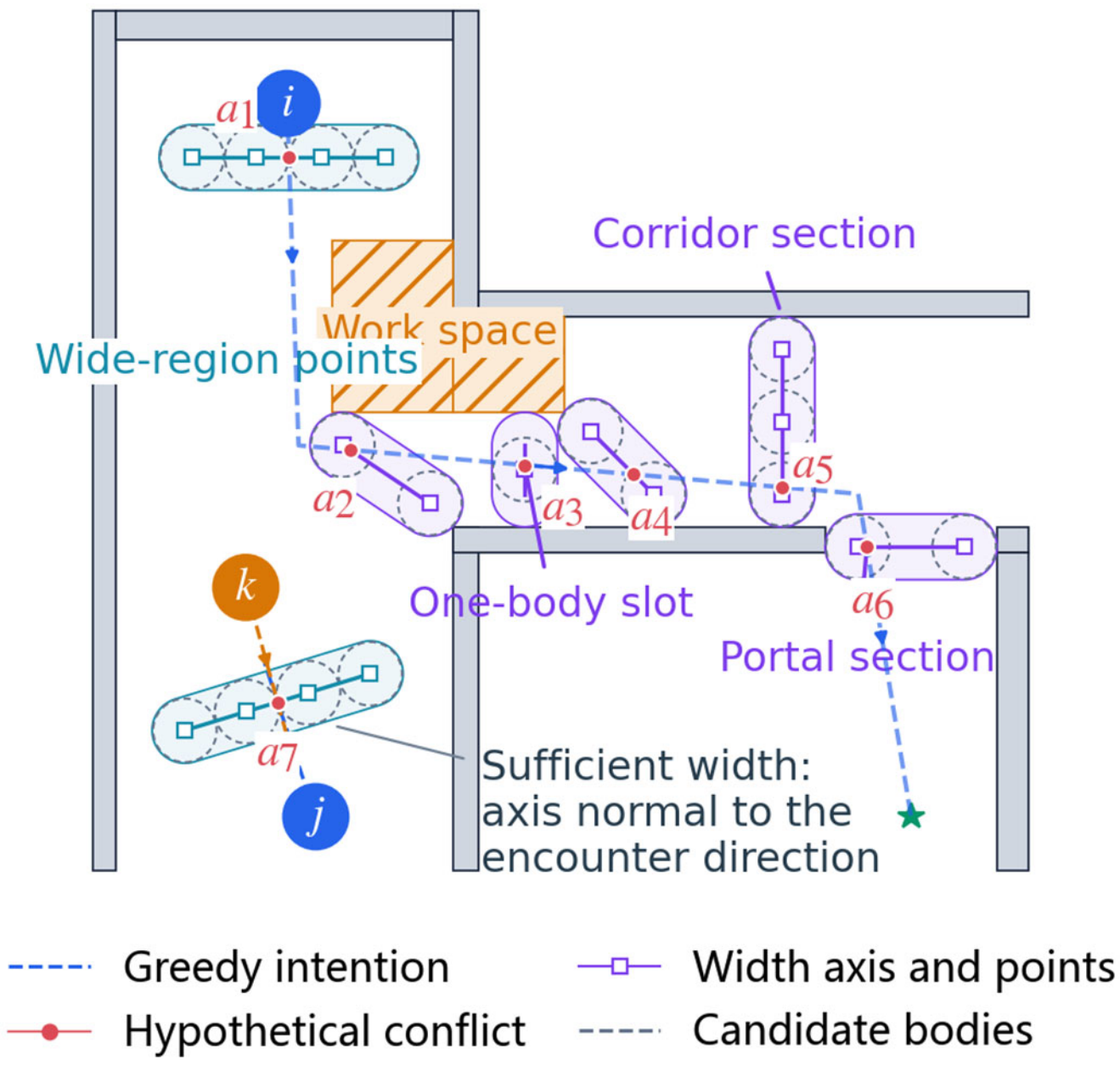

**Figure 10.** Local widths and shared channels. Points $a_1$–$a_6$ lie on an illustrative greedy intention and serve as hypothetical measurement origins; the planner creates grids only at detected encounters. Disk-sweep measurements select the narrowest candidate direction at $a_2$–$a_4$, with a vertical axis at $a_3$. Points $a_5, a_6$ correspond to corridor and portal sections. Squares are body centers; end caps and side envelopes indicate clearance. At wider point $a_7$, the width axis is perpendicular to the encounter direction, and disk clearance is rechecked. The encounter direction follows the unoriented bisector of the two source routes.

### 4.3.3 Compatible channel assignments for encounter relations

Channel use depends on how agents approach and leave the encounter. Same-direction following permits sequential passage through one point at a maintained distance. Opposing passage requires separate placements, and crossing departures also require lateral-order coordination. Figure 11 illustrates these relations using the source routes before avoidance.

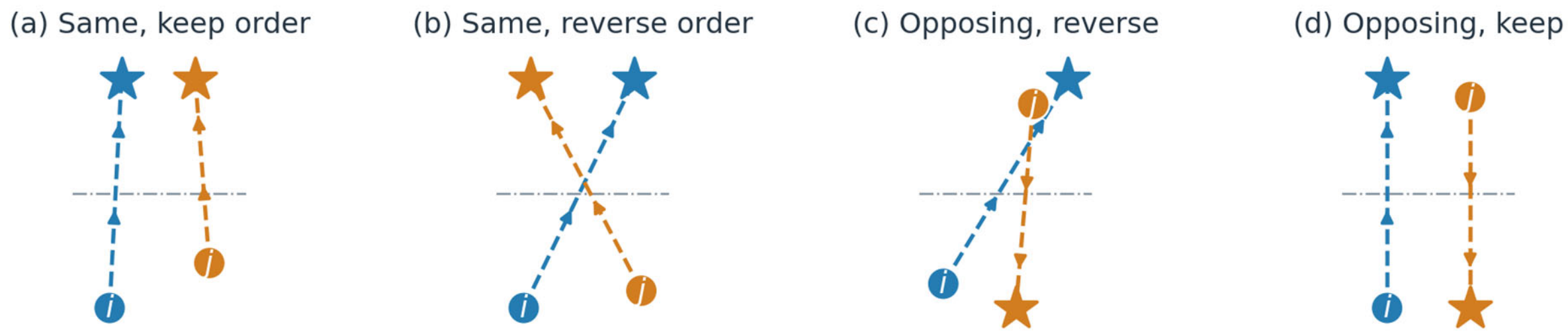

**Figure 11.** Four approach–departure relations of source routes. (a) Same-direction approach with preserved lateral order. (b) Same-direction approach followed by crossing departures. (c) Opposing approach with crossing passage. (d) Opposing approach with preserved lateral order. Disks show current bodies, dashed lines show source routes, and arrows and stars indicate motion directions and local departure positions.

Use the width axis through the conflict point as a common reference line and compare representative source-route positions before and after the encounter. Entry from opposite sides followed by crossing to the other side is **opposing**. Other local relations belong to the **same-direction** group, typically following and merging. Compare lateral order along the width axis: reversal is **side-changing**, and preservation is **side-preserving**. Reversing the axis changes only the left–right labels, leaving the classification unchanged. Motions containing multiple turns or reversals are treated as successive local encounters. After relocation of the measurement point, the original source routes still determine the encounter relation.

Distinct channel points separated by $d$ under Eq. (2) accommodate separated bodies. Same-direction agents may pass through one point after the leading body releases it. Opposing agents and wide-angle crossings that cannot form a following relation use separate points. Table 1 summarizes the four assignment rules.

**Table 1.** Four source-route encounter relations and admissible channel assignments. The symbols $S, O$ denote same-direction and opposing relations; subscripts 0 and 1 denote side preservation and side change. Section 4.5 uses these types to constrain joint assignments.

| Encounter type | Source-route relation | Channel assignment and passage |
|---|---|---|
| Same-direction, side-preserving $S_0$ | Following or merging with preserved lateral order | Sequential use of one point; preserve lateral order if using separate points |
| Same-direction, side-changing $S_1$ | Reversed lateral order within a same-direction relation | Sequential use of one point; preserve or reverse lateral order at separate points |
| Opposing, side-preserving $O_0$ | Entry from opposite sides and crossing to the other side with preserved lateral order | Two separated points with preserved lateral order |
| Opposing, side-changing $O_1$ | Opposing passage with reversed lateral order | Two separated points, with either preserved or reversed lateral order |

Side changes may occur at the channels or on approach and departure routes while preserving channel-side order. All connections undergo route and simultaneous-motion checks. Near a channel, the current lateral order is preferred to reduce crossing on approach (Figure 12).

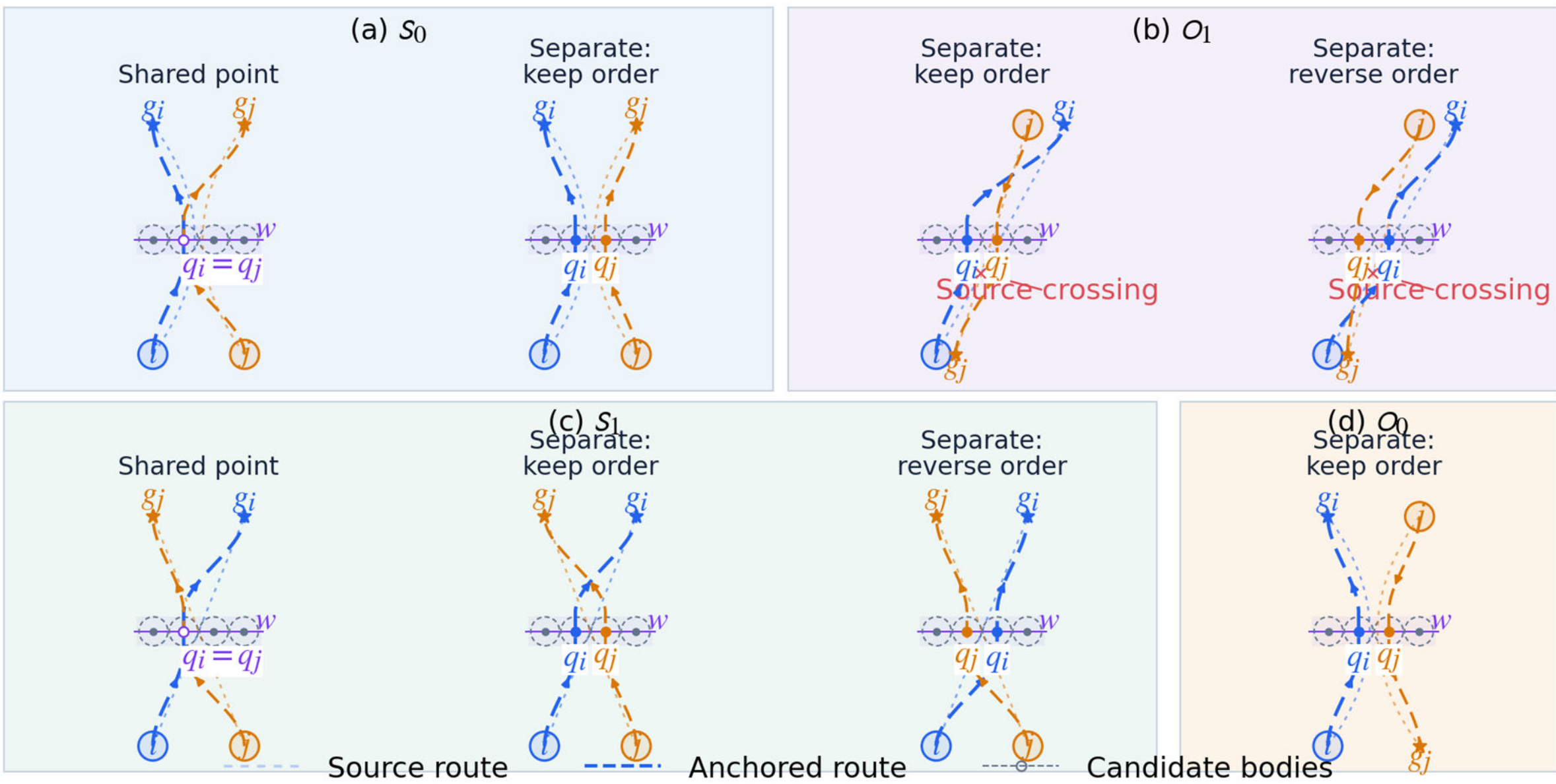

**Figure 12.** Four encounter types and admissible channel assignments. Pale bands identify candidate channel regions. Light dashed lines show source routes; dark dashed lines pass through anchors. Stars denote goals, and $q_i, q_j$ are channel anchors. (b) In an opposing, side-changing encounter, the separate-point choice determines whether the side change occurs before or after the channels. Shared points are used sequentially by the bodies.

### 4.3.4 Alternative encounter points when width is insufficient

Theorem 1 relates a complete fixed-point exchange to minimum center-chord width. Theorem 2 requires missing directional space to be available at reachable points along a composite motion. WP first seeks one alternative location that can organize passage, reducing joint search over multiple-reference-point compositions. During a local opposing passage, reversal of longitudinal order requires an instant of equal longitudinal position. The transverse center distance must then be at least $d$, so the candidate width axis must accommodate two separated bodies.

If the origin cannot provide two admissible separated points, the conflict point is projected onto a relevant agent's current intention. A bounded backward probe follows that intention, updating the measurement direction with the projection position. Channels are regenerated at each new measurement point, while the original conflict point retains the encounter's source identity. A candidate is accepted when both agents can approach from their actual positions and continue their departures. If none is found, the planner retains safe arrangements and continues updating them as space and encounters change (Figure 13).

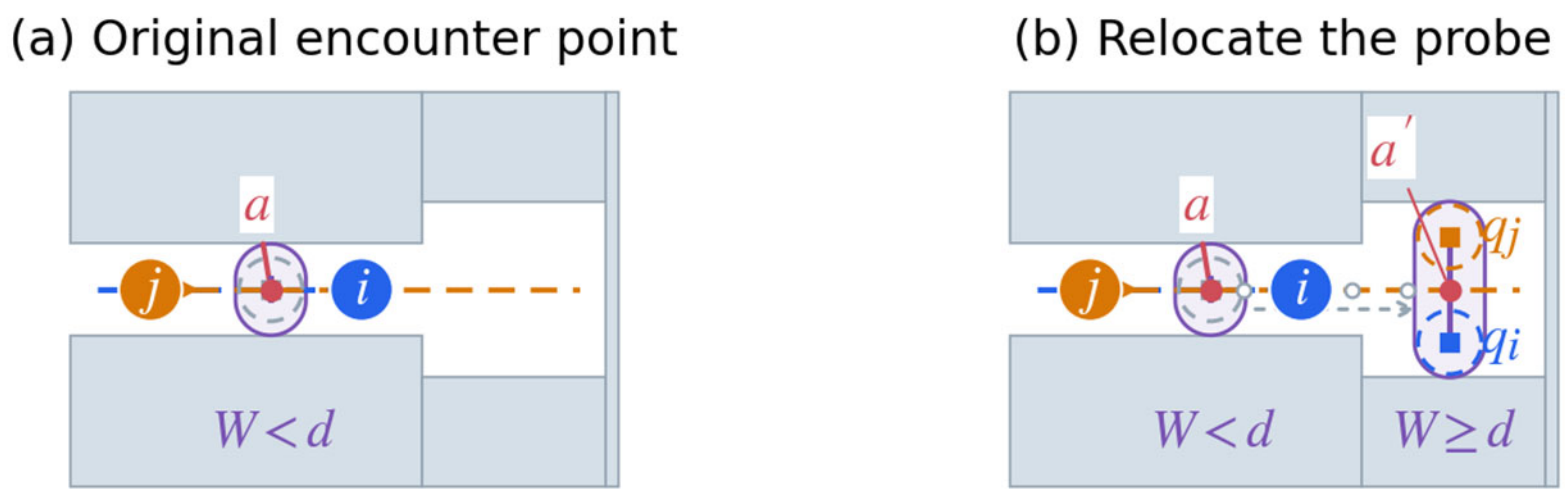

**Figure 13.** Probing from a narrow encounter point toward a wider location. (a) The available center width at the origin is less than $d$, providing only one channel point. (b) Backward measurement along agent $i$'s intention finds two separated candidates $q_i, q_j$ at $a'$. Purple capsules show planning clearance envelopes, transverse solid lines show admissible center intervals, and dashed disks show candidate bodies. Gray points and dotted lines represent geometric probing; actual body positions remain unchanged. The new location enters planning after approach and departure checks.

## 4.4 Routes and execution: spatial avoidance and velocity adjustment under anchor constraints

Channel assignments specify required passage positions, which WP connects into continuous routes. It first adjusts route shape while preserving encounter arrangements, then tests velocity deflection or waiting to coordinate the body motions. If continuation remains blocked, the relevant anchors are reassigned.

### 4.4.1 Anchored Routes and State Updates

Anchors comprise channel points and admissible original passage positions retained by the planner. Agent $i$'s $m_i$ remaining anchors form the ordered sequence $A_i = (q_{i,1}, \dots, q_{i,m_i})$. Starting at its current position $x_i$ when the route is constructed, the route connects these points to the goal $g_i$:

$$x_i \to q_{i,1} \to \cdots \to q_{i,m_i} \to g_i. \tag{27}$$

Shortest connections between consecutive required positions are found under the region, portal-order, and body-clearance constraints of Section 4.1. Merging consecutive duplicate positions gives vertices $p_0, \dots, p_K$, comprising the current position, anchors, obstacle-avoidance vertices, and goal; $K$ is the number of connecting segments. The final anchor connects to the goal through the current intention starting there. With no anchors, the intention starts at the current position.

Piecewise cubic Bézier curves smooth the connections. Fixed endpoints preserve required passage positions; nearby control points adjust shape, and adjacent segments share a tangent to maintain directional continuity. Let segment $k$ be $B_k(u)$, $u \in [0,1]$, with $B_k(0) = p_{k-1}$ and $B_k(1) = p_k$. Each segment lies entirely in disk-center free space $F_t$ and satisfies $B_k'(u) \neq 0$. Connecting the segments and parameterizing by arc length $s$ gives the implemented anchored route $\gamma_i(s)$. Let $S_k$ be the cumulative length of the first $k$ segments, with $S_0 = 0$ and total length $L_i = S_K$. Then

$$\gamma_i(s) = B_k(u), \qquad s = S_{k-1} + \int_0^u \| B_k'(v) \| \, \mathrm{d}v, \qquad S_{k-1} \le s \le S_k. \tag{28}$$

Here $v$ is the integration variable. The derivative $B_k'(v)$ is taken with respect to the segment parameter; its norm converts parameter increments into arc length. Denote the connection and smoothing construction by $G_t(x_i, A_i, g_i)$, with tangents, control lengths, and clearance conditions selected for the current connections. Local updates retain unaffected intervals and their boundary tangents, rebuilding only the modification interval defined in Section 4.5.2.

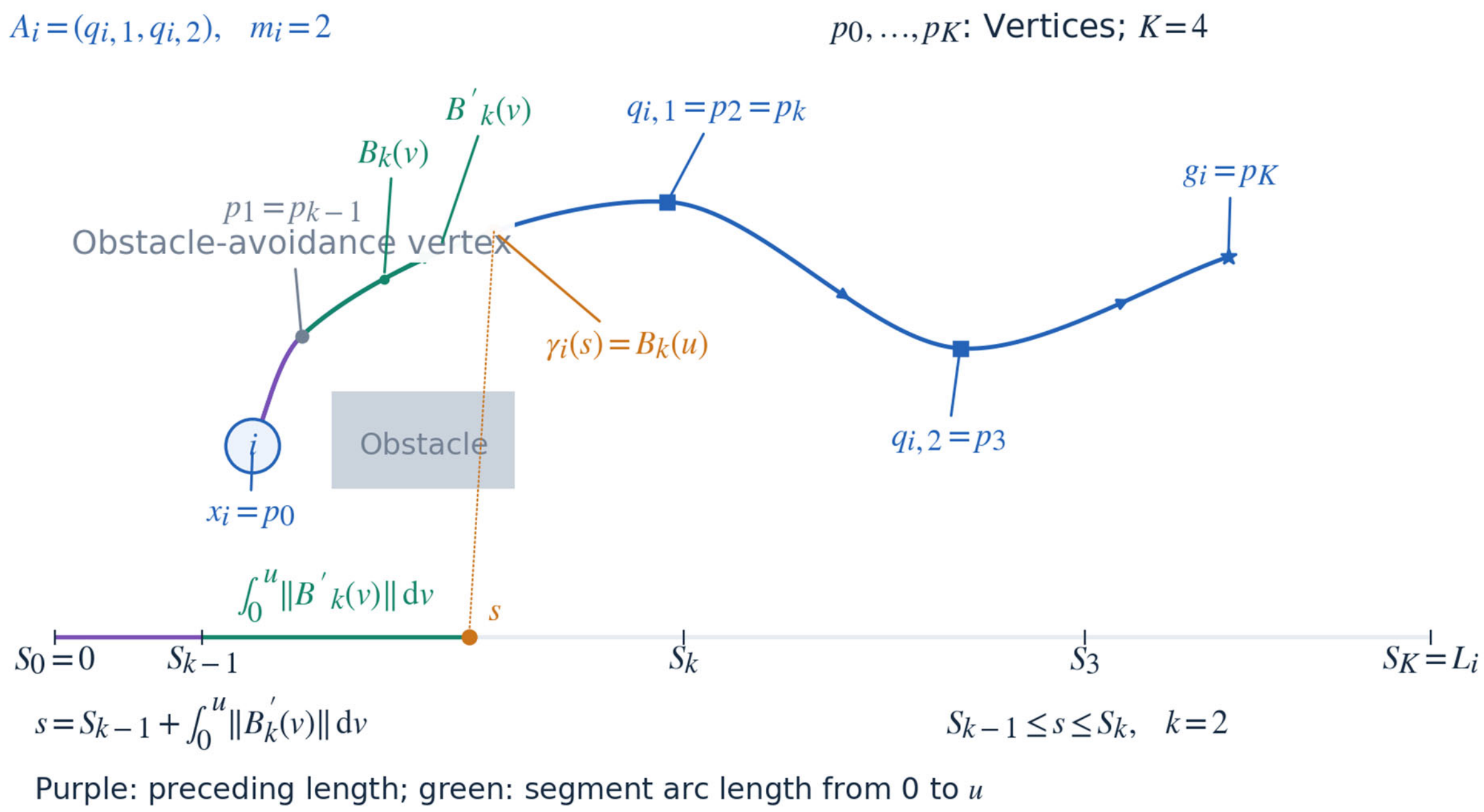

**Figure 14.** Anchors and route arc length. Squares, gray points, and stars mark anchors, obstacle-avoiding connection points, and the goal. The orange point on segment $k = 2$ represents both $B_k(u)$ and $\gamma_i(s)$. Adding purple cumulative length $S_{k-1}$ to the green within-segment arc length gives $s$.

Along-route motion changes only progress $s_i$; waiting preserves it. Reconnection from the actual position is required when velocity deflection moves the body off the route or when the curve, anchors, goal, or environment changes. Let $\eta_i = 0{,}1$ denote route retention and reconstruction, respectively, and use superscript $+$ for updated quantities:

$$(\gamma_i^+, s_i^+) = \begin{cases} (\gamma_i, s_i + \Delta s_i), & \eta_i = 0, \\ (G_{t^+}(x_i^+, A_i^+, g_i^+), 0), & \eta_i = 1. \end{cases} \tag{29}$$

The step progress is $\Delta s_i \in [0, L_i - s_i]$, with zero progress during waiting; $A_i^+, g_i^+$ are the updated anchors and goal. Replanning before execution uses $t^+ = t$ and leaves body positions unchanged. After execution, $t^+ = t + \Delta t$ and $x_i^+$ is the actual endpoint, where $\Delta t$ is the step duration. Visited anchors are removed from the sequence, but body-clearing requirements persist and the remaining original route may be retained. Successful reconstruction resets progress to zero. A failed candidate is discarded and adjustments continue. Prediction and execution thus use consistent routes and progress. Anchor insertion, replacement, and removal are described in Section 4.5.

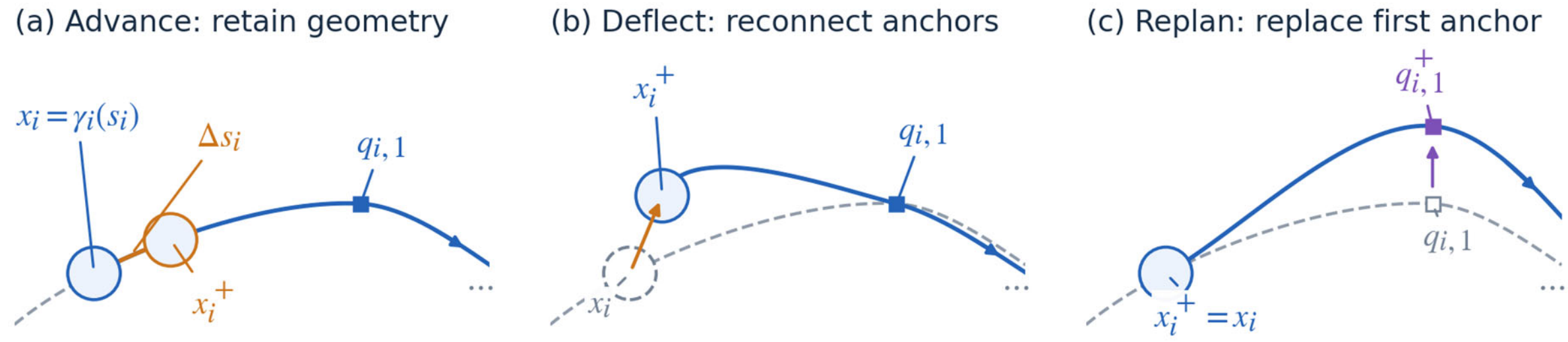

**Figure 15.** Updates to the implemented anchored route. (a) Advancing along the route preserves geometry; waiting preserves progress. (b) After velocity deflection, the original anchors are reconnected from actual endpoint $x_i^+$. (c) Replacing the first anchor within a planning step connects the updated sequence from the unchanged body position. Both reconstructions reset progress to zero. Arrows and ellipses indicate subsequent anchors and the goal. The goal is unchanged in all three cases.

### 4.4.2 Route avoidance with fixed anchors

Separating channel points does not eliminate simultaneous contact along approach and departure routes. WP first retains anchors and visit order, adjusting connection tangents and control lengths to change passage directions and turning extent. Let $d_j$ be the unit tangent at vertex $p_j$. For segment $k$, endpoint-to-control-point distances are $a_k, b_k$, and the control points are $p_{k-1} + a_k d_{k-1}$ and $p_k - b_k d_k$. At start $p_0 = x_i$, the reference tangent is the direction of the shortest initial connection satisfying region and portal constraints; it points to the first anchor when that anchor is directly reachable. Other vertices retain the current route tangents as references. Each tangent rotates by $\delta_j$ from its reference and satisfies

$$\begin{gathered} |\delta_j| \le \bar{\delta}_j \; (j = 0, \dots, K), \\ 0 < a_k, b_k \le \bar{h}_k (k = 1, \dots, K). \end{gathered} \tag{30}$$

The direction bound $\bar{\delta}_j$ is zero when the direction is fixed. The control-length bound $\bar{h}_k$ depends on adjacent lengths and clearance. The rotated unit tangent is still denoted by $d_j$. Start and end tangents affect the first and last segments, while each internal tangent couples its two adjacent segments. A finite set of candidates preserves anchors, visit order, assigned lateral relations, and unaffected connections. Obstacle clearance and simultaneous body separation are rechecked using Eq. (1).

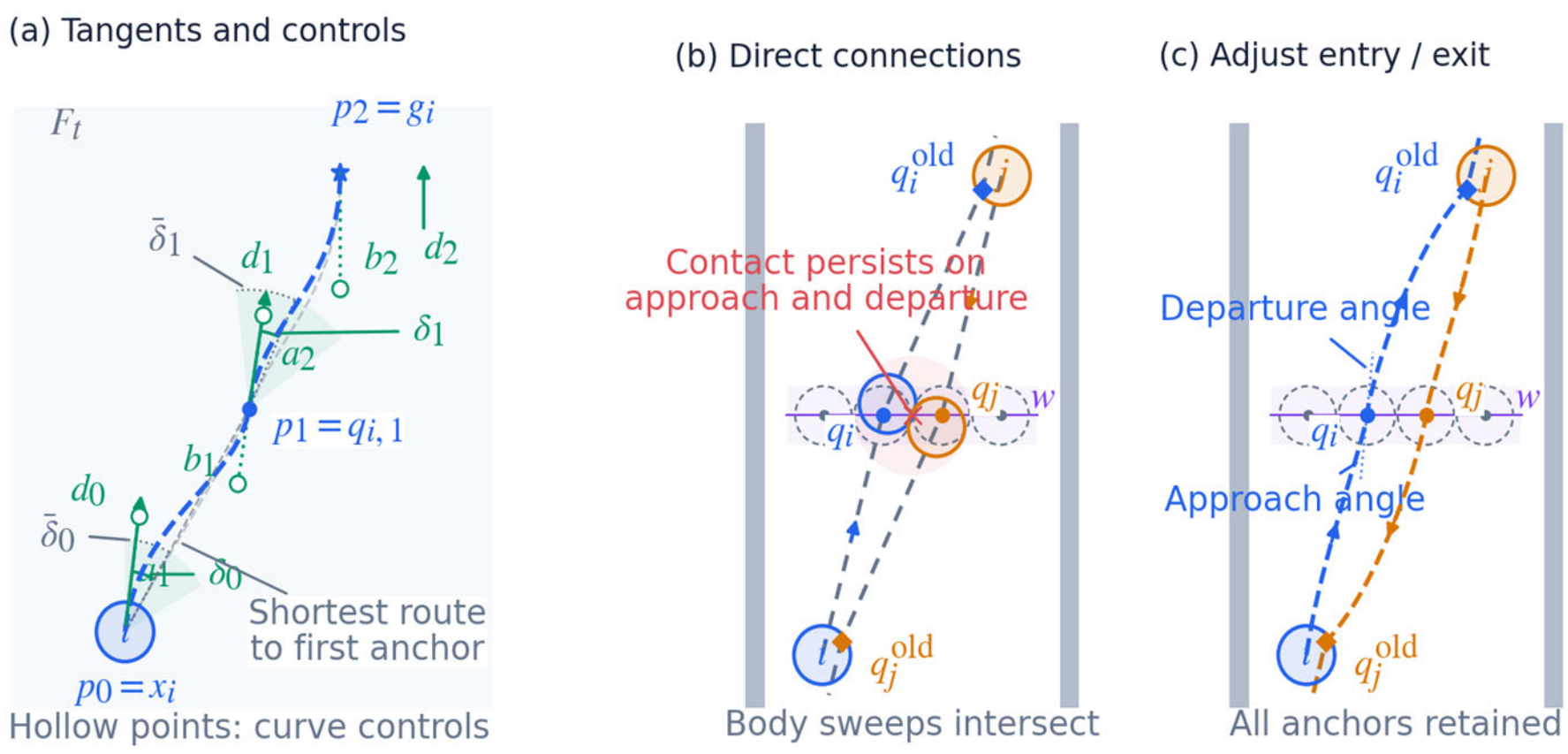

**Figure 16.** Curve adjustment with fixed anchors. (a) The points $p_0 = x_i$, $p_1 = q_{i,1}$, and $p_2 = g_i$ are fixed. Four hollow control points are determined by $a_1, b_1, a_2, b_2$ and $d_0, d_1, d_2$. The angle $\delta_0$ is relative to the shortest connection from the start; $\delta_1$ is relative to the anchor reference tangent. Pale sectors show allowed ranges. (b, c) With identical endpoints and anchors, curve adjustments change near-term simultaneous body separation.

#### 4.4.3 Velocity deflection

When route adjustment still provides no safe near-term motion, velocity is adjusted. WP uses a finite set of RVO-inspired angular candidates when the route tangents near the encounter are approximately aligned or opposed. Let their angle be $\theta_{ij} \in [0, \pi]$. Velocity deflection is permitted when $\min\{\theta_{ij}, \pi - \theta_{ij}\} \le \bar{\theta}$, where $0 \le \bar{\theta} < \pi/2$ is an angular tolerance. Other oblique crossings are handled by curve adjustment, waiting, or anchor replanning. With current route unit tangent $t_i$ and its left-normal unit vector $n_i$, construct

$$\begin{aligned} \boldsymbol{v}_i^R &= v_i(\cos\alpha_i\, t_i + \sin\alpha_i\, n_i), \qquad |\alpha_i| \le \bar{\alpha}_i, \\ z_i &= x_i + \Delta t\, \boldsymbol{v}_i^R. \end{aligned} \tag{31}$$

$v_i$ is nominal speed, superscript $R$ marks an RVO candidate, $\alpha_i$ is its deflection angle, and $\bar{\alpha}_i$ is the angle bound. The deflection side follows the encounter's lateral relation, accounting for both progress toward the anchor and body separation. Candidate motion $y_i(\tau) = x_i + \tau \boldsymbol{v}_i^R$ must satisfy Eq. (1) and permit reconnection from endpoint $z_i$ to the remaining anchors with the candidate velocity direction. Joint prediction with the other agents' current motions checks near-term contact over $H \ge \Delta t$. For the actual step, all body sweeps are checked with $H = \Delta t$. Reconnection follows Eq. (29) after execution; if the actual endpoint equals the checked $z_i$, the checked continuation is reused.

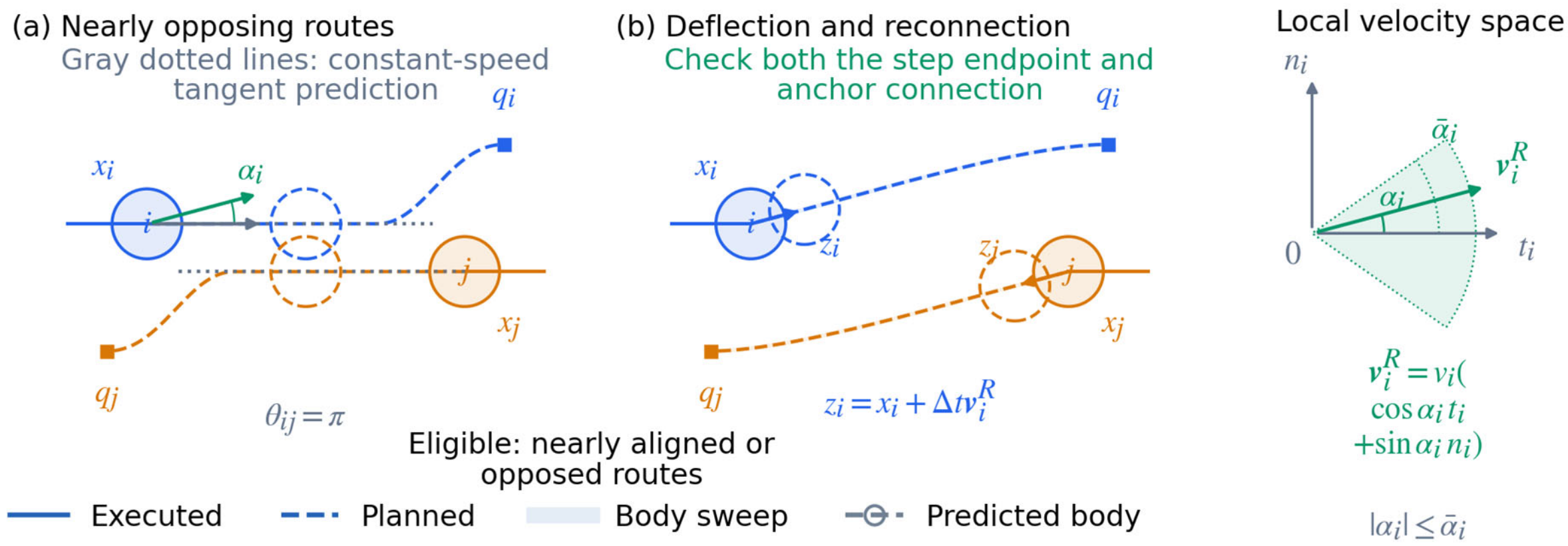

**Figure 17.** Encounter angle and velocity deflection. (a) Locally opposed routes have $\theta_{ij} = \pi$, and advancing along them would cause contact. Gray dotted lines show constant-speed predictions along the original tangents. The local velocity diagram locates $\boldsymbol{v}_i^R$ relative to $t_i, n_i$; the pale green sector indicates $|\alpha_i| \le \bar{\alpha}_i$. (b) An accepted candidate reconnects the original anchors from $z_i = x_i + \Delta t \boldsymbol{v}_i^R$, and similarly for $j$. Pale bands and dashed disks show body sweeps and predicted positions. Nearly aligned motion uses the same eligibility condition.

#### 4.4.4 Waiting to yield

If the encounter angle is ineligible, the required deflection exceeds its bound, or velocity candidates cannot provide both safety and continuation, the planner tests whether one agent waiting while the other moves can separate their passage times. Let $\chi_i = 0,1$ denote waiting and along-route advancement:

$$y_i(\tau) = \gamma_i(\min\{s_i + \chi_i v_i \tau, L_i\}), \qquad 0 \le \tau \le \Delta t. \tag{32}$$

Equation (32) caps progress at the route endpoint $L_i$. A waiting agent remains at $\gamma_i(s_i)$, with its route, anchors, and progress unchanged. The waiting body is checked together with all proposed moving bodies under Eq. (1). When its position leaves room for the other agent to pass, the separated-point or ordered shared-point arrangement can continue. Waiting is rechecked at every step and ends when along-route motion becomes safe.

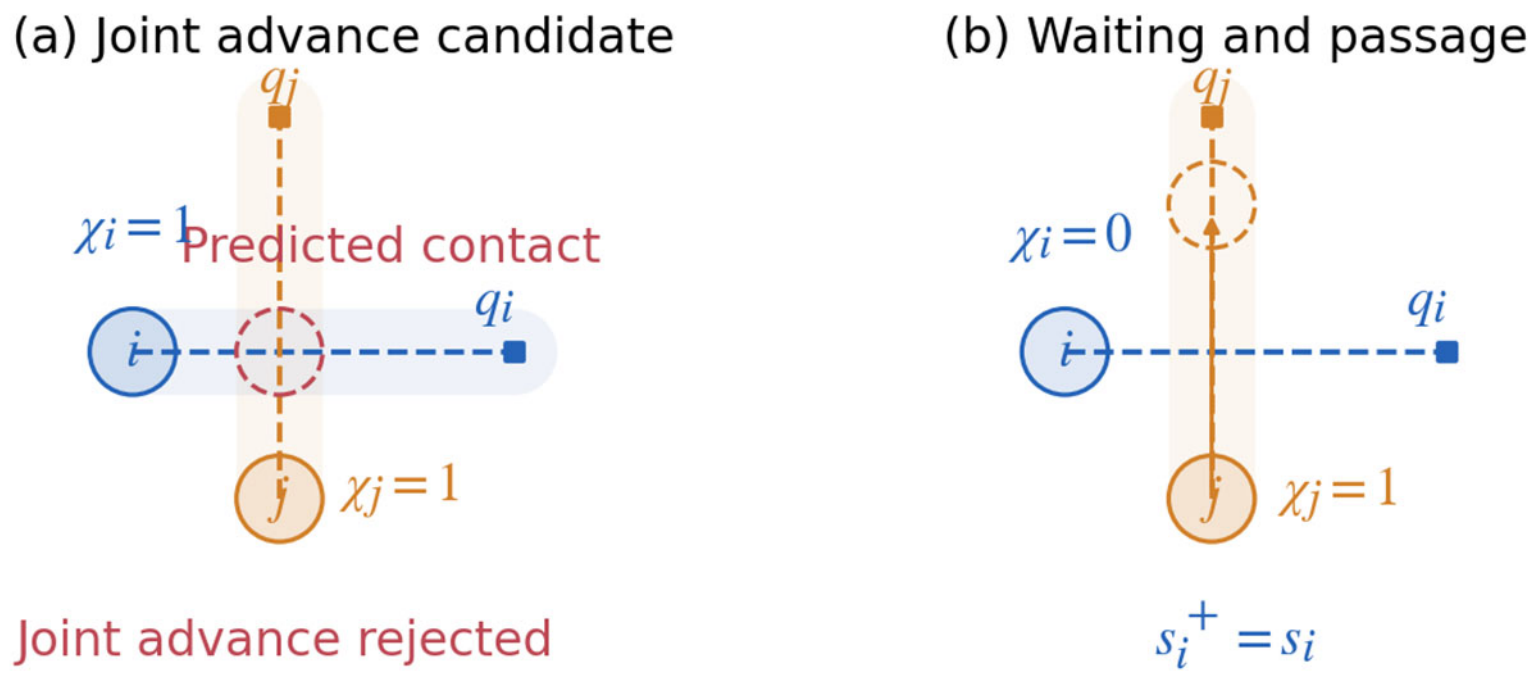

**Figure 18.** Waiting and route progress. (a) The choice $\chi_i = \chi_j = 1$ predicts body contact and is rejected. (b) Agent $i$ waits with $\chi_i = 0, s_i^+ = s_i$, while $j$ advances with $\chi_j = 1, s_j^+ = s_j + v_j \Delta t$, capped at the endpoint by Eq. (32). Both panels retain the same anchors and routes. Pale bands show body sweeps; the waiting body lies outside the other's passage region.

If waiting still fails to restore continuation, the shared arrangement is adjusted under Section 4.5.2 and coordination continues through the algorithm in Section 4.5.5.

## 4.5 Multi-agent coordination: shared channels and successive encounters

An avoidance route for two agents may affect a third, and a new participant may change an existing encounter. WP detects these conflicts along the current implemented routes. Encounters that can share channels are coordinated on one grid; successive encounters are connected through anchor sequences. Each route change triggers new prediction, prioritizing conflicts that obstruct current continuation while retaining other valid arrangements.

### 4.5.1 From a new conflict to a shared encounter grid

Let $\mathcal{P}$ be the set of conflicting pairs detected from current routes under Section 4.2. Conflicts that prevent continuation in the current step take priority. The rest are ordered by the smaller of the two agents' remaining travel distances to the predicted encounter, from near to far, with ties resolved by fixed agent identifiers. After goals, occupancy, or actual motion change, routes are first updated using Eq. (29), then $\mathcal{P}$ is recalculated. Agents with no predicted conflict continue along their routes.

A new conflict first attempts to reuse a nearby encounter grid's point set. If none is suitable, width is measured again under Section 4.3. Sweeping an existing grid $h$'s admissible center segment with the clearance radius $R$ defines its coverage:

$$D_h = \bigcup_{u_L \le u \le u_R} (a + uw + B_R). \quad (33)$$

The origin $a$, axis $w$, and interval $[u_L, u_R]$ come from Section 4.3, and $B_R$ is the radius-$R$ disk. $D_h$ is a capsule of thickness $2R$. A new conflict point $a' \in D_h$ makes $h$ a reuse candidate. Its points must retain disk clearance, permit an assignment under Table 1, and connect both agents' routes. Grids already used by either agent are tried before other nearby grids. If none is suitable, width is measured anew; insufficient width leads to relocation of the measurement point under Section 4.3.4.

When the conflicting pair reuses one grid, it selects channels jointly with the agents whose encounters there remain incomplete. Let $M_h$ be this member set. For $|M_h| \ge 3$, a two-agent encounter extends into a multi-agent encounter on the shared grid. Grids with different directions may remain separate, with their approach and departure routes coordinated by subsequent conflict detection (Figure 19).

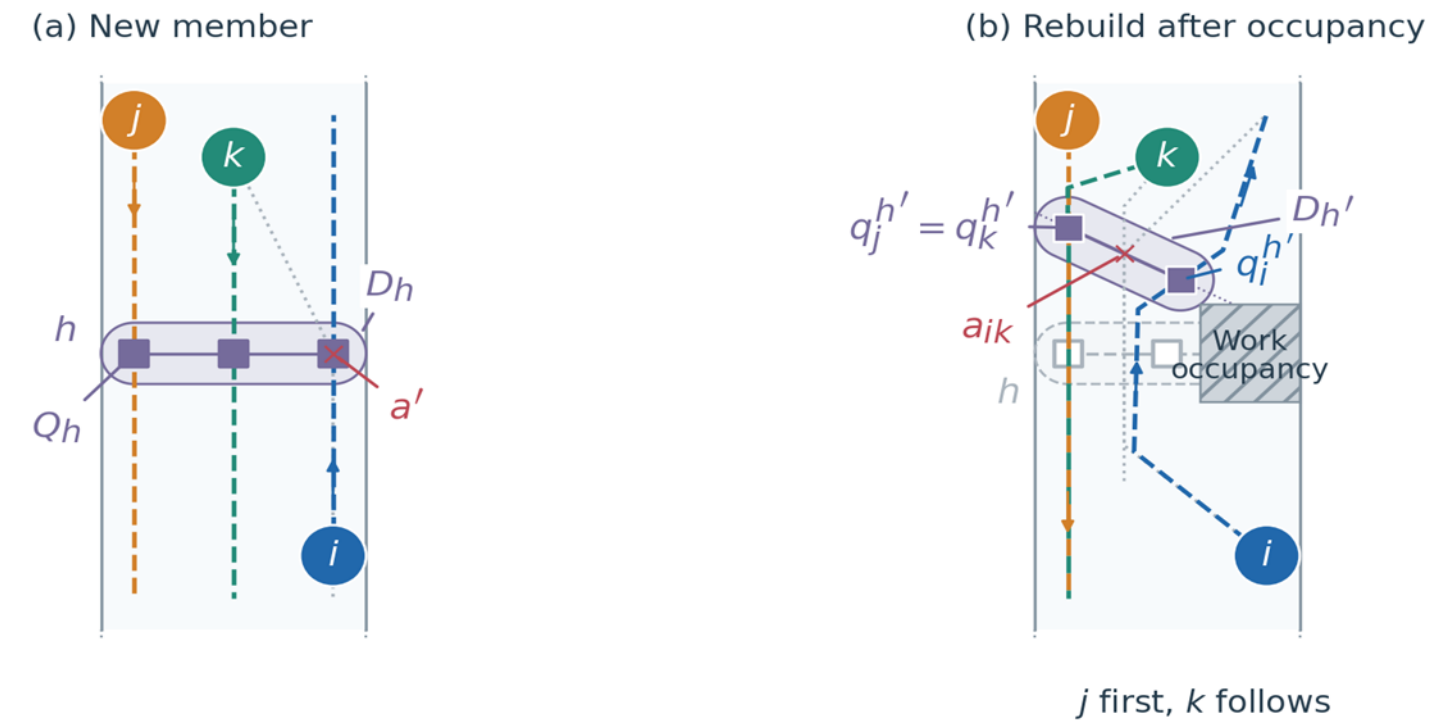

**Figure 19.** Formation of shared channels. (a) The original point set still supports the new encounter, so new and existing members are assigned jointly. (b) After work occupancy intrudes, width is remeasured at the new conflict point $a_{ik}$ between $i$ and $k$. The width axis extends from the left boundary to the upper edge of the occupancy. The new grid has two channel points: $j, k$ use the left channel sequentially, and $i$ uses the right one. Capsules show the coverage in Eq. (33); gray dashed lines show previous routes or grids.

### 4.5.2 Joint assignments and anchored route connections

**Joint assignment on one grid.** Let $q_i^h$ be member $i$'s passage position for its current visit to grid $h$. A new position is selected from $Q_h$. The original route's crossing of the measurement axis may be retained if it satisfies the same clearance, spacing, and Table 1 relations. A channel compatible with the existing arrangement is sought for a new member before relevant original positions are changed.

When a member participates in several pairwise encounters during the same visit, they share its passage position. Let $C_{ij}^h(q_i^h, q_j^h)$ be the admissibility test in Table 1: it equals 1 if the position pair satisfies the rules for the source-route encounter type and 0 otherwise. Joint assignment requires

$$C_{ij}^h(q_i^h, q_j^h) = 1, \qquad i, j \in M_h, \quad i < j. \tag{34}$$

The condition $i < j$ checks each pair once. Opposing encounters and crossings that cannot form a following relation use separated points. Side-preserving relations retain the original lateral order. Same-direction relations permitting following may share a point. Distinct positions remain at least $d = 2r + b$ apart, as ensured for new grid points by Eq. (24). Shared points are visited in front-to-back order, with along-route spacing and velocity checked under Section 4.4.

When reassignment is necessary, feasible continuations with smaller changes and shorter detours are preferred. Bodies already passing through remain in the occupancy checks, and subsequent assignments reserve space for them to clear.

**From shared positions to individual routes.** Insert $q_i^h$ into the anchor sequence $A_i$ of Eq. (27) in visit order. A visit involving multiple pairwise encounters creates one anchor. A later visit to the same grid creates a separate entry, distinguished by the sequence index in $q_{i,k}$. Each anchor records its encounter and participants so that local route changes identify the shared arrangements requiring rechecking.

The **insertion order** of a new anchor $q$ in $A_i$ is determined by its closest projection onto the remaining route:

$$s_q = \min \operatorname*{arg\,min}_{s_i \le s \le L_i} \| \gamma_i(s) - q \|. \tag{35}$$

The inner argmin gives the arc lengths of closest projections, and the outer min selects the earliest. Let $k$ be the number of existing anchors whose visit arc lengths do not exceed $s_q$. Then

$$A_i^+ = (q_{i,1}, \dots, q_{i,k}, q, q_{i,k+1}, \dots, q_{i,m_i}), \qquad 0 \le k \le m_i. \tag{36}$$

$m_i$ is the original number of remaining anchors. The cases $k = 0$ and $k = m_i$ insert at the beginning and end, respectively. Existing anchors with equal arc length retain their order, and the new point follows them. Reassigning a passage position removes the original visit entry before inserting the replacement according to its projection. Other anchors retain their relative order. Each agent orders its anchors using its own remaining route.

If a new anchor lies between retained endpoints $u, v$, replace $u \to v$ by $u \to q \to v$. Replacement or removal likewise rebuilds only adjacent connections. The connection and update rules in Section 4.4.1 preserve the remaining route and its boundary tangents; the modification scope expands only if local reconnection fails. If the change affects other encounters' anchors or connections, their member assignments are rechecked. A change is adopted only if it satisfies Table 1 and admits route continuation. Each agent's anchor sequence thus links the grids.

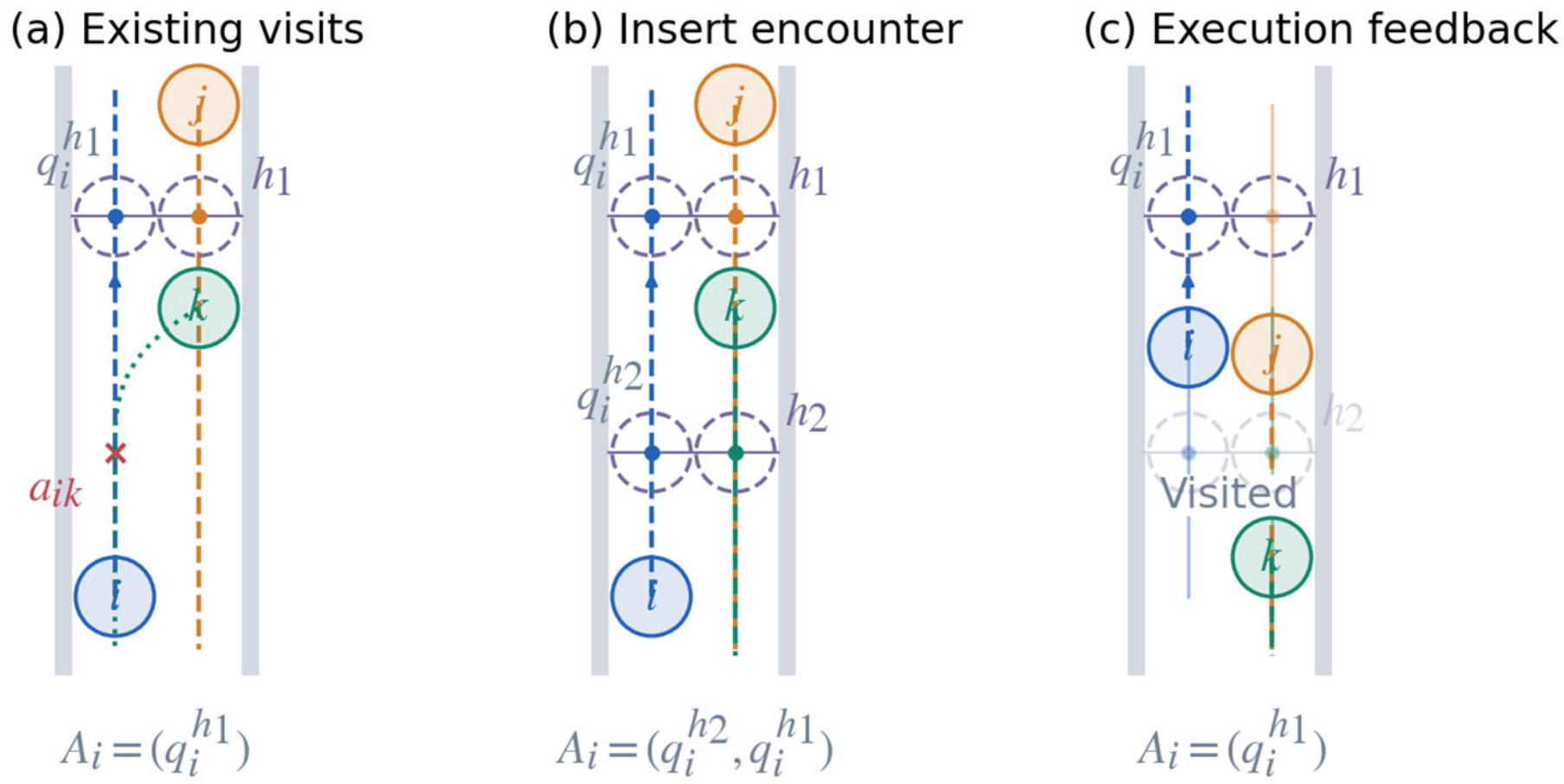

**Figure 20.** Route connections across successive encounters. The points $q_i^{h_1}, q_i^{h_2}$ are the current passage positions on grids $h_1, h_2$. Inserting a new point before an existing one rebuilds only the affected connections. After the point is visited, its anchor is removed; its occupancy is released after the body clears.

### 4.5.3 Conflict updates after rerouting and feedback from execution

**Prediction after rerouting.** After channel positions or anchor connections change, Eqs. (19)–(20) recalculate contacts between affected agents and all others. Arrival-time changes caused by upstream rerouting or speed adjustment propagate to downstream segments. If a candidate cannot connect, other admissible assignments or measurement points are tried.

Rerouting may introduce an earlier encounter. If either agent will contact the other before its own first anchor, channels and a new anchor are arranged under Sections 4.5.1–4.5.2. If both agents' contact positions lie beyond their first anchors, the current connections are executed first and detection continues with actual progress. If either agent has no remaining anchor, conflicts within its current prediction range are handled directly. Every executed step remains subject to the body checks in Eq. (1).

**Consistency after execution.** Along-route motion accumulates progress; waiting preserves it. Velocity deflection requires reconnection from the actual position to the remaining anchors under Eq. (29). Updated routes and arrival times are then used for conflict prediction. When a goal change or work occupancy invalidates a connection, invalid future arrangements are removed and affected connections rebuilt, while other valid arrangements remain.

A visited anchor is removed from the required-point sequence, but its occupancy persists until the entire body leaves the encounter region. Goal changes and anchor reassignments likewise retain checks for bodies that have not yet cleared. A member leaves $M_h$ after completing all its passage requirements there and clearing the region. The grid is removed when both membership and actual occupancy end.

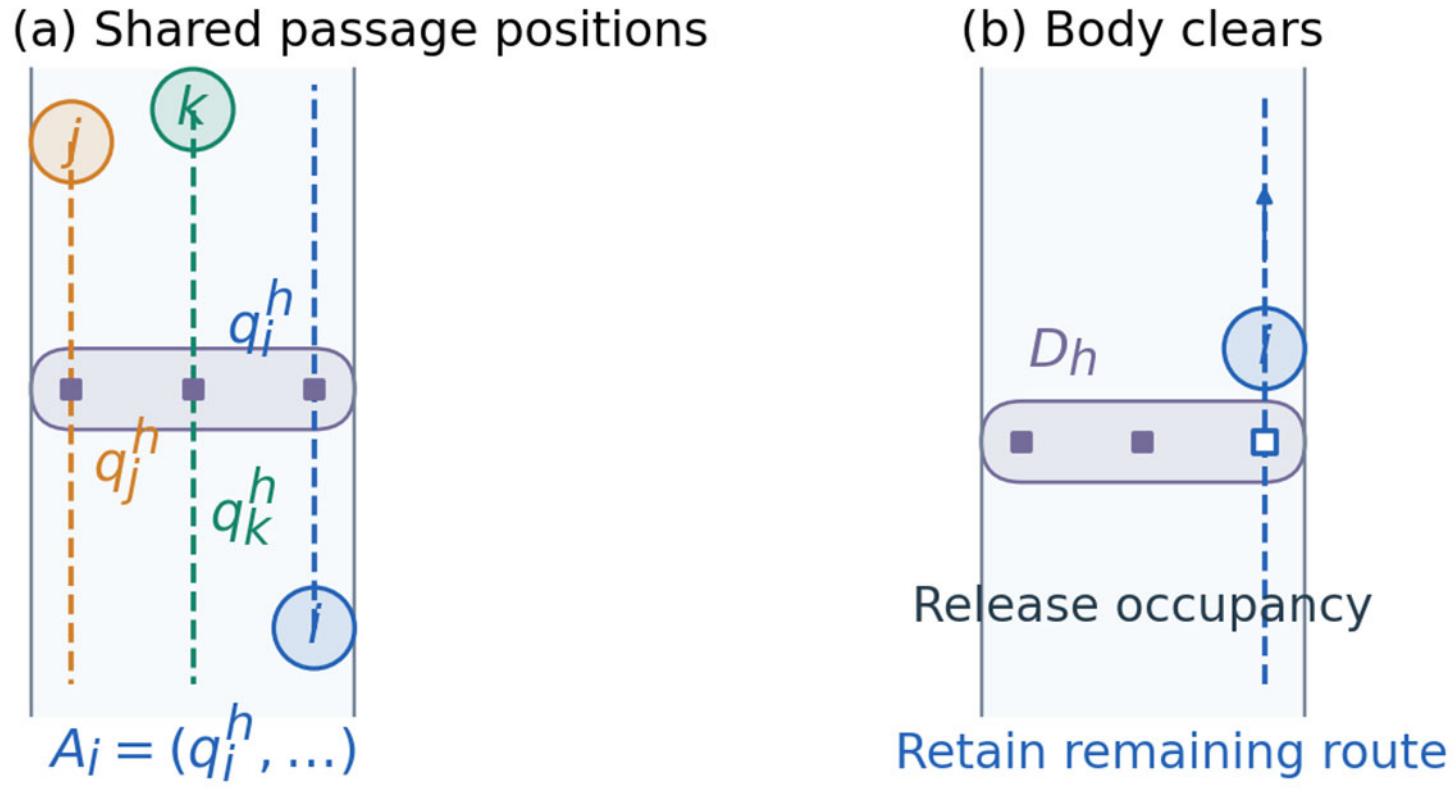

**Figure 21.** Shared arrangements and actual clearing. (a) Joint passage arrangement. (b) Occupancy release after the body clears. One passage position may support several pairwise encounters but appears only once as an anchor for this passage in an individual route. The agent continues along its route after its center passes the anchor; occupancy is released only after the disk fully clears.

If an encounter changes so that the first anchor no longer serves an avoidance purpose, a direct connection to the next anchor or goal may be attempted. This shortcut must preserve other valid arrangements and pass clearance and simultaneous-body checks over the same prediction horizon. Once accepted, the point is released and the changed route reenters prediction. With no anchors, the greedy intention of Eq. (18) resumes.

### 4.5.4 Channel clearing when approach is blocked

Shared channels provide separated positions at the encounter, yet bodies approaching their anchors may still block one another. If route adjustment, velocity deflection, and waiting cannot restore progress, WP organizes motion in the prescribed direction of the currently occupied channels. After clearing, the remaining valid anchors are reconnected from actual positions. A body not yet occupying a channel generates an approach candidate in the direction of its nearest admissible channel. If a body that has not cleared is blocked by another agent ahead, that agent is included in the same-direction clearing group. Further blockers are added recursively until an exit region that can accommodate the bodies is reached. If a body receives incompatible clearing requests from different channels, its adopted direction is retained and the incompatible arrangements are recoordinated under Sections 4.5.1–4.5.2.

At channel point $q$, let $n$ be the unit vector perpendicular to width axis $w$ and pointing toward the prescribed exit. For each agent included in clearing along this direction, retain the transverse component of its proposed displacement $u$ and supplement its forward component within a disk-clear route segment:

$$u' = u - (u \cdot n)n + \max\{u \cdot n, \beta \parallel u \parallel\} n, \qquad 0 < \beta < 1. \tag{37}$$

The scalar $u \cdot n$ is the original along-channel component, and $u - (u \cdot n)n$ is its transverse component. The parameter $\beta$ controls the candidate's progress toward the exit. For zero original displacement, a forward candidate within the speed limit is generated directly. Candidates are capped by the step's speed limit and jointly checked with other agents' same-step motions under Eq. (1). Accepted motions reconnect to anchors through Eq. (29).

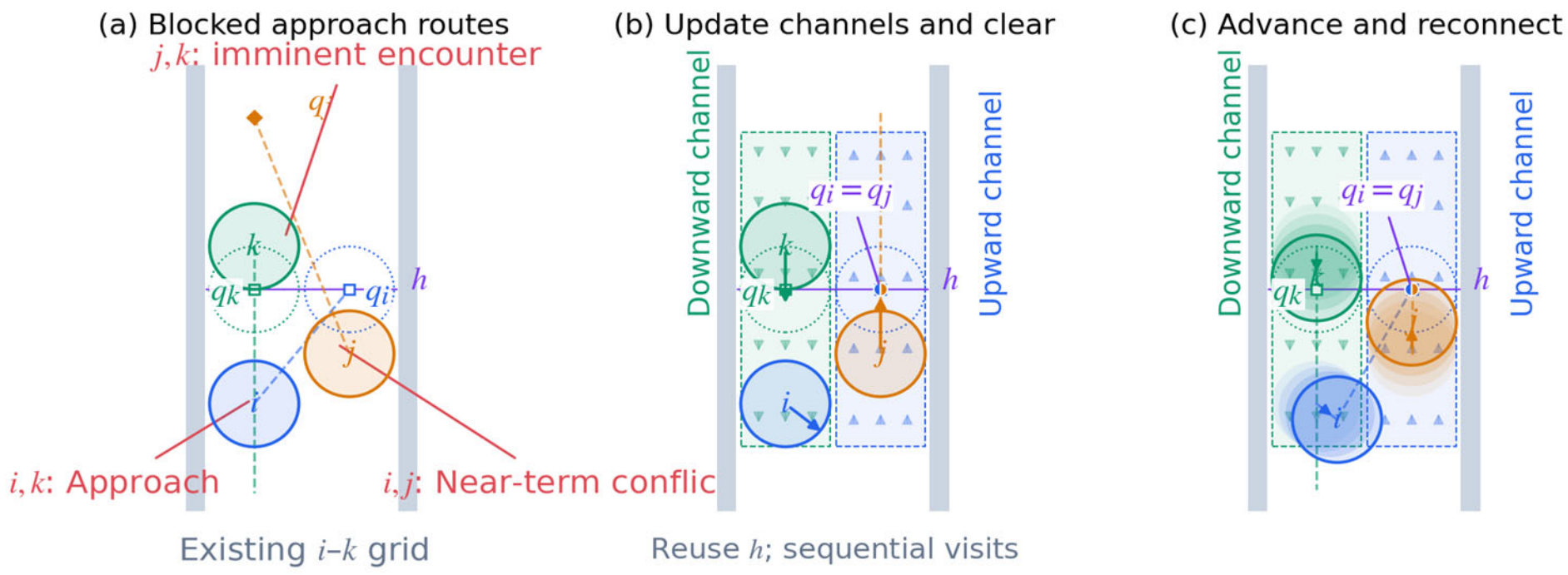

**Figure 22.** Forward motion in occupied channels. Blocked bodies advance toward the exit along their current channels and reconnect to their anchors when space permits. Pale bodies and solid outlines indicate the start and candidate end of the step, respectively.

Theorem 1 characterizes the spatial compatibility required for local exchange, and Eq. (26) directly guarantees body separation on separated axial channel segments. Bodies on different channel axes remain transversely separated; bodies in the same channel follow those ahead in order. Mutual blocking is thereby reduced to distance coordination between adjacent bodies. Once forward blockers are included, the leading body enters available exit space and successive bodies gain room to advance.

For a clearing-time bound, fix the finite member set after the current expansion and take the available exit ahead of each queue as its clearing boundary. Suppose channels and exits remain available during clearing, members do not move backward, and at each step at least one uncleared queue leader advances by at least $\min\{\delta, D_i(k)\}$. $\delta > 0$ is the minimum progress scale for this phase, and $D_i(k)$ is member $i$'s remaining longitudinal distance, after step $k$, until its entire body crosses the exit boundary in the prescribed direction; it is zero after clearing. Define

$$\Phi(k) = \sum_i \left\lceil \frac{D_i(k)}{\delta} \right\rceil, \qquad \Phi(k+1) \le \Phi(k) - 1 \quad \text{when } \Phi(k) > 0. \tag{38}$$

The operator $\lceil \cdot \rceil$ rounds upward, and the sum runs over the members to be cleared in this phase. The quantity $\Phi$ converts their remaining clearing distances into progress counts. At least one term decreases at each step while the others do not increase, so this group clears within at most $\Phi(0)$ steps. If additional forward blockers join during clearing, the remaining distances are recalculated for the expanded group.

Recursively included bodies advance together in the blocked channel's prescribed direction. After clearing, they reconnect the remaining valid anchors from actual positions under Eq. (29). If work occupancy cuts off the channel, width is remeasured and connections are adjusted; such spatial changes are handled separately from blockage by bodies ahead. Every candidate remains subject to Eq. (1). Bodies unable to advance safely wait, while others execute feasible motions.

### 4.5.5 Automatic planning of repeated multi-agent encounters

Before each physical step, Algorithm 1 processes urgent conflicts on current routes within the scope specified in Section 4.5.3. Predict detects encounters using Eqs. (19)–(20), and Coordinate selects grids, assigns shared positions, and reconnects routes as defined in Sections 4.5.1–4.5.2. Adjustments in Section 4.4 take precedence when anchors can be retained. Each step permits finitely many trials, excluding failed or repeated proposals at the same state. New earlier conflicts are coordinated while untried candidates and trial budget remain. When the budget is exhausted or no untried candidate remains, the inner loop ends and tests route-following, velocity deflection, and waiting. If near-term conflicts still block entry, forward blockers join channel clearing. Unresolved conflicts carry into the next step through the actual state. Only joint motion satisfying Eq. (1) is executed.

**Algorithm 1. Repeated multi-agent encounter planning**

**Input:** Current positions, goals and free space; routes $\gamma_i$, progress $s_i$, timing $\ell_i$, anchors $A_i$ and active grids.

**Output:** Admissible joint motion and updated routes and encounter arrangements.

```
while the lifelong task continues do
    Update anchors and occupied space from actual motion
    Reconnect changed routes by (29)
    Try first-anchor release while preserving other arrangements // §4.5.3
    P ← Predict current routes and timing by (19)–(20)
    while an eligible conflict in P and a trial remain do
        (i, j) ← Select the most urgent eligible conflict with an untried candidate
        proposal ← Coordinate(i, j)                         // §§4.5.1–4.5.2
        if proposal is feasible and has not been tried then
            Adopt its common allocation and route connections
            P ← Predict changed routes and timing          // §4.5.3
        end if
        Record the trial
    end while
    Test joint route-following, velocity and yielding motions // §4.4
    If access remains blocked, include forward blockers and test directed clearing // §4.5.4
    Execute a safe advancing motion, or remain at rest
    Update progress and actual positions; t ← t + Δt
end while
```

Body positions remain fixed during the inner planning loop. The outer execution loop updates routes, occupancy, and encounter relations.

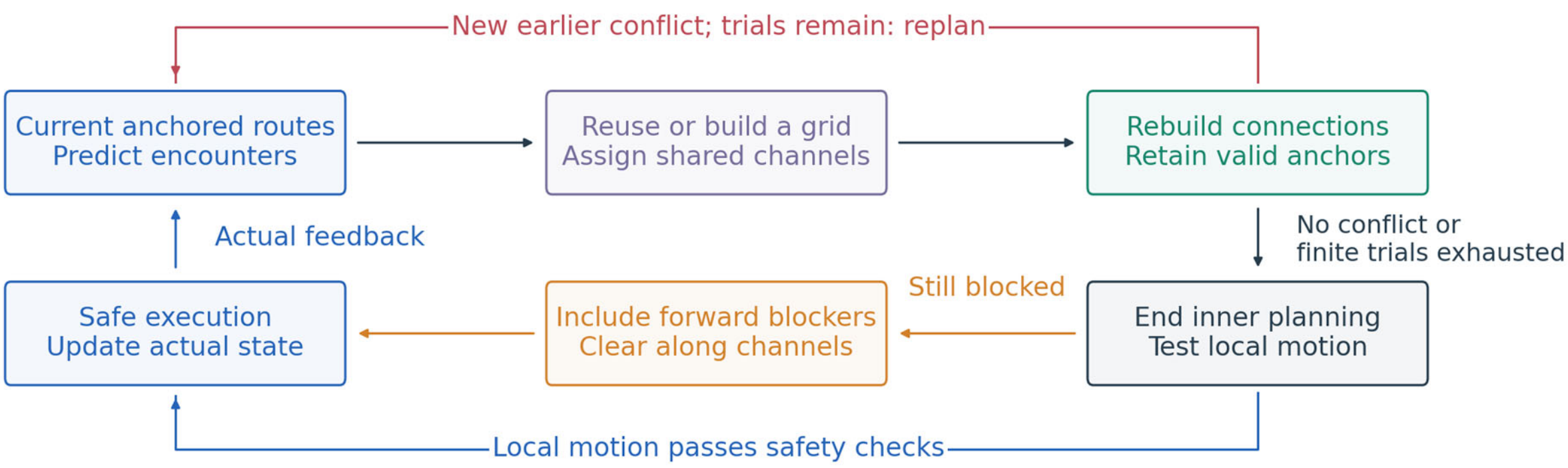

**Figure 23.** Finite retries and execution feedback. A new earlier conflict returns to joint planning only while an untried candidate and trial budget remain. After the inner loop ends, local motion is tested; persistent blockage invokes channel clearing. Every executed candidate passes the safety check. Actual positions and clearing states initialize the next physical step.

## 5. Experiments and results

WP is evaluated through encounter progression, sustained goal arrivals, and computational cost. Encounter replays and motion-state statistics examine how agents continue moving after goal updates, the arrival of additional agents, and changes in work occupancy. Comparisons with four representative methods assess goal arrivals, trip durations, and computational effort over complete observation windows.

## 5.1 Case study and experimental setting

A computational lifelong multi-agent planning scenario was constructed from the typical-floor layout and construction demands of a building. Work on vertical components in the mixed precast and cast-in-place system generates both motion goals and occupied spaces. Workers from different trades travel between material storage and work locations and move between neighboring components. Their task-driven journeys share portals and connecting passages, while operating spaces restrict neighboring movement. Encounter positions and available widths therefore change with construction state.

Figure 24 shows the floor layout and worker model. Walls (W), columns (C), and precast components (PC) are the main construction objects; openings between barriers and components connect the activity regions. Workers belong to five trades: reinforcement (R), formwork (F), precast joint treatment (JC), precast grouting (G), and hoisting (HI). Each worker selects tasks compatible with their trade and is represented by a disk of the same radius. Ongoing work occupies its required operating space, while work locations and storage areas alternately serve as movement goals.

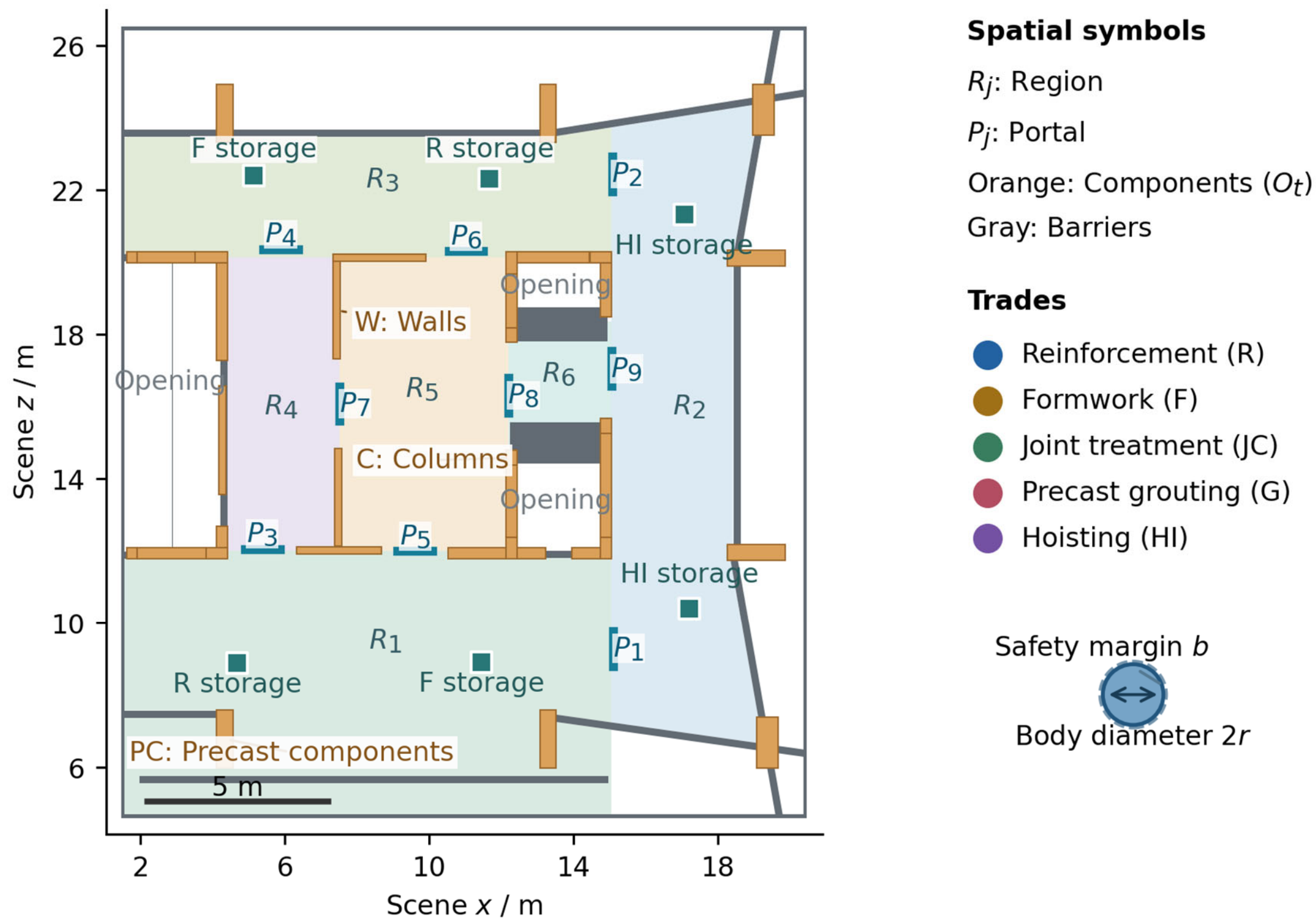

**Figure 24.** Construction layout and disk-motion model. Regions $R_j$ are connected by portals $P_j$. Orange denotes components, and gray denotes barriers. The right-hand panel lists the trades. The narrow ring around the disk illustrates the safety margin $b$ defined in Section 3.1; its thickness is schematic.

Work progression and material replenishment were accelerated, and the quantity carried per trip was reduced, to generate repeated encounters within a finite observation window. These settings increased travel between work locations and storage areas. Subsequent goals were selected from reachable locations compatible with each worker's trade. When task-generated transfers were insufficient, reachable locations of the same type supplied additional movement demands. The scenario retained the effects of work occupancy on passage geometry while providing more observations of sustained movement and repeated encounters.

The experiments use a common observation window of 1,000 simulation steps (200 s), combining encounter replays with statistics from multiple complete runs. Comparative experiments cover six configurations of 5–10 agents, with repeated runs for every method at each configuration.

## 5.2 Passage through dynamic encounters and motion states

The saved planning states and stepwise motion records provide examples of two-agent passage, goal changes, occupancy changes, and multi-agent continuation. Figures 25–29 each present three successive stages showing the channel arrangements, route changes, and resulting passage.

### 5.2.1 Two-agent passage and continuation after goal changes

In Figure 25, two agents pass through a connecting area in opposite directions using distinct channel points on a shared width. Both have moved away by the final frame, illustrating body separation through distinct channel assignments.

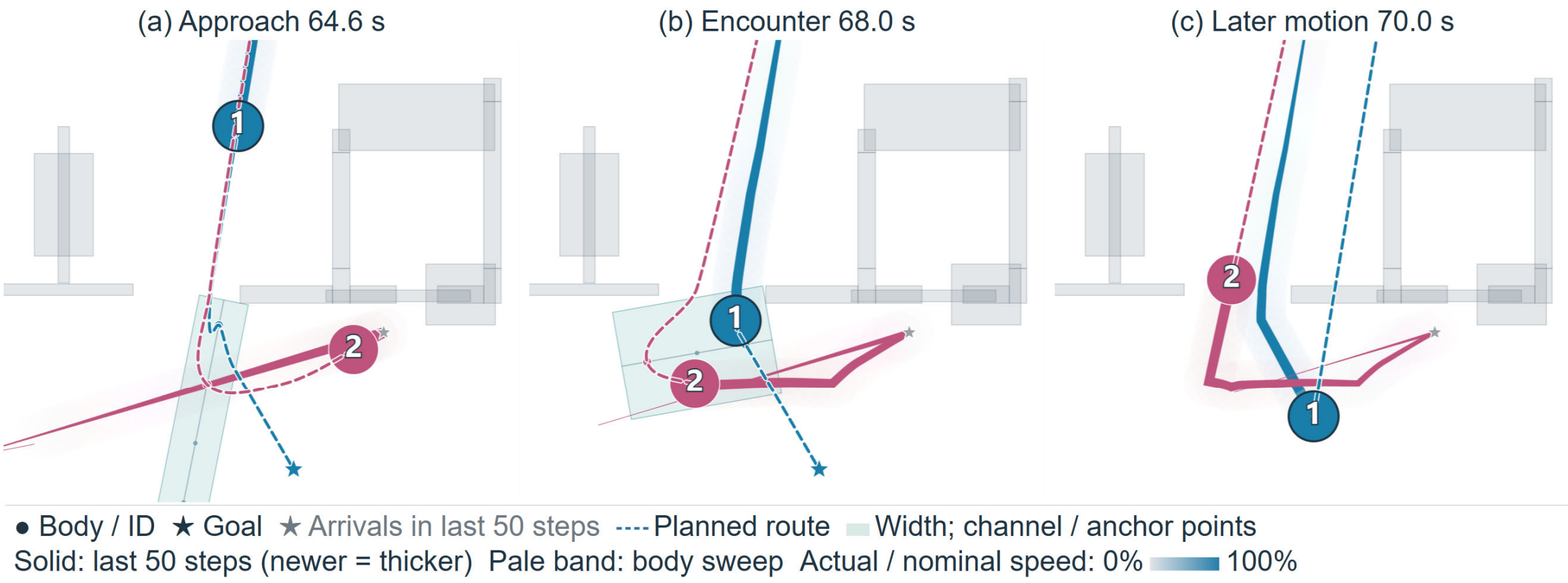

**Figure 25.** A two-agent encounter near an entrance at 64.6, 68.0, and 70.0 s. The agents approach the shared width and pass through distinct channel positions. In the final frame, agent 1 continues downward and agent 2 upward. The gray star marks a position previously reached by agent 2.

In Figure 26, agent 2 reaches its goal and turns downward, forming a new passage relationship with agents 1 and 3 approaching from below. The shared channels and connecting routes are then updated, allowing the three agents to move away in separate directions. The sequence illustrates how a three-agent encounter continues after a goal change.

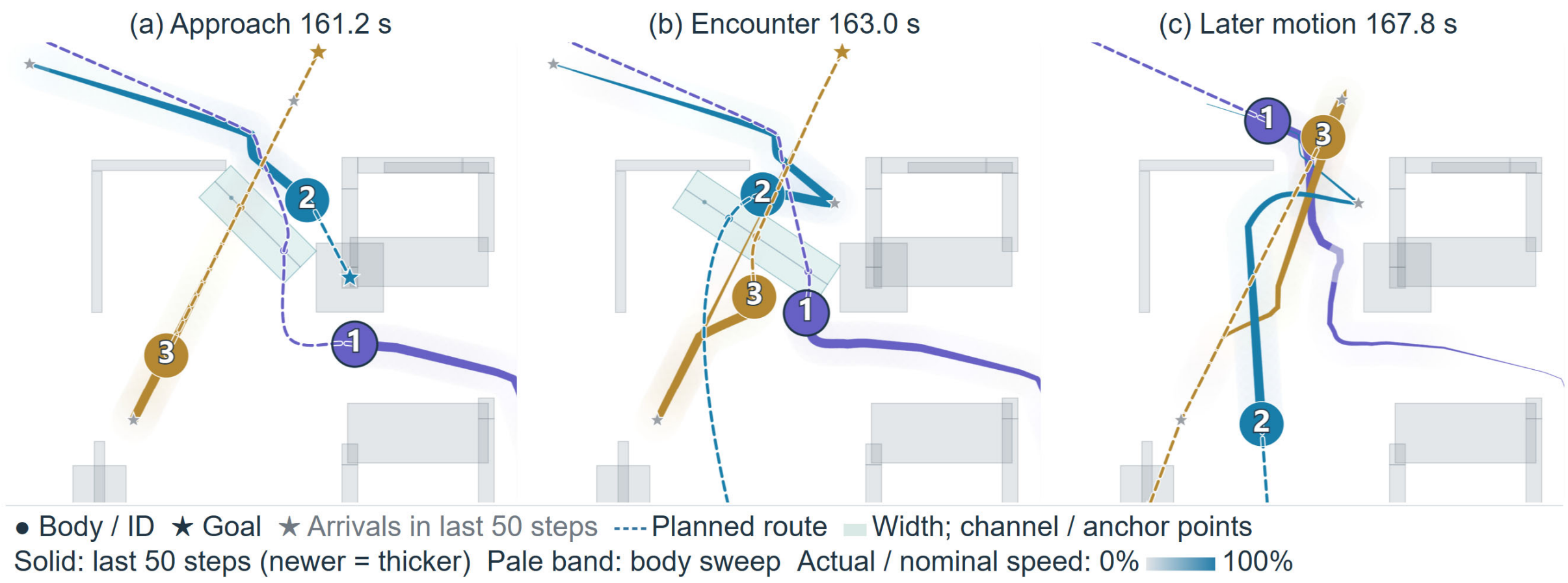

**Figure 26.** Coordination of successive movements on a shared grid at 161.2, 163.0, and 167.8 s. Agent 2 turns downward after reaching its goal, while agents 1 and 3 approach from below and pass upward. The middle frame shows the shared width and updated route connections; the final frame shows the agents continuing in separate directions. Gray stars identify turns caused by goal arrivals.

### 5.2.2 Route recovery after new work occupancy

In Figure 27, agent 2 starts work, and its newly occupied operating space intersects the body-swept regions along the original routes of agents 1 and 3. Both retain their goals, update the affected connections from their actual positions, and continue downward and upward, respectively. The same occupancy change is thus accommodated by local route adjustments for trips in opposite directions.

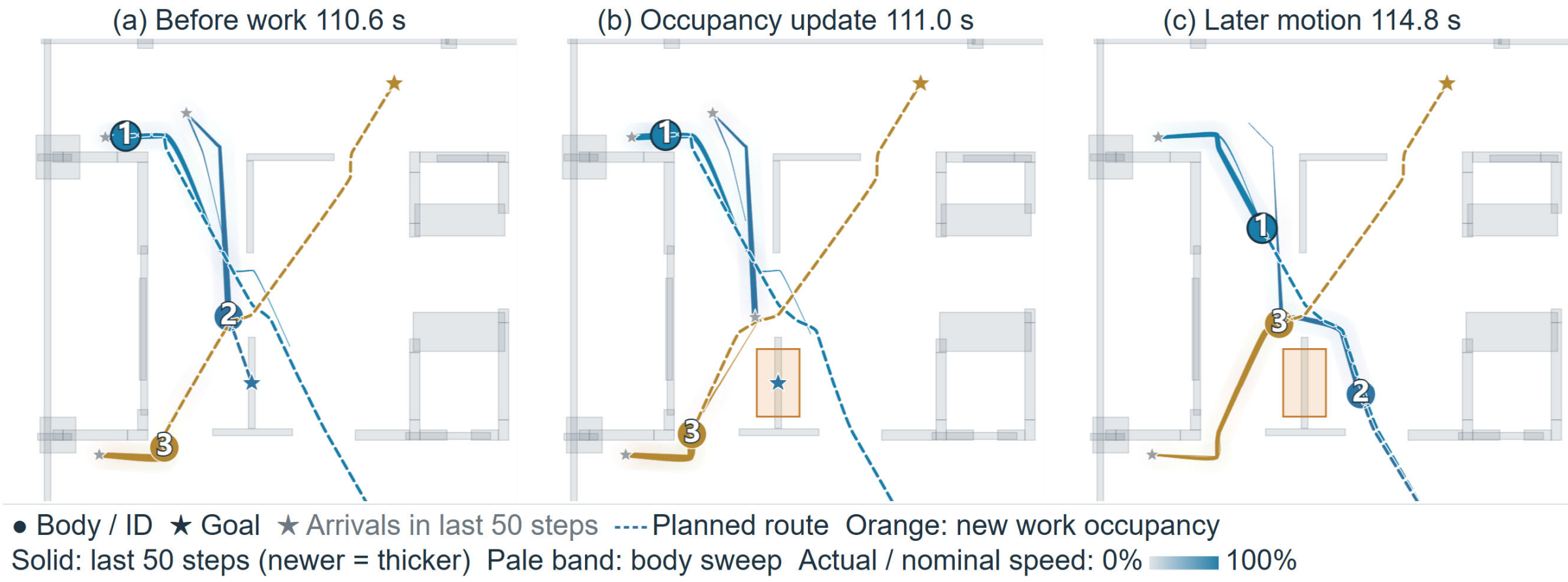

**Figure 27.** Effects of the same new work occupancy on two routes. (a) At 110.6 s, agent 2 approaches its work location, while agents 1 and 3 approach from above and below. (b) At 111.0 s, an orange border marks the newly occupied operating space. Agent 2 has temporarily left the navigation plane, and agents 1 and 3 update their connections. (c) At 114.8 s, both agents continue toward their original goals along the adjusted routes. Agent 2 has returned to the navigation plane and is moving downward. The goals of agents 1 and 3 remain unchanged throughout the sequence.

### 5.2.3 Additional agents and encounter continuation

In Figure 28, a two-agent encounter expands to four agents as others approach. After it ends, agents 1 and 2 continue away, while agents 3 and 4 enter subsequent local encounters. Agents join an encounter when current conflicts require coordination and leave after their bodies clear the area. Passage arrangements are consequently established, released, and connected as trips continue.

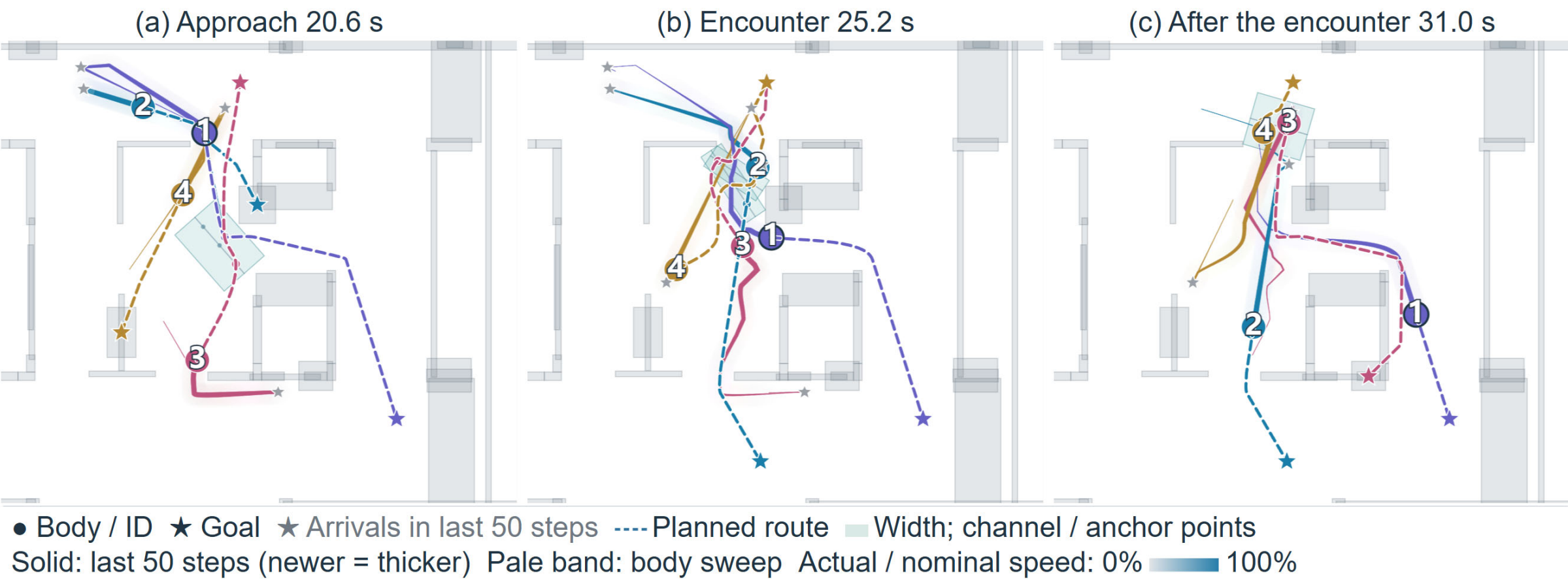

**Figure 28.** A four-agent encounter formed by extending an initial two-agent relationship. The agents approach at 20.6 s and coordinate in the connecting area at 25.2 s. By 31.0 s, the encounter has ended: agents 1 and 2 continue away, while agents 3 and 4 begin subsequent local encounters. The three frames were selected to represent successive event stages.

In Figure 29, five agents converge for coordinated passage at 195.0 s. The encounter ends at 201.8 s, after which the agents disperse. The sequence follows the five-agent arrangement from concentrated passage until the bodies clear the encounter area.

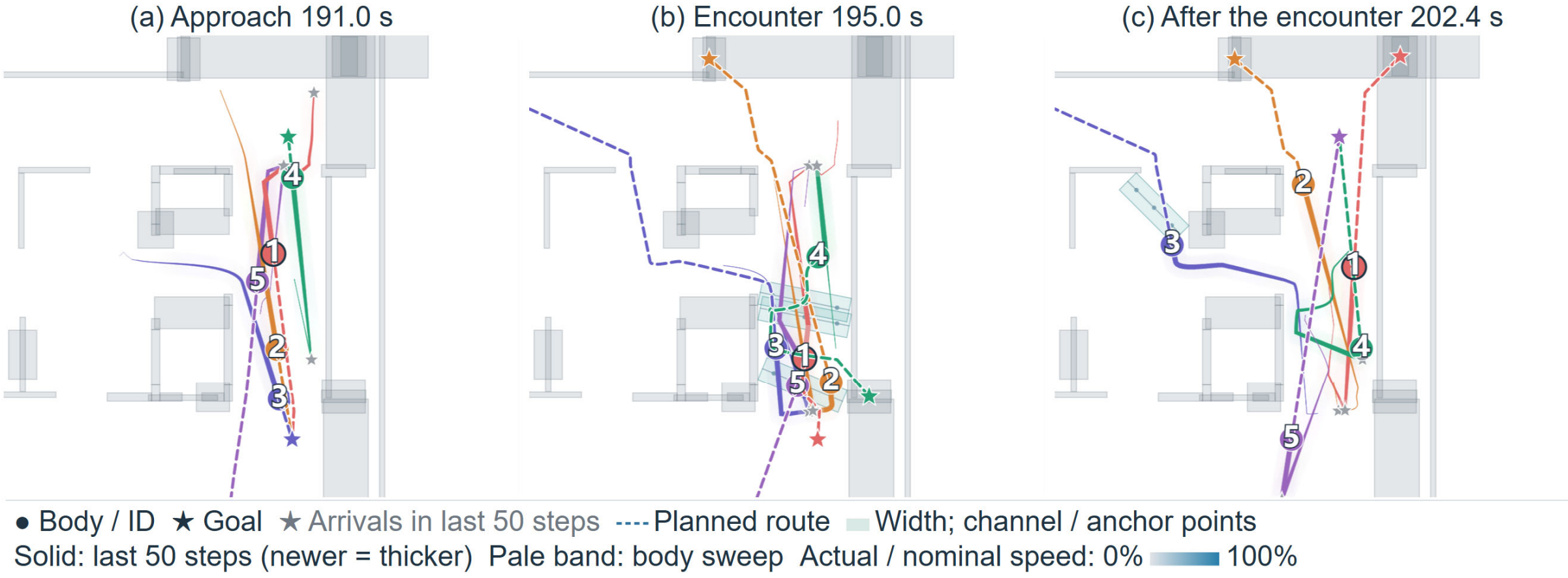

**Figure 29.** Entry, concentrated passage, and dispersal in a five-agent encounter at 191.0, 195.0, and 202.4 s. Local tracking extends beyond the common observation window to show the state after the encounter ends at 201.8 s. Efficiency statistics still use only the first 200 s.

### 5.2.4 Near-stop frequency and duration

Local yielding and execution constraints can temporarily slow an agent; their effects depend on how often and for how long this occurs. One agent's state during one motion step constitutes an *agent-step*. Navigation states are classified by the ratio of actual distance traveled to the distance allowed by the nominal speed during that step. A near-stop has a ratio no greater than 1%; low-speed motion has a ratio greater than 1% but below 20%; all remaining navigation states have ratios of at least 20%. Work activity is counted separately.

Under accelerated work and material replenishment, navigation accounted for 97.15% of all agent-steps across multiple complete runs. The 126 near-stop agent-steps represented 0.393% of navigation agent-steps, while low-speed motion accounted for 0.128% (Figure 30). Within the same agent's trip to the same goal, the longest consecutive near-stop lasted five steps, or 1.0 s. Most near-stops were attributed to execution constraints or collision-avoidance adjustments; only a small fraction were explicitly recorded as channel yielding. Near-stops were therefore infrequent and brief during the repeated encounters and goal updates examined, and the complete method generally maintained ongoing movement.

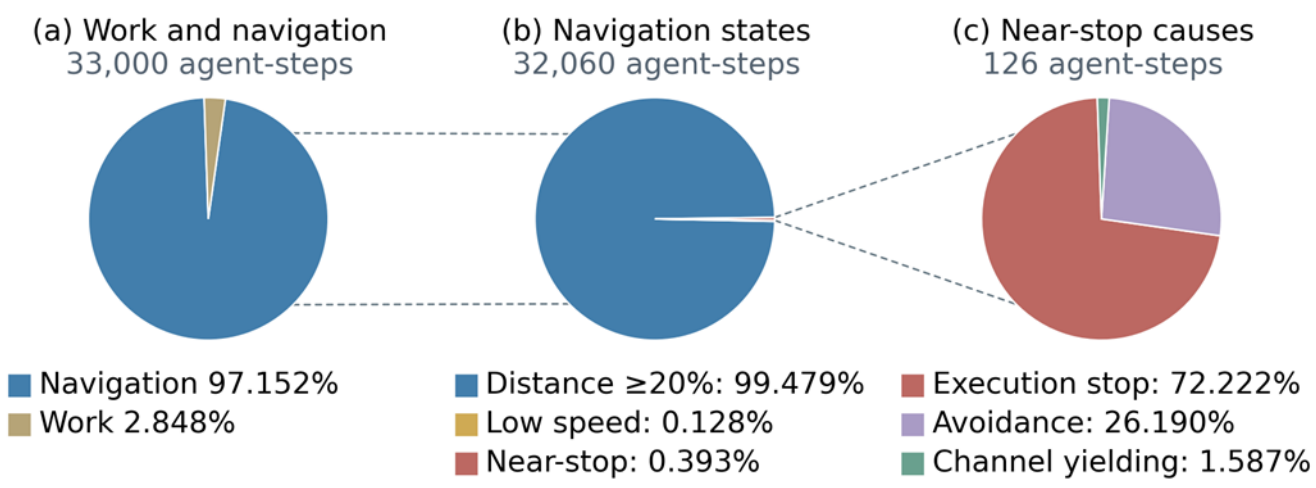

**Figure 30.** Motion states and brief stops. Panel denominators are 33,000 total agent-steps, 32,060 navigation agent-steps, and 126 near-stop agent-steps. (a) Navigation and work account for 32,060 and 940 steps. (b) Motion at or above 20% of nominal step distance, low-speed motion, and near-stops account for 31,893, 41, and 126 steps. (c) Execution-constrained stops, collision-avoidance adjustments, and explicit channel yielding account for 91, 33, and 2 steps. Dashed lines expand the corresponding sector into the conditional distribution in the next panel.

Figures 25–30 document passage under channel arrangements, route recovery after occupancy changes, and coordinated continuation as additional agents enter. Complete-run statistics show that near-stops were infrequent and brief. This evidence supports WP's ability

to maintain ongoing movement in the dynamic encounters examined. We next compare arrival efficiency and computational cost over complete observation windows.

## 5.3 Comparative experiments

### 5.3.1 Objectives and experimental design

The comparison covers discrete space–time search, path-guided reciprocal collision avoidance, and continuous trajectory search. Reciprocal collision avoidance was evaluated both alone and with local MAPF recovery to examine how different recovery mechanisms affect operation under obstruction. Table 2 summarizes the four baselines, their principles and previously demonstrated applications, and the differences examined here.

**Table 2.** Rationale for baseline selection.

| Method | Principle and previously demonstrated applications | Difference examined in this study |
|---|---|---|
| RHCR + PBS [5,23] | RHCR converts continuing goal demands into MAPF problems within rolling time windows, while PBS searches agent-priority relationships in response to conflicts. The original work demonstrated continuous task handling and scalability in automated warehouse simulations. | Discrete space–time coordination versus local continuous encounter organization. |
| Theta* + ORCA [4,19,31] | Theta* generates any-angle paths, and ORCA selects collision-avoiding velocities through reciprocal responsibility. ORCA has demonstrated efficient local collision avoidance for mobile robots and dense crowd simulations; their combination provides reactive navigation guided by paths. | Whether individual paths and velocity adjustments can accommodate repeated encounters in constrained spaces. |
| ORCA + local MAPF [31,32] | When reciprocal collision avoidance is obstructed, a local discrete planning problem is constructed for the agents involved. The original work improved goal-reaching success through local coordination in narrow and constrained settings. | Reactive navigation with local replanning during repeated encounters. |
| Rolling-horizon K-CBS adaptation [6] | Conflict-tree constraints are combined with single-agent sampling-based planning to generate continuous trajectories directly. The original work demonstrated continuous coordination for car-like robots in open, narrow, and cluttered environments. | Continuous trajectory search versus channel-anchor decisions in motion efficiency and computational cost. |

All five methods were run repeatedly in six configurations with 5–10 agents, sharing the floor layout, body dimensions, nominal speeds, initial goals, and task-update rules. Subsequent goals were updated at each method's own arrival times. Arrival was defined by reaching a safe grid point for grid-based methods and by contact for continuous methods. The comparison therefore concerns the methods as adapted under these conditions.

The evaluation measures are **the number of goal arrivals, the duration of completed trips, and the mean per-step computation time**. Goal arrivals count visits to task areas and storage areas, measuring sustained transfers within a fixed observation window. Trip duration is the simulated time taken to complete a goal-directed trip, and per-step computation time is the online computation time for one motion step. For a complete run containing at least one completed trip, the latter two means are

$$\bar{T} = \frac{1}{L}\sum_{\ell=1}^{L} T_\ell, \qquad \bar{w} = \frac{1}{K}\sum_{k=1}^{K} w_k. \tag{39}$$

$L$ is the number of completed trips, $T_\ell$ is the simulated duration of trip $\ell$, $K$ is the number of simulation steps, and $w_k$ is the computation time for step $k$. The mean $\bar{T}$ covers completed trips within a run, whereas $\bar{w}$ covers all steps. Group summaries give equal weight to each complete run's metric. Arrival counts and computation times use runs that complete the 200 s window; trip durations use the trips completed within those runs. The results therefore describe efficiency conditional on complete observation.

Computation time includes planning, execution checks, and task updates at every step. All measurements were taken on the same AMD Ryzen 9 9950X3D host, with records obtained under different timing conditions marked separately.

### 5.3.2 Actual trajectories in dense encounters

Figures 31–35 show dense encounter regions from eight-agent runs of the five methods. Each figure presents three frames spanning 9 s, using the same display conventions for actual trajectories. The comparison examines sustained progress, local reversals under an unchanged goal, and recovery after stopping. Gray stars identify normal turns caused by goal arrivals; stepwise positions, traveled distances, and goal records distinguish reversals and near-stops. Because subsequent goals and local states diverge during execution, these excerpts compare motion patterns. Overall efficiency is evaluated over complete observation windows.

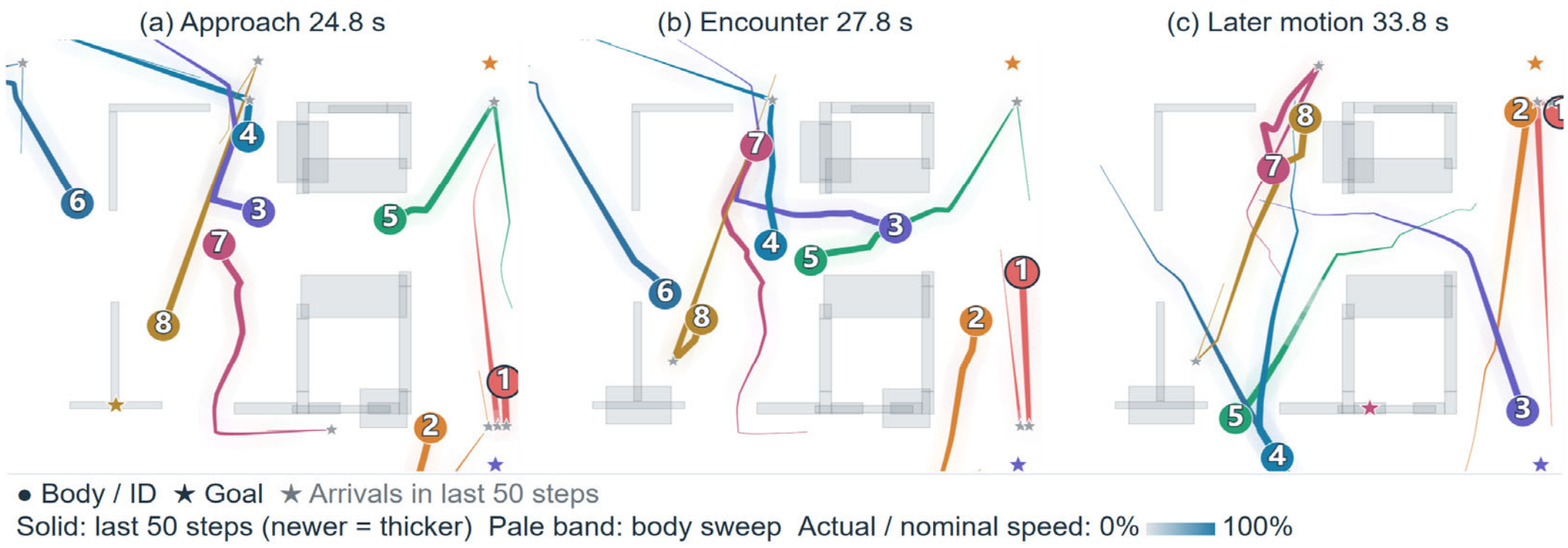

**Figure 31.** Local motion under WP at 24.8, 27.8, and 33.8 s. Agent 3 passes to the right, and agents 4 and 5 gradually leave downward. Agent 8 turns upward after arriving below, while agent 7 turns downward after arriving above. Gray stars mark actual arrival positions within the preceding 50 steps. Only actual trajectories are displayed.

In the WP excerpt, movement converges from several directions at the central connection and then disperses. Agents 3, 4, and 5 retain their goals and leave through the spaces to the right and below. Agents 8 and 7 reach their goals at 26.0 and 30.2 s, respectively; their reversals begin new trips. Other local direction adjustments accompany sustained net displacement. The principal agents continue to leave the encounter location, showing multidirectional passage amid goal updates and dense encounters.

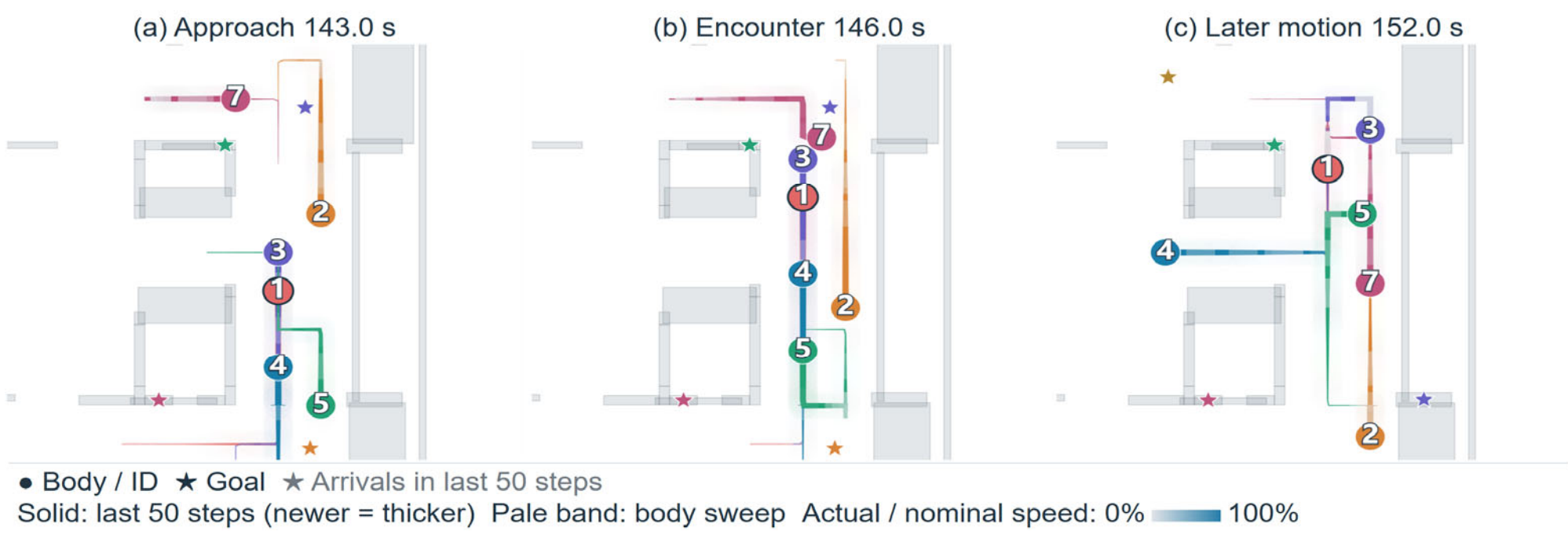

**Figure 32.** Local motion under RHCR + PBS at 143.0, 146.0, and 152.0 s. Agents 1, 3, 4, and 5 move successively through the right-hand passage, while agent 7 enters from above. The trajectories primarily consist of connected horizontal and vertical segments.

RHCR + PBS trajectories follow discrete passage directions, joining straight segments at corners. In Figure 32, agent 3 first approaches upward and then moves laterally to an adjacent position, while agent 4 turns left from the longitudinal passage. Some agents resume forward motion after waiting. This orderly continuation reflects discrete space–time coordination; the observed right-angle turns primarily correspond to route transitions in the graph.

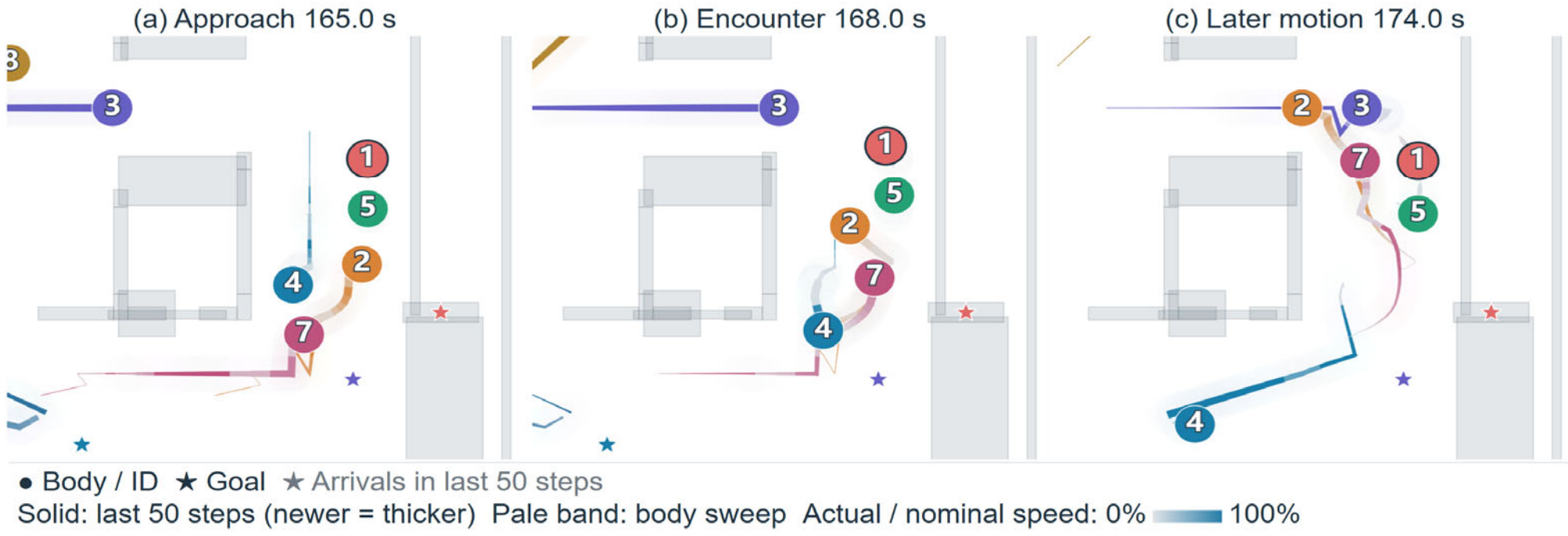

**Figure 33.** Local motion under Theta* + ORCA at 165.0, 168.0, and 174.0 s. Agents 1 and 5 remain close to their previous positions on the right, while agents 2, 3, and 7 gradually converge. Agent 4 leaves downward.

Theta* + ORCA adjusts collision-avoiding velocities along individual guide paths. In Figure 33, agents 1 and 5 repeatedly make small direction changes under unchanged goals, yet their accumulated movement produces little net progress. Agent 4 leaves the neighboring group downward. Individual avoidance thus coexists with local stagnation: continued velocity adjustment does not promptly clear the obstructed agents on the right in this excerpt.

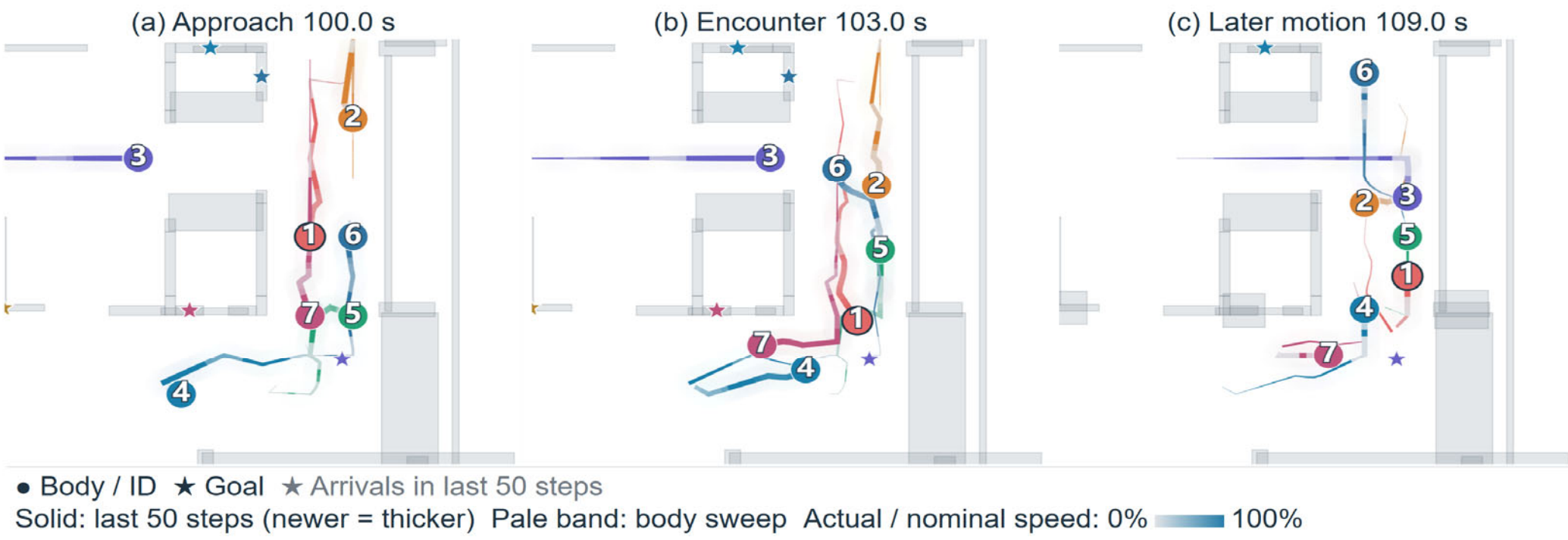

**Figure 34.** Local motion under ORCA + local MAPF at 100.0, 103.0, and 109.0 s. Agent 6 advances upward, and agent 3 enters from the left. Agents 1, 2, and 5 occupy the passage, while agents 4 and 7 adjust their motion at the lower connection.

ORCA + local MAPF uses local coordinated planning to recover from obstruction. In Figure 34, agent 4 moves outward at the lower connection before turning back toward the passage. Agent 5 remains near-stationary for consecutive steps, while agent 6 advances upward. Progress, local reversals, and consecutive near-stops occur together, showing that some agents remain temporarily obstructed despite local recovery planning.

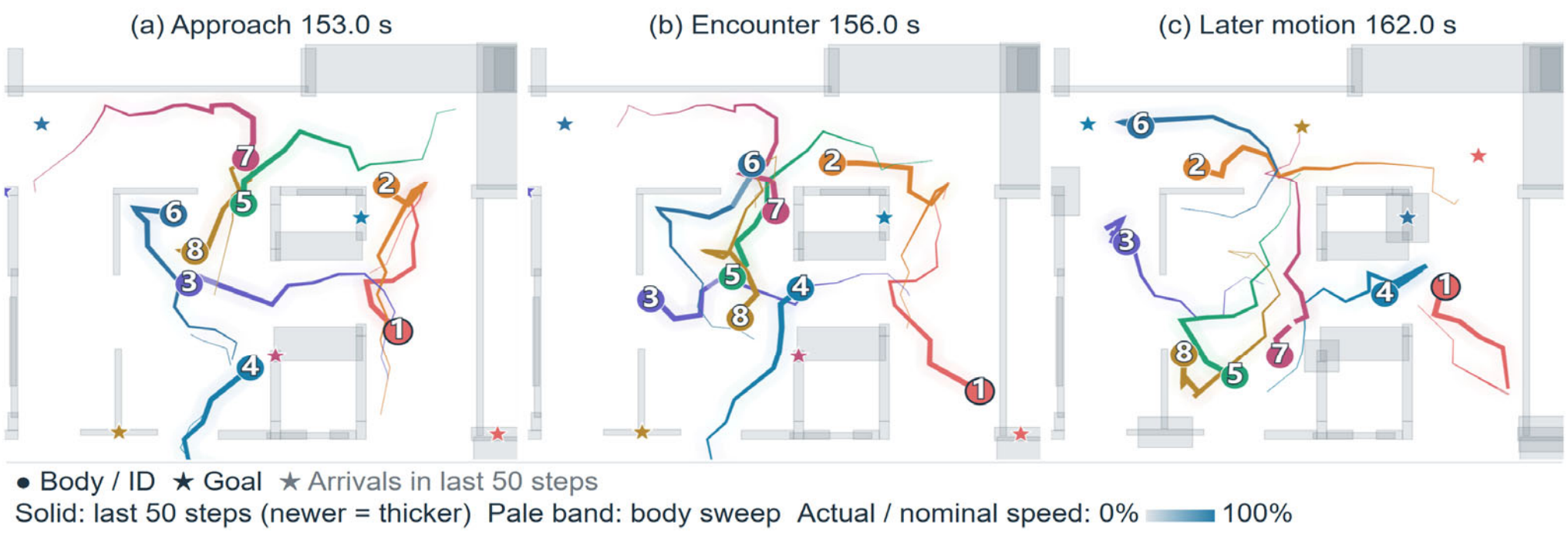

**Figure 35.** Local motion under the rolling-horizon K-CBS adaptation at 153.0, 156.0, and 162.0 s. Agents 2 and 6 pass through the upper space, while agents 5, 7, and 8 move successively through the central connection. Lines connect the actual positions recorded at consecutive steps.

The rolling-horizon K-CBS adaptation coordinates compatible trajectories in continuous space, allowing oblique and multidirectional connections. In Figure 35, most agents continue moving, and their spatial distribution changes as they pass through the central region. Agents 3, 7, and 8 nevertheless make local reversals under unchanged goals. The excerpt shows coordinated multi-agent continuation together with execution-time direction adjustments. Polylines connecting successive recorded positions reveal those changes; continuous curvature requires the original trajectory representation.

Local turns in these excerpts arise from goal updates, collision-avoidance adjustments, or persistent obstruction. They must therefore be interpreted alongside net progress, duration, and clearance of shared space. Multidirectional passage under WP, orderly continuation under discrete coordination, and local stagnation in the two reciprocal-avoidance baselines help explain the complete-window efficiency results. Overall performance is assessed using the arrival counts and trip durations reported next.

### 5.3.3 Goal arrivals and sustained progress

Table 3 reports the mean number of goal arrivals within 200 s and the observed range across complete runs. Figure 36 shows the corresponding cumulative arrival curves.

**Table 3.** Goal arrivals within 200 s: mean [minimum, maximum].

| Agents | WP | RHCR + PBS | Theta* + ORCA | ORCA + local MAPF | Rolling-horizon K-CBS adaptation |
|---|---|---|---|---|---|
| 5 | 96.7 [95, 98] | 65.7 [65, 66] | 49.3 [41, 55] | 57.3 [54, 61] | 76.0 [70, 80] |
| 6 | 111.0 [106, 115] | 72.7 [72, 73] | 46.7 [40, 59] | 68.0 [66, 69] | 88.7 [82, 94] |
| 7 | 136.0 [126, 143] | 85.3 [84, 86] | 57.3 [52, 62] | 71.0 [66, 76] | 104.3 [102, 107] |
| 8 | 153.0 [148, 160] | 94.3 [91, 98] | 57.3 [51, 63] | 78.0 [71, 88] | 110.7 [110, 111] |
| 9 | 164.3 [162, 166] | 98.7 [97, 100] | 65.7 [62, 70] | 70.7 [68, 74] | 108.0 [73, 126] |
| 10 | 189.0 [185, 194] | 109.7 [108, 112] | 67.3 [65, 71] | 79.0 [76, 83] | 133.7 [132, 137] |

WP achieves the highest mean number of goal arrivals in all six configurations, exceeding the strongest baseline in each—the rolling-horizon K-CBS adaptation—by 25.2%–52.2%. Gains of 25.2%–30.4% for 5–7 agents rise to 38.3%–52.2% for 8–10 agents, showing a larger relative advantage in the denser configurations. In the eight-agent group with paired initial states, WP averages 153.0 arrivals: 38.3% more than the K-CBS adaptation's 110.7 arrivals and 1.62 times the RHCR + PBS result. WP produces more arrivals for every matched initial state, with differences of 37–50 arrivals. Its advantage therefore extends over the complete observation window.

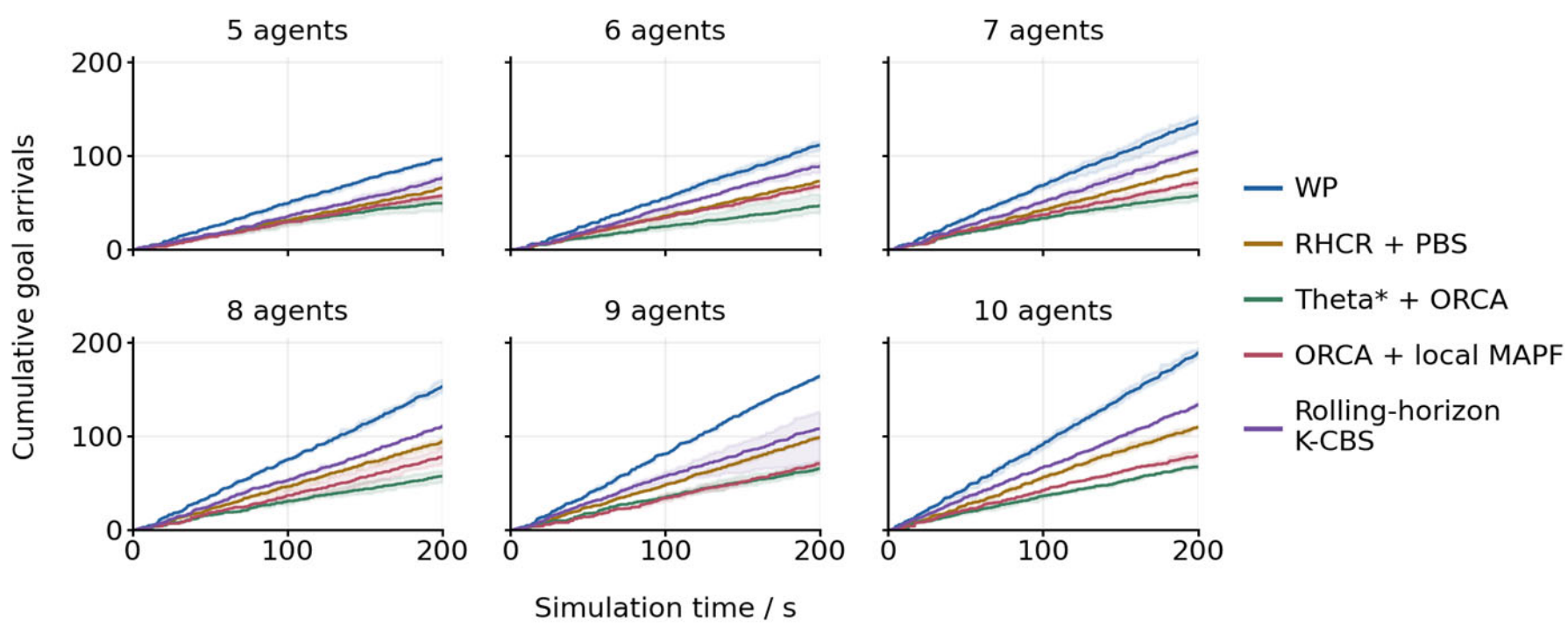

**Figure 36.** Cumulative goal arrivals for different numbers of agents. Curves show means over complete runs, and shaded bands show the observed minimum and maximum. All panels use the same axis scales.

In Figure 36, WP steadily accumulates goal arrivals, and its mean advantage over the other methods generally grows through the runs and persists to the end of the observation window. Across the 5–10-agent configurations, WP records 18.3–19.4 arrivals per agent on average.

Total arrivals increase with the number of agents, while arrivals per agent remain similar. This trend is limited to the present floor layout, task rules, and complete-run sample. Because some configurations use different sets of initial states, it cannot be attributed solely to changes in agent count.

### 5.3.4 Completed-trip durations and mean computation times

Table 4 reports the mean duration of completed trips and the mean per-step computation time, measuring simulated travel time and online computational effort, respectively. Group summaries give equal weight to each complete run's metric.

**Table 4.** Completed-trip durations and mean computation times for different numbers of agents.

| Agents | Method | Mean duration / s | Mean per-step computation time / ms |
|---|---|---|---|
| 5 | **WP (proposed)** | 9.66 | 19.7 |
| 5 | RHCR + PBS | 14.44 | 7.5 |
| 5 | Theta* + ORCA | 16.20 | 76.0 |
| 5 | ORCA + local MAPF | 15.70 | 89.7 |
| 5 | Rolling-horizon K-CBS adaptation | 12.51 | 139.9 |
| 8 | **WP (proposed)** | 9.83 | 65.5 |
| 8 | RHCR + PBS | 15.81 | 12.1 |
| 8 | Theta* + ORCA | 18.89 | 87.1 |
| 8 | ORCA + local MAPF | 19.14 | 116.7 |
| 8 | Rolling-horizon K-CBS adaptation | 13.32 | 227.1 |
| 10 | **WP (proposed)** | 9.86 | 107.8 |
| 10 | RHCR + PBS | 16.01 | 14.5 |
| 10 | Theta* + ORCA | 22.74 | 104.9† |
| 10 | ORCA + local MAPF | 21.60 | 121.2† |
| 10 | Rolling-horizon K-CBS adaptation | 13.00 | 357.9 |

† Concurrent replay preparation affected supplementary-batch timing. These means were obtained under different timing conditions and are provided for reference only.

In the eight-agent group, the mean duration of completed trips is 9.83 s for WP and 13.32 s for the rolling-horizon K-CBS adaptation, with mean per-step computation times of 65.5 and 227.1 ms, respectively. WP achieves more arrivals, shorter completed trips, and lower mean computation time. RHCR + PBS achieves 94.3 arrivals at 12.1 ms per step and has a lower computational cost. The methods complete different sets of subsequent trips, however, so duration comparisons concern the demands each method actually completes and cannot be converted directly into time savings for identical trips.

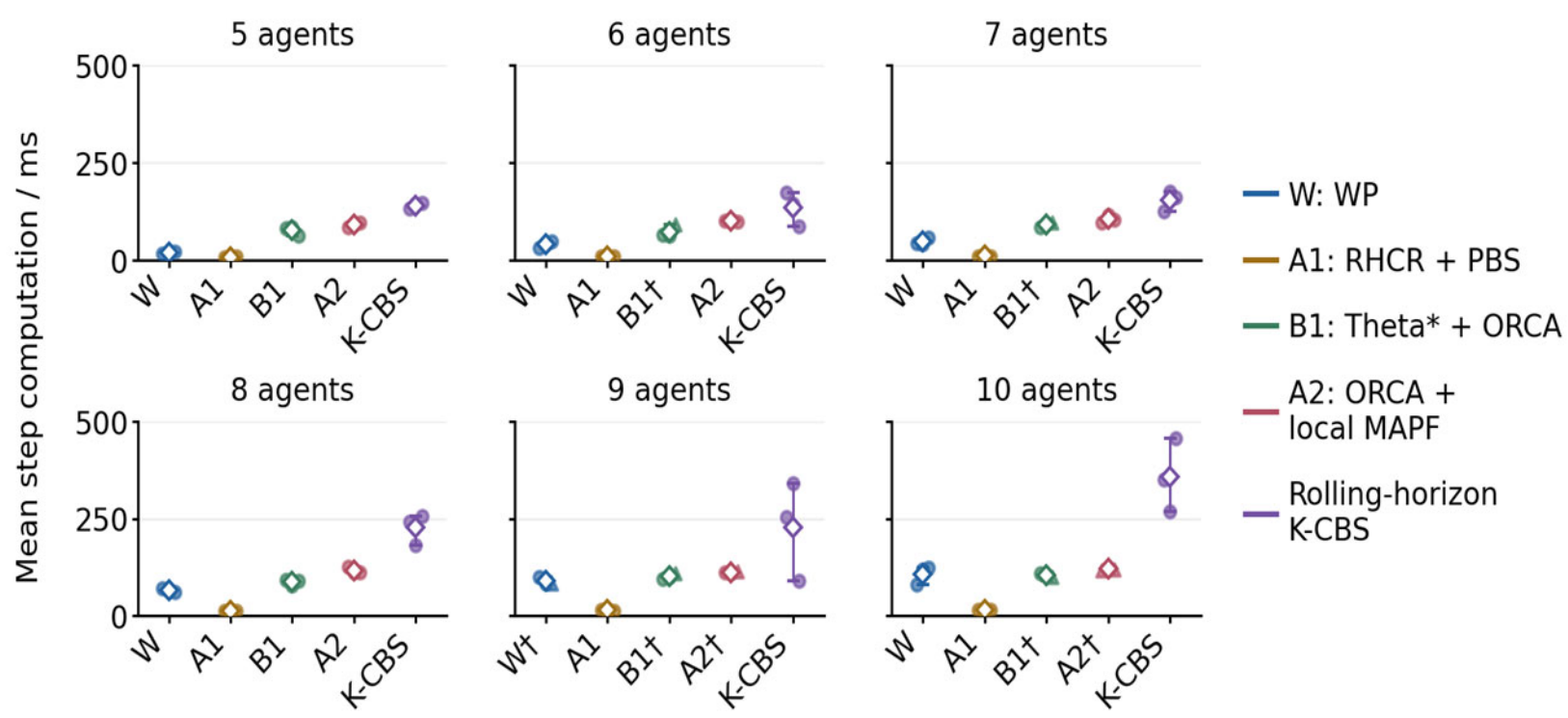

**Figure 37.** Mean per-step computation time and between-run variation. Circles and triangles represent complete runs from different batches; open diamonds denote group means, and whiskers indicate observed ranges. All panels use the same millisecond scale. † marks reference means obtained under different timing conditions. W, A1, B1, A2, and K-CBS denote WP, RHCR + PBS, Theta* + ORCA, ORCA + local MAPF, and the rolling-horizon K-CBS adaptation, respectively.

Tables 3 and 4 show how WP balances sustained arrivals and computational cost in the tested setting. It achieves more arrivals than RHCR + PBS at greater computational cost, while outperforming the rolling-horizon K-CBS adaptation in both arrivals and mean computational cost in the paired eight-agent group. These results support the planning efficiency of the complete method under the stated movement demands and adaptation conditions.

Figure 37 shows that WP's mean per-step computation time increases with the number of agents. Across the 5–10-agent configurations, its group means remain below those of the rolling-horizon K-CBS adaptation and above those of RHCR + PBS. Continuous channel planning thus retains a computational burden that grows with the configuration. Mean timing characterizes cost over an entire run; compliance with real-time deadlines also depends on the distribution of individual step latencies.

# 6. Discussion

Work activation and completion change local clearance, while rerouting affects later encounters, coupling current passage to subsequent continuation. Local conflict resolution, width guidance, spatial separation, and dependency maintenance have established foundations [7–14,18,25,28,29]. WP connects these functions through actual encounter demands: it constructs shared passage relationships from real-time space, retains continuous greedy connections between encounters, and updates affected arrangements as space and membership change. Relative to representations organized by grid positions, partition boundaries, or sampled connections [6,18,19], this architecture specifies where shared constraints arise, how they connect, and when they change, while retaining regional topology and geometric path solving.

Spatial compatibility criteria provide the geometric basis. Under the stated local-space conditions, Theorem 1 gives the necessary and sufficient width condition for fixed-point exchange, and Theorem 2 identifies the separation directions that composite exchange must cover. They guide channel construction and encounter relocation, respectively. Agents coordinate positions on shared grids using Eq. (34), insert them into individual anchored routes through Eqs. (35)–(36), and repeat prediction after rerouting. All executed motion satisfies Eq. (1). Local spatial conditions therefore govern both initial arrangements and their subsequent retention or reconstruction.

Shared channels provide a common spatial basis for yielding and clearing. Separated positions keep passage space available, yielding separates arrival times, and unidirectional channel clearing recursively brings forward blockers into the same direction of motion. Lateral separation along channel axes follows from Eq. (26); finite-step clearing also requires exit space to receive agents and actual progress, as specified in Eq. (38). Records of encounters involving two to five agents and of occupancy changes show how passage and clearing continue through these events. Across complete WP runs, near-stops accounted for 0.393% of navigating agent-steps, and the longest continuous near-stop for the same agent within a trip toward the same goal lasted 1.0 s. These results support the complete method's ability to maintain motion, although the benefits of individual mechanisms remain to be isolated.

In complete-run comparisons with 5–10 agents, WP achieved 25.2%–52.2% more mean goal arrivals than the best baseline. More arrivals were accompanied by shorter mean durations of completed trips, providing consistent evidence of arrival performance. Mean computation time was lower than that of the rolling-horizon K-CBS adaptation, supporting sustained arrivals with manageable computation in the tested setting. These system-level results provide computational evidence for combining shared coordination at encounters with route freedom between them.

The contribution is a planning principle that constructs local passage relationships from real-time space and maintains them during continuous motion. Geometric criteria support construction; shared channels and anchor connections embed local arrangements in multi-agent coordination; urgency-based updates and actual clearing determine when relationships are retained or released. Together, these mechanisms address work–motion space coupling and provide a methodological basis for engineering settings with both task occupancy and continual travel demands.

## 7. Conclusions

We proposed WP to construct and maintain local encounter passability under work–motion space coupling. It organizes joint decisions around predicted body conflicts and integrates real-time compatible spatial allocation, continuous route connections, and blockage clearing in an architecture generated on demand and updated during execution. The necessary and sufficient width condition for fixed-point exchange and the necessary directional condition for composite exchange provide geometric grounds for channel construction and relocation. Shared channels and ordered anchors connect multi-agent encounters; urgency-based updates retain valid relationships and rebuild affected parts. This yields a planning principle for continuous collective motion organized by local passage relationships, combining shared coordination with greedy route freedom and extending lifelong planning to changing work occupancy and continuing movement demands.

Computational experiments on a building floor recorded route continuation after occupancy changes and clearing in multi-agent encounters. Complete WP runs exhibited infrequent, brief near-stops. Across complete 200 s runs with 5–10 agents, WP achieved 25.2%–52.2% more mean goal arrivals than the best baseline at each configuration, with lower mean computation time than the rolling-horizon K-CBS adaptation. These findings provide computational evidence for the architecture in engineering production settings. Local clearing remains subject to the stated geometric and execution conditions; the independent effects of its mechanisms and applicability across settings require controlled evaluation.

# 作业动态占用的拥挤运动空间中基于交会空间相容分配的多代理路径规划

刘伯达　合肥工业大学智能制造技术研究院，合肥，中国　wlbdwww@163.com
杨学志*　合肥工业大学计算机与信息学院，合肥，中国　xzyang@hfut.edu.cn
杨彬*　同济大学土木工程学院，上海，中国　yangbi@tongji.edu.cn
李舒　清华大学合肥公共安全研究院，合肥，中国　lishu@tsinghua-hf.edu.cn
* 通讯作者

## 摘要

楼层施工中的多代理终身路径规划需应对持续更新的目标和作业占用变化。已有规划可在线调整路径与时序，但作业与运动空间耦合会改变未完成交会的空间依据，改线又影响后续交会，因而需要构造并持续保持局部可通过性。本文提出了基于交会空间相容分配的路线规划方法（WP），创新在于按实际交会需求和实时宽度构造共同通过关系，以共享通道、有序锚点及紧迫更新组织连续协同，同时保留交会之间的贪婪运动自由。建立的定点换位充要宽度条件与组合换位必要方向条件，为通道构造和交会迁移提供了几何依据。在拥挤楼层计算实验中，WP 完成了占用变化后的路线接续与多方清出，近停比例较低、持续时间较短；与四种方法的完整运行比较显示，5—10 代理各配置的平均目标到达量较最佳对照提高 25.2%—52.2%，平均计算耗时低于 K-CBS 滚动适配。本文的贡献在于建立了依据实时空间构造并持续维护局部通过关系的规划原则，将共同空间分配与有条件的局部疏通统一于同一架构，扩展了终身路径规划对作业与运动空间耦合情景的处理。

## 1. 引言

楼层中的受限活动区域通过狭窄入口相连。工人或移动设备持续往返于作业位置和材料堆场，到达当前目标后继续接受新的运动目标。这构成多代理终身路径规划问题：代理需要在不断更新的目标需求下选择路线，并在共享空间中反复协调相互通过。

多代理路径规划通过协调个体路径与通行时序处理运动冲突，已用于仓储搬运和持续取送任务[1,2]。持续到来的目标、邻居运动及局部受阻要求规划能够在线调整。已有研究利用互惠避碰响应邻居运动[3,4]，利用滚动规划和连续冲突搜索更新路径与时序[5,6]，并通过拓扑引导、走廊推理、安全域及行为规则组织拥挤空间中的通行与受阻恢复[7－14]。

然而，施工任务在产生新运动需求的同时，还通过作业启停占用或释放周边运动空间，形成作业与运动空间耦合。运动目标持续更新，尚未完成交会所依赖的空间条件也可能改变。具体而言，新增占用会压缩速度避让的可用空间，或使原路径、通过位置及次序安排失效；绕行又改变与其他代理的交会，局部修复因而还需接续后续通过关系。本文回顾的研究已分别处理持续任务、环境配置和动态导航[2,5,15－17]，但对作业启停与终身运动共同改变交会条件的过程，针对性讨论仍较少。持续受阻会妨碍施工活动中的运动接续，因此规划需要解决**变化的受限空间中，局部交会可通过性的构造与持续保持**：依据当前空间建立相容的通过安排，在条件变化时保留有效关系、重建受影响部分。

本文据此提出了**基于交会空间相容分配的路线规划方法**（Width-based Planning with Compatible Encounter-Space Allocation，简称 WP）。其创新在于将共同决策集中于当前路线上的预测身体冲突，依据实时宽度、身体尺度与相对运动构造通过关系，再以共享通道和有序锚点接续多方运动。出现新的紧迫交会时，更新受影响安排并保持有效关系，交会之间保留贪婪引导下的连续路线。为建立这种空间分配的几何依据，本文给出了定点换位的充要宽度条件与组合换位的必要方向条件，分别用于通道构造和交会位置迁移。相容通道既组织正常通过，也在接入受阻时为时间让行和单向疏通提供空间基础，使可通过性的构造与持续保持纳入同一规划过程。

在含复杂走廊与狭窄入口的拥挤楼层计算实验中，WP 完成了作业占用变化后的原目标行程接续及两至五人交会的通过与清出。多次完整 WP 运行中，近停占导航代理步 0.393%，同一目标行程内最长连续近停为 1.0 s；交会过程与运动记录表明，WP 在所考察的拥挤场景中实现了局部阻塞疏通。与四种代表性适配方法的完整运行比较显示，WP 在 5—10 代理各配置下的平均目标到达量均高于对照，较各配置最佳对照提高 25.2%—52.2%，且较高密度配置的相对优势更大；平均计算耗时低于 K-CBS 滚动适配。这些结果表明，WP 在所考察的占用变化与持续运动条件下，兼顾了持续通行、到达效率和计算代价。

本文贡献在于将群体协调与个体贪婪运动自由相结合，为网格位置、空间分区和采样连接等基本表达[6,18,19]补充了按实际交会需求生成共同决策结构的方式。其中，几何推导给出了局部构造的空间条件，动态交会记录展示了运动接续，完整运行比较提供了整体性能证据。这一架构扩展了终身路径规划对作业与运动空间耦合过程的处理，并为其他兼有任务占用与持续转移需求的工程场景提供了方法依据。该方法也可帮助探索机械臂等系统在动态构型空间中的局部避碰依赖；相应的身体分离、运动可达与疏通条件需结合其运动学重新建立。

## 2. 相关研究现状

多代理路径规划（MAPF）通过协调共享位置的占用与通行时序，支持代理在各自目标之间运动；空间表达、冲突定义与优化目标共同确定求解要求[20]。针对作业与运动空间耦合，本文从三个递进问题回顾相关研究：共同决策如何集中于必要交互，局部空间如何支持相互通过，以及通过关系如何随运动和占用变化延续。

### 2.1 共同决策的局部化与空间表达

反应式避碰直接调整速度以响应邻居运动。速度障碍方法排除可能导致碰撞的相对速度[21]；互惠速度障碍（RVO）由双方分担避让，最优互惠避碰（ORCA）将速度选择转化为线性约束问题[3,4]。这类方法可与绕障路径结合形成分层导航，但狭窄处仍需安排多个身体共同通过。

搜索方法提前协调路径与占用次序。冲突搜索（CBS）逐次添加约束以消除路径冲突[22]，优先级搜索（PBS）比较代理间的先后关系[23]。图上粒度同时影响几何余量与求解规模：粗图可能舍弃侧移空间，细图增加搜索和更新成本。安全区间路径规划（SIPP）以无碰撞时间区间表示通行时机[24]，运动学与动力学冲突搜索 K-CBS 直接协调连续轨迹[6]，扩展了时空选择范围，也增加了共同求解的负担。

局部耦合仅联合处理相互影响的代理或路径。M*随新交互扩大联合搜索范围[25]；经验规划方法 EMP 在窄道冲突处调用离线协调解，其复用受预存形态限制[9]；拓扑引导方法 CDR-RRT 以通行骨架和容量引导连续搜索[7]。这些方法分别借助交互范围、既有解或空间结构缩小搜索。

施工新增占用会压缩避让方向、改变连接的可行性，并使局部协同范围需要重建。因此，在线调整还需依据当前净空确定侧向位置与通过次序。当前路线已包含连通、绕障和前进方向信息；沿路线预测同步身体冲突，可限定共同决策的位置，为按实时几何构造身体占位选择提供依据。

## 2.2 空间分离与交会完成的条件

空间分离方法先划定可用区域，再生成身体运动。缓冲 Voronoi 区域方法（BVC）在邻近代理的分区中扣除身体余量，并以右手规则恢复受阻运动[18]。这类表达简化碰撞约束，但当分区限制朝目标前进时，仍需调整局部配置。

安全约束也可能与目标驱动共同形成受阻平衡[26]。模型预测控制方法 IMPC-DR 滚动调整空间分离，在相应目标条件下排除稳定死锁，并分析规划的递归可行性[11]；DLSC-CDPOP 结合时变安全域、邻居轨迹交换与分布式优化，再以右让和优先退让处理受阻[13]。这些研究分别联系了身体分离、持续推进与受阻恢复，三者的保证条件不能相互替代。

通行规则还需适应可用几何。社会规则协调利用门的位置、宽度和出入方向安排离开者优先[14]；窄路会车方法 SM-NR 按宽度占用寻找间隙，并选择会车位置与运动方案[27]。道路具有明确纵向，一般二维交会则需结合当前路线确定测宽方向。

施工净空变化后，原有分区、优先方向和容量均需复核。独立目标带来不同运动方向，单一容量值不足以确定通过方式：对向需要分离位置，同向可跟随，换侧涉及接入与离开。减速、改向和等待还须为其他代理保留推进及清出空间；若停留身体封堵通道，仅延长等待不能疏通。因此，局部可通过性需共同联系净空、身体尺度、相对运动与时间让行，明确占位间隔、路线接续和身体清出的条件。

## 2.3 动态变化中的关系保持与重建

通过安排形成后，速度、目标和占用仍会改变，规划须兼顾响应新条件与延续未完成协同。滚动时窗协调方法 RHCR 将持续目标需求分解为有限时域内的冲突求解[5]；窗口末端状态成为下一轮起点，因而当前安排还需考虑后续释放瓶颈空间。

执行偏差未必要求重建全部关系。固定路径上的同伦协调允许通过暂停或延迟调整时序[28]；可切换动作依赖图（SADG）滚动调整动作次序，并以避免循环等待等条件维持执行[29]。分布式协调还可传播运动约束以衔接多任务行程[30]。这些方法为保持有效关系、调整具体运动提供了依据；用于身体交会时，相应依赖仍须落实为共同可用的通过空间。

环境更新的对象也有所不同。可配置环境路径规划 C-MAPF 联合选择布局与路线[16]；路径与动作规划 PAPO 协调施工搬运中的宽度、载荷及方向约束[2]；施工动态导航则结合地图、实时感知与多约束规划更新运动[17]。主动配置改变布局，感知修正通行判断，作业启停直接改变占用。

在施工中，旧避让位置可能失效，并行区可能缩为顺序通行区；新成员、延迟与改线又改变交会紧迫性。因此，保持依赖须复核其几何依据，更新运动须兼容其他未完成交会。整体替换可能打断有效安排，无条件沿用则可能保留失效约束。规划需要明确保持、重建和解除的条件，使新冲突及时进入规划，并在身体清出后释放空间。

## 2.4 研究总结与科学问题

已有研究通过动态避碰、局部求解、空间分离和依赖维护，分别支持安全运动、受阻恢复与连续执行。作业与运动空间耦合使这些要求同时作用于未完成的交会：新

增占用改变几何依据，改线重新关联参与者，目标与实际进度改变后续次序。因此，单条路线恢复可行后仍需接续多方通过关系，局部安排还须具有可执行的接入与清出条件。

本文据此提出科学问题：**变化的受限空间中，局部交会可通过性的构造与持续保持。** 其一，净空、身体尺度和相对运动如何确定相容安排，并为受阻疏通提供空间依据？其二，空间、成员和运动改变后，哪些关系应保留、重建或解除，如何接续后续交会？WP 以实时宽度构造空间相容分配，以锚定路线和紧迫交会更新维持接续；实验评估通行、持续到达与计算代价。下文先建立空间相容条件，再据此组织规划与多代理协同。

# 3. 交会基元与空间相容条件

本章以双体围绕交会点的运动及其接续为基本构造，在理论上建立身体尺度、局部空间与交会完成之间的关系。

## 3.1 场景与运动约束

建筑构件围合出相连的活动区域，工人在构件附近作业，并往返于作业位置与材料堆场。作业开始时，其操作空间从可运动空间中扣除，见图 1。施工任务和材料需求持续产生运动目标，代理到达当前目标后接收下一目标，形成多代理终身路径规划。该场景中，代理建模为半径均为 $r > 0$ 的全向运动圆盘。代理 $i$ 的当前位置、当前目标和额定速度分别记为 $x_i, g_i, v_i$，其中 $v_i > 0$。时刻 $t$ 的障碍集 $O_t$ 包含固定构件与已启用的作业空间；扣除这些占用和场地外部后，可供身体占据的物理空间记为闭集 $\Omega_t$。中心与障碍及场地边界至少相距 $r$ 的位置构成自由空间 $F_t$。设当前有 $M$ 名代理，$y_i(\tau)$ 为当前环境下的一段绝对连续候选中心运动，其速度不超过 $v_i$；$\tau$ 为从规划时刻 $t$ 起计的运动时间，$H > 0$ 为考察时长。运动、停留及其组合统一满足

$$\begin{cases} y_i(0) = x_i, \quad y_i(\tau) \in F_t (1 \le i \le M), \\ \| y_i(\tau) - y_j(\tau) \| \ge 2r \qquad (1 \le i < j \le M). \end{cases} \qquad 0 \le \tau \le H. \tag{1}$$

式(1)依次规定运动起点、障碍净空和同步身体分离，$\|\cdot\|$ 为欧氏范数。停留对应 $y_i(\tau) \equiv x_i$。预测时按路线和时序确定 $H$；实际执行时取运动步长 $\Delta t$，环境改变后重新检查。

在当前路线与时序上预测将发生身体冲突、需要共同组织通过的局部运动过程，称为**交会**；用于该次预测的路线称为**来源路线**。

规划通过位置时预留净距 $b \ge 0$，相应的障碍净空半径为 $R = r + b$，双体安全中心距为 $d = 2r + b$。在交会 $e$ 的闭且有界局部范围 $U$ 内，定义合法中心集和双体相容配置集：

$$\begin{aligned} F_e(t) &= \{x \in U : x + B_R \subseteq \Omega_t\}, \\ \mathcal{Q}_e(t) &= \{(x_i, x_j) : x_i, x_j \in F_e(t), \| x_i - x_j \| \ge d\}. \end{aligned} \tag{2}$$

$B_R$ 为以原点为圆心、半径为 $R$ 的闭圆盘，$x + B_R$ 表示其中心移至 $x$ 后的占位。$U$ 限定分析范围，$\Omega_t$ 决定身体可用空间；$R$ 约束障碍净空，$d$ 约束双体间距。双方沿有界速度产生的绝对连续路径运动，始终保持相容配置，并完成规定的接入、相互通过和身体清出，称为一次**相容通过**。这些规划净距用于局部空间与通道构造，实际身体运动按式(1)检查。 图 1 对应展示了场景占用和两类距离约束。

连续测宽还要求两身体之间具有直通净空。在当前环境下，这一**共同可见条件**写为

$$[y_i(\tau), y_j(\tau)] \subseteq F_e(t), \qquad 0 \le \tau \le H. \tag{3}$$

其中 $[y_i(\tau), y_j(\tau)]$ 为同一时刻两中心之间的闭线段。式(3)要求线段上各点均具备半径 $R$ 的障碍净空。在非凸空间中，两个合法中心之间的直线段仍可能穿过障碍，因此连续宽度构造采用式(3)，一般双体运动采用式(2)。需要严格净距余量时，双方中心距大于 $d$，各中心至障碍的距离大于 $R$。

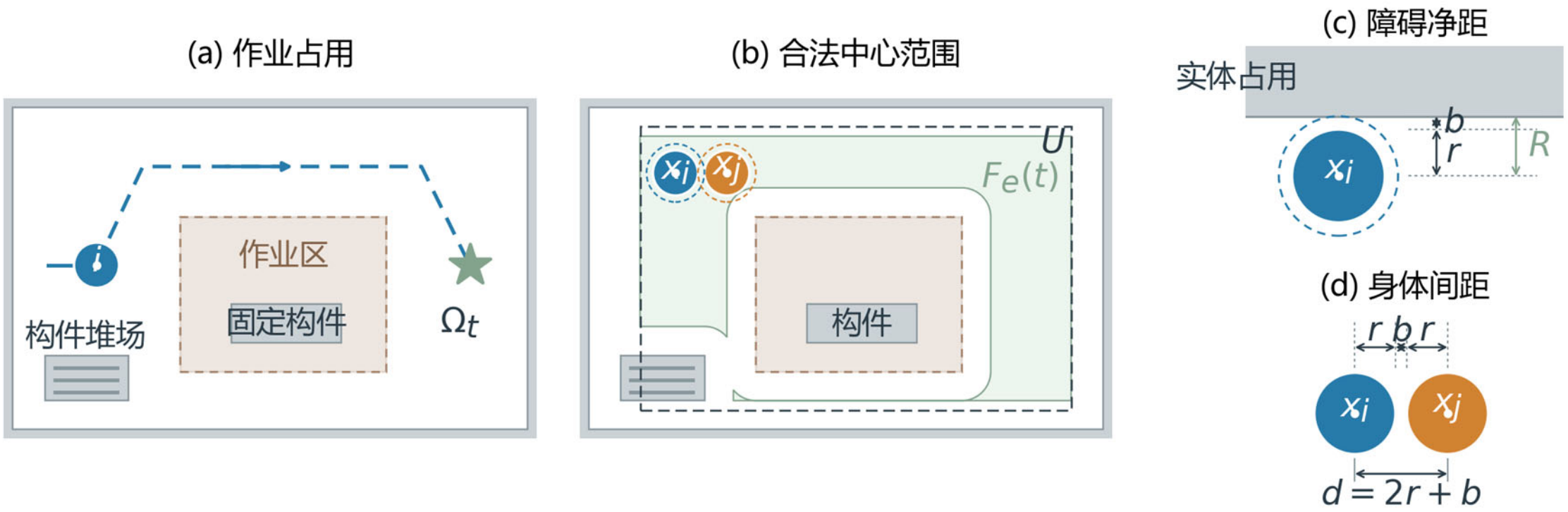

图 1　场景与统一运动约束。作业占用改变物理空间 $\Omega_t$；在局部范围 $U$ 内计入身体尺度与障碍净距，得到合法中心域 $F_e(t)$。$r$ 为身体半径，$R = r + b$ 为规划障碍净空半径，$d = 2r + b$ 为双体安全中心距。实际身体按式(1)保持分离，规划占位满足式(2)的净距要求。

环境占用在步内固定、步间更新；新增占用和加入的代理均需满足初始身体净空。作业占用改变后，重新检查受影响的连接。

## 3.2 定点交会基元与空间条件

两代理保持身体分离，由初始位置 $p, q$ 分别到达 $q, p$，称为一次**完整换位**。将两中心连线始终经过同一固定点的一段相容运动称为**定点交会基元**；完整换位可由一个基元完成，也可由多个基元接续完成。以下固定当前环境，简记合法中心域为 $F_e$、相容配置集为 $Q_e$；所用空间在运动期间保持可用，环境改变后重新规划。

### 3.2.1 定点交会基元及其组合

记双体配置为 $X = (x_i, x_j)$。在两中心连线段上选取参照点 $a$，以 $\theta$ 表示由 $j$ 指向 $i$ 的方向角，单位方向为 $e(\theta) = (\cos\theta, \sin\theta)$，则

$$\begin{gathered} x_i = a + \alpha e(\theta), \qquad x_j = a - \beta e(\theta), \\ \alpha, \beta \ge 0, \quad \alpha + \beta \ge d, \quad X \in Q_e. \end{gathered} \tag{4}$$

$\alpha, \beta$ 分别为 $i, j$ 的中心至 $a$ 的距离，两者之和为中心间距。定点交会基元保持 $a$ 固定，使 $\theta, \alpha, \beta$ 随运动变化且始终满足式(4)。图 2 中，双方持续运动，而各时刻的中心连线经过同一个 $a$；一方停留时，可取其中心为参照点。

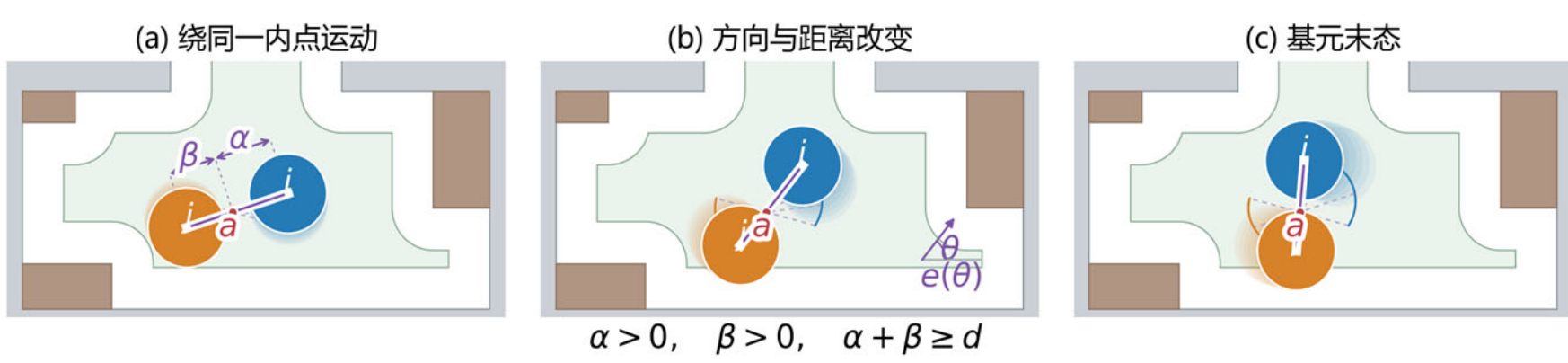

图 2　定点交会基元。(a—c)依次展示同一宽区中的连续运动。两中心连线始终经过固定点 $a$，方向 $\theta$ 与两侧距离 $\alpha, \beta$ 随运动变化。蓝、橙圆盘为当前身体，浅色拖影为近期运动，细实线为中心轨迹。

用 $X_k(u)$ 表示第 $k$ 个基元的双体配置，$u \in [0,1]$ 为段内参数，$a_k$ 为该段固定参照点。以 $m$ 个基元的端态依次连接完整换位：

$$\begin{aligned}\Gamma = X_1 \to \cdots \to X_m, \qquad & X_k(1) = X_{k+1}(0), \\ X_1(0) = (p,q), \qquad & X_m(1) = (q,p).\end{aligned} \tag{5}$$

$\Gamma$ 为组合运动，箭头表示顺序接续；中间等式对 $k = 1, \dots, m-1$ 成立。各段满足式(4)，段间参照点可以改变，而两身体位置连续，首末端态完成位置互换。 图 3 以三个基点的绕转展示这一接续。

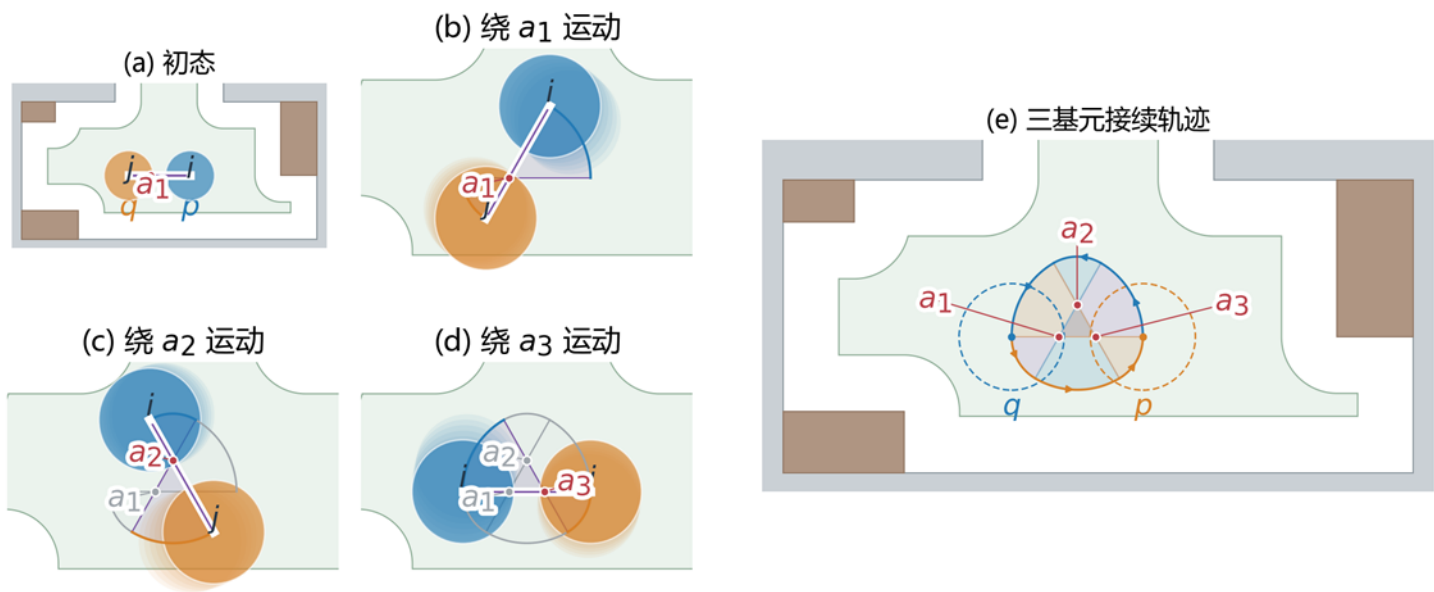

**图 3**　定点交会基元组成完整换位。(b—d)依次突出当前参照点 $a_1, a_2, a_3$，已完成段的参照点、标号和轨迹转为灰色。浅色扇形由当前参照点展开至两条中心轨迹，对应其负责的运动段；当前身体深于拖影。(e)汇合三个基元，初始位置 $p, q$ 在末态分别由另一代理占据。扇形表示参照点与轨迹的对应关系，实际身体扫掠用于检验安全。

**引理 1（定点交会基元的组合表达）。** 在有限时段内，两中心始终位于合法中心域内部、彼此间距始终大于 $d$ 的可行运动，可由有限个定点交会基元连接相同的起止位置，并按运动次序以任意小的空间偏差逼近原轨迹。

**证明。** 双体位置的局部变化可分为连线方向、中心间距和中点位置的变化。绕中点转动改变方向，调整两侧径向距离改变间距和沿线位置；横向位移由两次反向小转动构造。记中点为 $c = (x_i + x_j)/2$，半中心距为 $s = \| x_i - x_j \|/2$，在连线内取 $a = c - \kappa e(\theta)$，其中 $\kappa$ 为有符号偏置，$|\kappa| < s$。先取偏置 $\kappa_1$ 转过 $\varphi$，再按更新后的中点与方向取 $\kappa_2 \neq \kappa_1$，反转 $\varphi$。两次转动后的中点净位移为

$$\Delta c = (\kappa_1 - \kappa_2)[e(\theta + \varphi) - e(\theta)]. \tag{6}$$

方向与间距随之恢复，横向位移 $(\kappa_1 - \kappa_2)\sin\varphi$ 可由小转角连续调节，伴随的沿线位移再由径向移动消去。因此，这些基元可以实现任意足够小的双体位置变化。将原路径作有限细分，再逐段连接相邻配置，即得有限组合。

引理 1 以基元接续表达复杂双体运动。规划可先构造局部相容占位，再连接其前后运动，使交会位置迁移与多方扩展共用这一表达。

### 3.2.2 定点交会的宽度条件

固定交会点后，身体能否随相对方向改变而保持分离，取决于各方向可容纳的中心位置。对 $a \in F_e$，以 $\ell$ 为沿射线的测距长度，定义

$$\begin{aligned} h_a(\theta) &= \max\{\ell \geq 0 : [a, a + \ell e(\theta)] \subseteq F_e\}, \\ W_a(\theta) &= h_a(\theta) + h_a(\theta + \pi). \end{aligned} \tag{7}$$

$[a, x]$ 表示闭线段。$h_a(\theta)$ 是从 $a$ 沿方向 $e(\theta)$ 连续到达合法边界的最大中心距离；$W_a(\theta)$ 为两侧长度之和，称为合法中心弦宽。相反方向给出同一测量轴，故 $W_a(\theta + \pi) = W_a(\theta)$。所有过 $a$ 的连续合法中心弦的并集称为该点的**宽度空间**。式(4)

的两中心位于这一空间时，径向距离分别满足 $\alpha \le h_a(\theta)$、$\beta \le h_a(\theta+\pi)$，因而该方向存在中心距至少为 $d$ 的双体占位，当且仅当

$$W_a(\theta) \ge d. \tag{8}$$

式(8)称为**方向占位条件**，表示中心弦能够容纳一对安全分离的位置。定点完整换位还需经过半周内的全部无向方向，即将相反的两个方向视为同一测量轴；其中最短的中心弦决定该点能否承担换位。

**定理 1（定点交会的宽度条件）。** 设初态 $(p,q) \in \mathcal{Q}_e$、$[p,q] \subseteq F_e$，基点 $a \in [p,q]$ 位于 $F_e$ 内部，径向函数 $h_a$ 关于方向角 Lipschitz 连续。则两代理能够在该宽度空间内以一个定点交会基元完成 $(p,q) \to (q,p)$，当且仅当

$$W_{\min}(a) = \min_{0 \le \theta \le \pi} W_a(\theta) \ge d. \tag{9}$$

$W_{\min}(a)$ 为该点的**最小中心弦宽**，式(9)称为**定点换位条件**。Lipschitz 连续表示径向长度变化不超过转角的固定倍数，使构造轨迹连续且长度有限，可通过调整运动时间满足速度上限。 图 4 对比同一点各方向足宽与某一方向不足的情形。

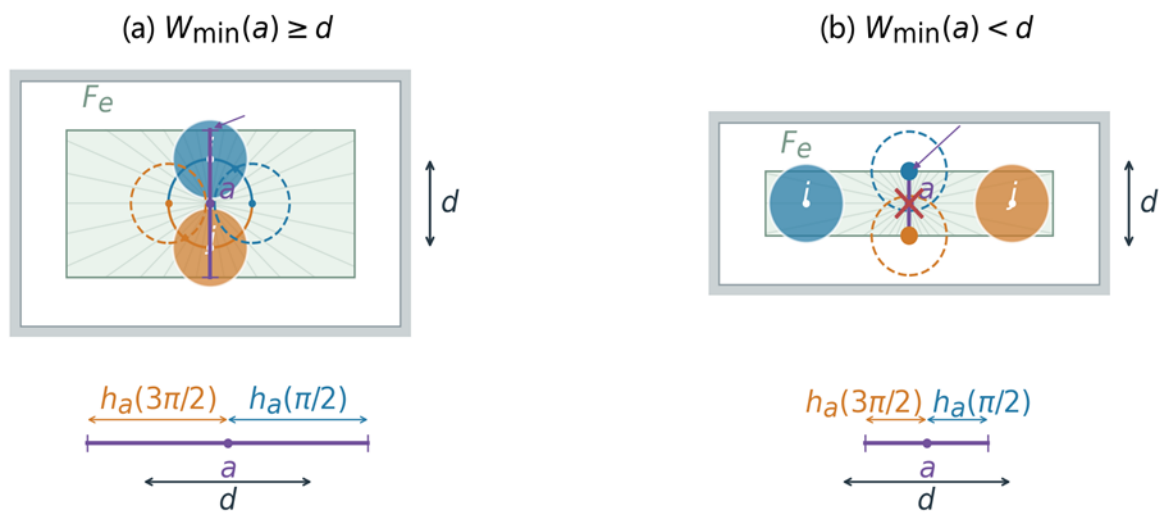

**图 4** 定点换位条件。(a)各方向均满足 $W_a(\theta) \ge d$，两身体可连续转过半周；(b)最窄方向满足 $W_a(\theta) < d$，基元无法经过该方向，因而无法在同一基点完成换位。粗线为最窄中心弦，淡线为其他方向，圆盘为身体；中心间距与中心弦宽按相同尺度绘制。

**证明。** 先证充分性。若式(9)成立，按两侧可用长度的比例分配安全中心距 $d$：

$$\alpha(\theta) = \frac{d\,h_a(\theta)}{W_a(\theta)}, \quad \beta(\theta) = \frac{d\,h_a(\theta+\pi)}{W_a(\theta)}, \\ \alpha + \beta = d, \quad \alpha(\theta+\pi) = \beta(\theta). \tag{10}$$

由 $d/W_a \le 1$，两中心均处于合法射线内且相距 $d$。先沿初始射线将原占位线性调整为式(10)，再连续转过半周，最后沿射线接至互换位置 $q,p$。两段径向连接的中心距均在原间距与 $d$ 之间，转动期间则恒为 $d$；半周后两侧距离交换，因而全过程合法。径向函数的 Lipschitz 连续性保证构造路线长度有限，延长执行时间即可满足速度上限。

再证必要性。中心距不小于 $d > 0$，故连线方向始终有定义，角度可沿运动连续累积。首末位置互换要求

$$e(\theta(T)) = -e(\theta(0)) \quad \Rightarrow \quad |\theta(T) - \theta(0)| \ge \pi. \tag{11}$$

其中 $T$ 为完成时刻。角度连续变化必覆盖半周内全部无向方向，各方向均有 $d \le \alpha + \beta \le W_a(\theta)$，故式(9)成立。充分性与必要性得证。

### 3.2.3 组合交会的方向条件

单个位置不足以承担换位时，可按式(5)切换基点，由不同位置共同承担方向变化，并以相同双体端态衔接。为确定可寻找这些位置的空间范围，定义从 $p$ 出发的单体可达集合

$$\mathcal{A} = \{a \in F_e : \exists\, \eta : [0,1] \to F_e,\ \eta(0) = p,\ \eta(1) = a\}. \tag{12}$$

$\eta$ 为绝对连续的中心路线，沿途半径 $R$ 的圆盘满足障碍净空。任意两点若属于 $\mathcal{A}$，均可经各自到 $p$ 的路线相连，因此 $\mathcal{A}$ 描述单体能够连续通行的范围。

**定理 2（组合交会的方向条件）。** 任一完成 $(p,q)\to(q,p)$ 的相容运动，包括式(5)的基元组合，均须满足 $q\in\mathcal{A}$，且

$$\forall\theta\in[0,\pi),\quad \exists\, a\in\mathcal{A},\ \ell\geq d:\quad a+\ell e(\theta)\in\mathcal{A}. \tag{13}$$

式(13)要求每个无向方向都能在同一可达范围内找到两个相距至少为 $d$ 的合法中心，其中 $a$ 为一端，$\ell$ 为两端距离。全过程满足式(3)时，还得到连续宽度条件

$$\forall\theta\in[0,\pi),\quad \exists\, a\in\mathcal{A}:\quad W_a(\theta)\geq d. \tag{14}$$

**证明。** 安全换位的中心距始终非零，连线方向连续；首末位置互换按式(11)要求其覆盖半周内的全部无向方向。代理 $i$ 的轨迹连接 $p,q$，故 $q\in\mathcal{A}$；沿该轨迹到达 $q$ 后，再接上代理 $j$ 的轨迹，可知两条轨迹上的中心均属于 $\mathcal{A}$。在每个方向出现时，选一方中心为 $a$，另一方为 $a+\ell e(\theta)$，必要时交换两端，身体分离给出 $\ell\geq d$，即式(13)。共同可见时，整段中心连线合法，其长度不超过过 $a$ 的弦宽，故有式(14)。基元组合按式(5)保持端态连续，属于上述安全换位。

式(13)给出一般换位必须覆盖的分离方向；中心连线始终合法时，同一证明进一步得到式(14)。将这些方向上的占位连接为完整运动，还需满足双体连续接续。以一方候行、另一方移动的交替安排为例，第 $k$ 段若以候行身体中心 $a_k\in F_e$ 为基点，另一身体沿绝对连续路线 $\eta_k$ 运动，则接续满足

$$\eta_k([0,1])\subseteq F_e,\qquad \min_{0\leq u\leq 1}\parallel\eta_k(u)-a_k\parallel\geq d,\qquad X_k(1)=X_{k+1}(0). \tag{15}$$

前两项分别保证移动身体的障碍净空与双方间距，末项要求相邻段的双体端态一致；双方共同运动时按式(4)、(5)连接。图 5 以三条窄通道展示式(13)所需的方向覆盖，以及满足式(15)的接续：先由 $i$ 在 $a_1=p$ 候行，$j$ 从通道 2 的 $q$ 移至通道 3 的 $a_2$；再由 $j$ 候行，$i$ 从通道 1 的 $p$ 移至通道 2 的 $q$；最后由 $i$ 在 $a_3=q$ 候行，$j$ 从 $a_2$ 移至 $p$。双体端态依次为 $(p,q)$、$(p,a_2)$、$(q,a_2)$、$(q,p)$；相对方向 $\theta$ 从 $\pi$ 依次变为 $5\pi/4$、$7\pi/4$、$2\pi$，三个基元连续完成半周换位。

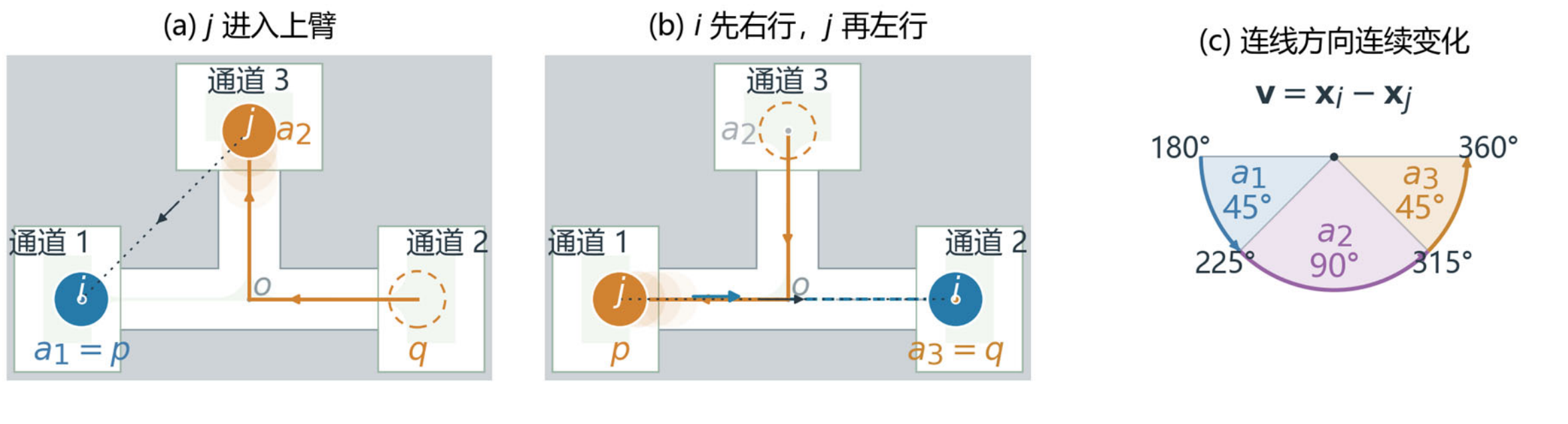

**图 5** 三处基点接续完整换位。(a) $i$ 在左臂候行，$j$ 进入上臂；(b) $i$ 先进入右臂，随后 $j$ 从上臂进入左臂，实圆盘表示最终位置，虚圆盘表示中间状态；(c) 三段相对方向连续覆盖半周。三条通道允许单体依次通行，$o$ 为中央连接位置。$a_1,a_2,a_3$ 分别承担 45°、90°、45° 的相对转角，合计 180°。细点线 $v=x_i-x_j$ 表示中心相对方向，圆盘沿通道路线运动；各段满足式(15)，相邻段以相同双体端态接续。末幅的颜色对应各基元负责的角域。

此例通过交替候行把各段方向变化接成完整换位，展示了定理 2 的方向条件与实际接续之间的关系。绕角时的中心连线可以穿过障碍，因此以一般配置约束式(2)和方向条件式(13)检验；式(14)则适用于下一章中心连线连续合法的宽度空间构造。

定理 1 确定了单点完成换位的宽度条件，定理 2 刻画了组合换位必须覆盖的分离方向，为交会位置选择及其迁移提供几何依据。WP 据此优先在预测冲突处构造相容通道，以连续路线接入、通过并连接相继交会；局部空间不足时，沿可达范围调整交会位置。共同规划由此集中于必要的分离位置，交会之间保留个体路线的选择空间。

## 4. 基于宽度空间的 WP 规划方法

第 3 章以定点交会基元及其组合描述局部运动，并给出身体相容通过的宽度条件。WP 据此以交会组织规划：在预测冲突处安排相容通道，以通道锚点连接贪婪引导的连续路线，并随实际运动和作业占用变化更新相关安排。

WP 的整体架构沿“意图—交会—通道—路线—执行—实际状态反馈”展开（图 6）。目标与障碍确定贪婪意图，当前实施路线及到达时序确定交会，实时测宽与来源路线关系共同确定通道分配。锚点约束路线的必经位置，曲线和速度调整协调接近与离开过程；接入受阻时，沿已建立的通道方向疏通。实际运动决定锚点访问与身体清出，并成为下一轮预测的起点。

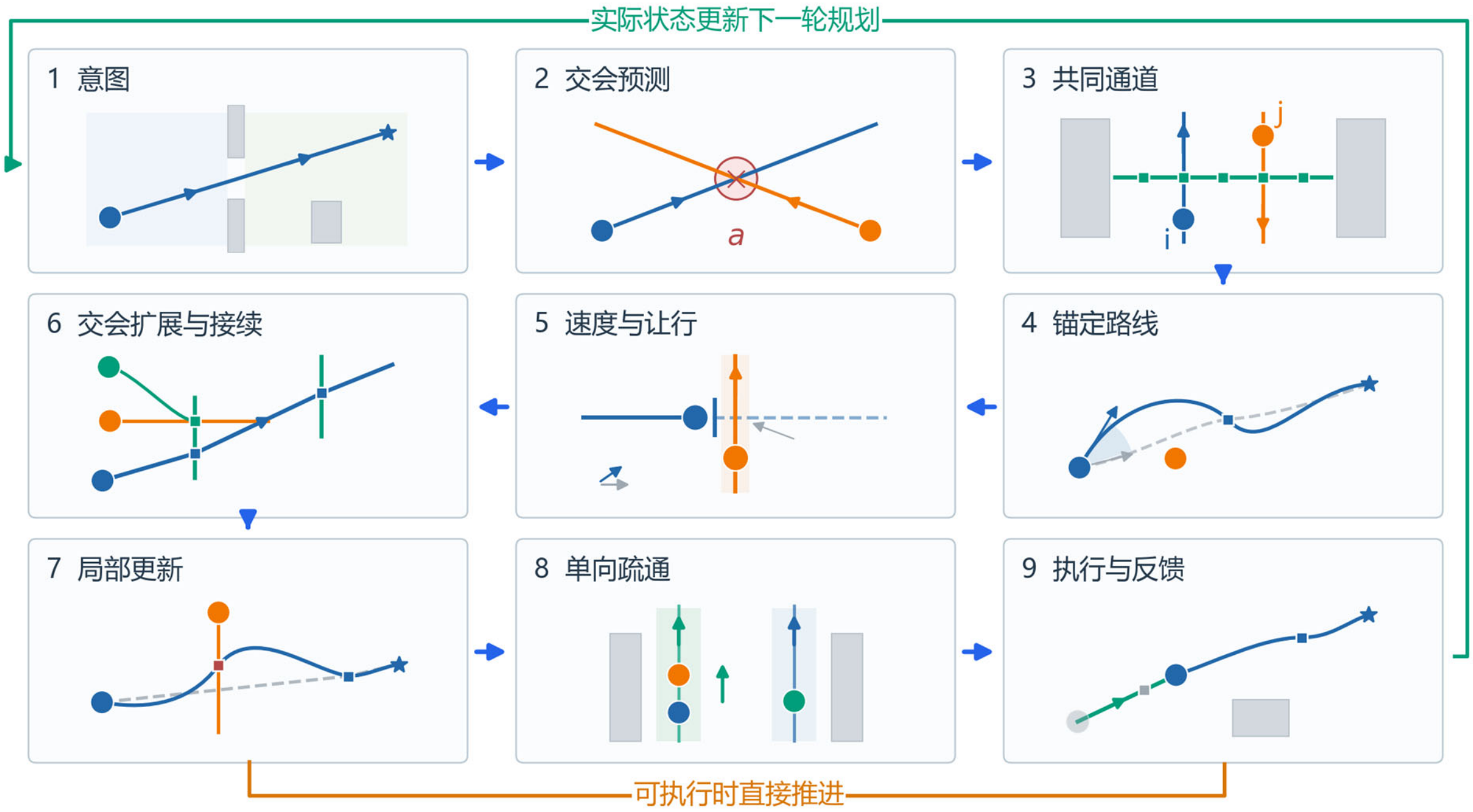

**图 6** WP 规划架构与持续反馈。

### 4.1 意图：随位置和空间更新的贪婪运动参考

代理首先依据固定围挡、当前作业占用和目标构造贪婪路径。沿围挡及门洞、走廊口等开口截面，将运动空间划分为内部连通的区域；相邻区域间可穿越的开口称为入口，其有效范围预留身体净空。以区域为顶点、入口连接为边建立拓扑图，按当前位置、入口代表位置和目标之间的估计行程，选择区域及入口序列

$$R_0 \xrightarrow{P_1} R_1 \xrightarrow{P_2} \cdots \xrightarrow{P_m} R_m. \tag{16}$$

$R_0, R_m$ 分别为当前位置 $x_i$ 和目标 $g_i$ 所在区域，$P_j$ 连接 $R_{j-1}$ 与 $R_j$，$m$ 为剩余入口数。式(16)确定入口及其通过次序，区内几何确定具体穿越位置。作业占用启用或释放后，重新检查区内连接与入口可通行性。

记入口 $P_j$ 的穿越位置为 $p_j$，并令 $p_0 = x_i, p_{m+1} = g_i$。各 $p_j$ 由入口中心初始化，再依据前后连接与入口的交点调整；交点受占用阻断时，取入口上的最近合法点。在区域 $R_j$ 内，以 $p_j, p_{j+1}$ 和预留身体净空的绕障边界顶点建立可见图：可容纳身体通过的直线段构成边，边代价为长度。记连接两端的图上路径集合为 $\mathcal{C}_j$，路径 $c$ 的长度为 $\ell(c)$，区内最短连接为

$$L_j \in \operatorname*{argmin}_{c \in \mathcal{C}_j} \ell(c). \tag{17}$$

其中 $j = 0, \ldots, m$，argmin 表示取得最小长度的路径集合。两端可直达时取直线，否则沿可见边界绕行。各段按式(16)接续，在净空约束下消除多余折点并平滑转弯；绕障调整保留入口及其次序，更新实际穿越位置 $p_j$。当前位置与目标间的直线可行且依次穿过所选入口时，直接取该直线与入口的交点，以保持通过方向；同区目标仅需一段连接。

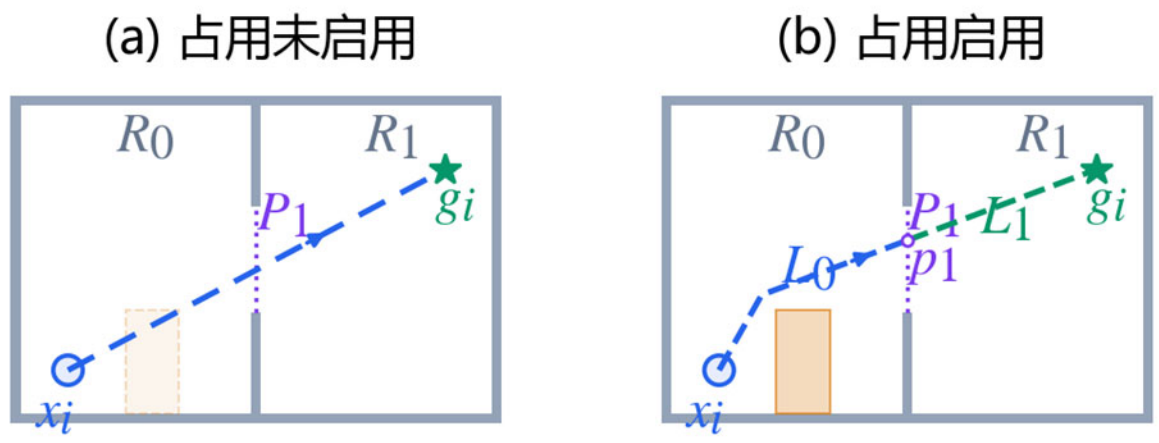

**图 7**　区域与入口约束下的意图。固定围挡分隔 $R_0, R_1$，入口 $P_1$ 连通两者，$p_1$ 为实际穿越位置。(a) 作业区未占用时，路线由 $x_i$ 直达 $g_i$。(b) 作业占用启用后，区内连接绕障并经同一入口趋向目标；$L_0, L_1$ 为入口两侧的连接。

这些局部最短连接经净空约束平滑后形成的贪婪路径称为**运动意图**。当起点 $p$ 与目标 $g$ 之间存在可行连接时，将时刻 $t$ 的意图记为 $\rho_t(p, g)$；代理 $i$ 的当前意图为

$$\rho_i = \rho_t(x_i, g_i). \tag{18}$$

意图随起点、目标和可运动空间更新。入口次序与穿越位置限定式(17)的局部求解范围；等长候选按固定次序选择，使相同输入对应同一选定意图。

代理当前实施的规划路线记为 $\gamma_i$。没有交会通过要求时采用意图；需要经过指定位置时，依次连接这些位置，再从最后一处接续朝目标的意图。

### 4.2 交会：沿路时序预测与共同冲突点

两条路线经过相邻空间时，身体能否安全通过还取决于到达时刻。WP 根据当前路线、沿路进度和速度安排预测同步身体接触，据此确定需要共同协调的局部空间。

将 $\gamma_i$ 按弧长参数化，记全长为 $L_i$、当前进度为 $s_i \in [0, L_i]$，则 $x_i = \gamma_i(s_i)$。根据分段距离、速度和停留时长计算到达时序。以 $\tau \geq 0$ 表示从当前时刻起的预测时间，$\ell_i(\tau)$ 表示这段时间内增加的沿路距离，预测中心为 $\hat{x}_i(\tau) = \gamma_i(s_i + \ell_i(\tau))$。其中 $\ell_i(0) = 0$，$0 \leq \ell_i(\tau) \leq L_i - s_i$；到达目标后进度保持不变。预测假定考察期间目标、路线和环境不变，其有限时域由剩余可推演路段及局部预测时长限定。

在双方预测共同有效的时域内，检验是否存在

$$\| \hat{x}_i(\tau) - \hat{x}_j(\tau) \| \leq 2r. \tag{19}$$

式(19)表示同一时刻两预测圆盘接触或重叠，对应代理对进入交会规划。按路线分段与速度变化时刻划分预测时段，逐段计算相对运动，得到从首次接触持续到分离或预测终点的区间 $I$。取 $I$ 内身体距离最小的时刻 $\tau_*$，以两中心的中点定义冲突点 $a$：

$$\tau_* \in \operatorname*{argmin}_{\tau \in I} \parallel \hat{x}_i(\tau) - \hat{x}_j(\tau) \parallel, \\ a = \frac{\hat{x}_i(\tau_*) + \hat{x}_j(\tau_*)}{2}. \tag{20}$$

式(20)并列最小时取最早时刻。$a$ 由同一时刻的预测中心确定，作为后续测宽原点。首次接触时刻、沿路前向距离、最近接触时的身体位置及来源路线分别用于交会排序、关系判断和宽度测量。 图 8 展示几何路线相交与同步身体接触的区别。

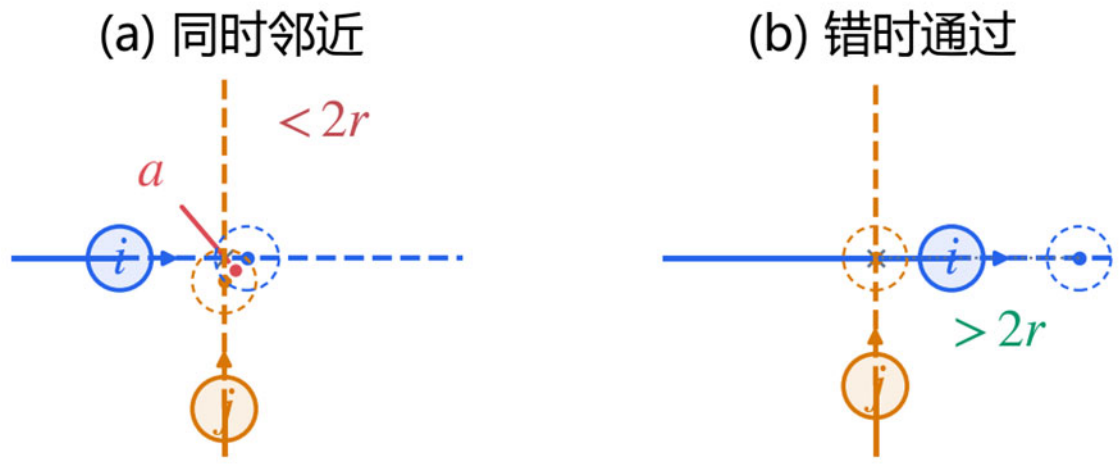

**图 8**　路线交点与身体交会的关系。实线截至当前身体，彩色虚线和虚线圆盘分别表示预测路线与同一未来时刻的身体。(a) 在首次接触区间的最近时刻，预测双中心的中点 $a$ 为测宽原点。(b) 代理 $j$ 恰好到达几何交点时，代理 $i$ 已经通过，预测身体保持分离。

按当前路线与预期沿路速度计算的结果称为**名义预测**。尚无交会约束时优先采用名义预测，未检出接触时再用速度扰动补充候选；已有协调路线则先检验当前时序，再评估速度变化的影响。每个候选保留相互对应的接触时刻、身体位置和来源路线，按风险权重筛选、排序。

速度误差通过沿路进度影响预测位置。设双方在考察时段内沿预测路线运动，实际与预测沿路速度之差的绝对值分别不超过 $\varepsilon_i, \varepsilon_j$，并从已知的当前身体位置出发。由于同一路线上的两点距离不超过其间弧长，经过时间 $\tau$ 后有

$$\parallel x_i(t+\tau) - \hat{x}_i(\tau) \parallel \leq \varepsilon_i \tau, \\ \parallel x_i(t+\tau) - x_j(t+\tau) \parallel \geq \parallel \hat{x}_i(\tau) - \hat{x}_j(\tau) \parallel - (\varepsilon_i + \varepsilon_j)\tau. \tag{21}$$

$x_i(t+\tau)$ 为实际未来位置。式(21)先由速度误差累积界定位置误差，再用三角不等式给出实际双体间距的下界。该下界在考察时段内始终不小于 $2r$ 时，两身体保持分离，从而将预测有效时段与速度误差联系起来。

运行中随实际进度更新预测。沿路避碰改变行程或速度时，重算到达时序；实际运动偏离路线时，先从当前位置重建连接，再预测交会。目标、路线或作业占用改变后，同样重新计算。一名代理阻挡另一名代理下一步运动时，也按式(19)—(20)定位交会；近端受阻和较远冲突由此采用同一空间描述，并由逐步身体检查约束实际执行。

## 4.3 共同通道：从定点交会到相容分配

冲突预测确定需要协调的位置，定理 1 给出该点完成定点换位的宽度条件。WP 据此测量可用空间，构造共同通道，再按来源路线的进出关系分配通过位置；原点空间不足时，探测替代交会点。

### 4.3.1 圆盘测宽与最窄方向

圆盘中心沿一条线段移动时，身体扫过两端为半圆的条带。测量原点须满足 $a \in F_e$，使半径 $R$ 的圆盘能够在该处放置。沿方向 $e(\theta)$，取 $d_R(\theta) = h_a(\theta)$、$d_L(\theta) = h_a(\theta + \pi)$，即两侧首次到达不可用边界前的最大连续中心距离。对应的规划净空包络为

$$\bigcup_{-d_L(\theta)\le u\le d_R(\theta)} (a+ue(\theta)+B_R)\subseteq \Omega_t. \tag{22}$$

式(22)的并集表示沿中心线段放置半径 $R=r+b$ 的圆盘，所得圆头条带厚度为 $2R$；实际身体扫掠半径为 $r$。$d_L, d_R$ 量到端帽圆心，二者之和为合法中心弦宽，已计入障碍净空。完整圆头条带的外包长度还包括两端的 $R$，见图 9。冲突点不具备该净空或无法形成所需合法中心区间时，按第 4.3.4 节选择替代测点。

定理 1 的定点换位条件式(9)以最窄方向检验单点空间。在线计算在非空有限方向集 $\Theta\subset[0,\pi)$ 上生成候选，选取

$$\theta_*\in\arg\min_{\theta\in\Theta}[d_L(\theta)+d_R(\theta)],\qquad w=e(\theta_*). \tag{23}$$

$w$ 为单位宽度轴。并列方向依次按其与来源路线切向的夹角关系、固定方向次序选择；观测范围内未触障的带端记为开放端。式(23)从有限方向中选择候选，式(9)则在全方向上给出定点换位条件。实际候选继续接受圆盘净空与连续路线检查。

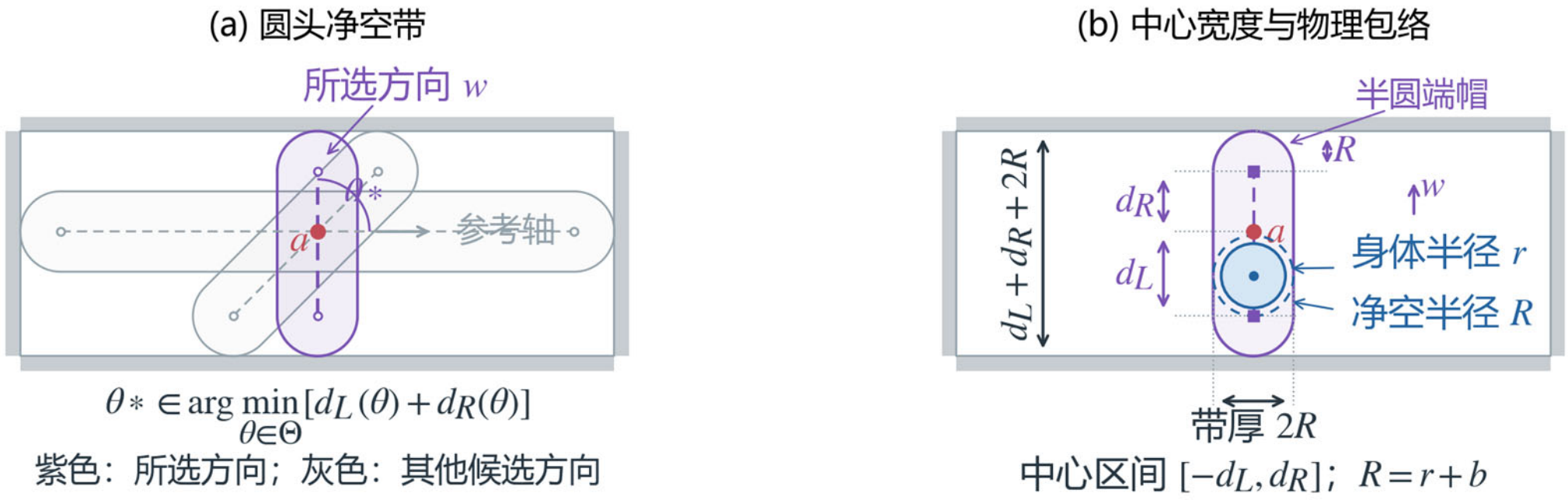

**图 9**　圆盘测宽。(a) 在同一原点比较有限方向的圆头条带，选定轴为 $w$。(b) $d_L, d_R$ 分别量到两端圆盘的中心，端帽半径和条带半厚均为 $R$；身体半径为 $r$。中心可用长度为 $d_L+d_R$，候选中心直接在此范围内分配。

### 4.3.2 从宽度空间到交会栅格

两代理选定的通过位置还可能影响第三方。为使同处交会共享身体占位和左右次序，WP 将测得的合法中心区间离散为共同通道点。

在一段连续合法中心区间 $[u_L, u_R]$ 上，按双体安全中心距 $d=2r+b$ 排列通道点：

$$N=1+\left\lfloor\frac{u_R-u_L}{2r+b}\right\rfloor,\qquad q_k=a+u_k w,$$
$$u_k=u_L+\frac{k-1}{N-1}(u_R-u_L)\quad (N\ge 2). \tag{24}$$

完整双向测量对应 $[u_L,u_R]=[-d_L,d_R]$。$N$ 为该区间可容纳的最大静态占位数，$\lfloor\cdot\rfloor$ 表示向下取整，$k=1,\ldots,N$；$N=1$ 时取中点。任意 $N$ 个有序中心的相邻间距之和至少为 $(N-1)d$，故式(24)达到点数上界；等间距排列又使最小相邻间距最大。

将测量原点、方向与点集组成**交会栅格**：

$$h=(a,w,Q_h),\qquad Q_h=\{q_1,\ldots,q_N\}. \tag{25}$$

取两个分离通道点 $q_k, q_l$，并令 $n$ 为垂直于 $w$ 的单位通行方向。在两条轴向中心路线均满足净空的范围内，以纵向位移 $s$ 构造

$$a_{kl}=\frac{q_k+q_l}{2},\qquad x_i(s)=q_k+sn,\quad x_j(s)=q_l-sn,$$
$$\|x_i-x_j\|^2=\|q_k-q_l\|^2+4s^2\ge d^2. \tag{26}$$

两中心连线始终经过中点 $a_{kl}$，式(26)因而给出一个定点交会基元片段。通道点间距保持横向分离，接入与离开路线确定该片段的方向变化及前后接续。分配给代理的通道点称为其**锚点**。 图 10 展示不同测点的宽度及候选占位。

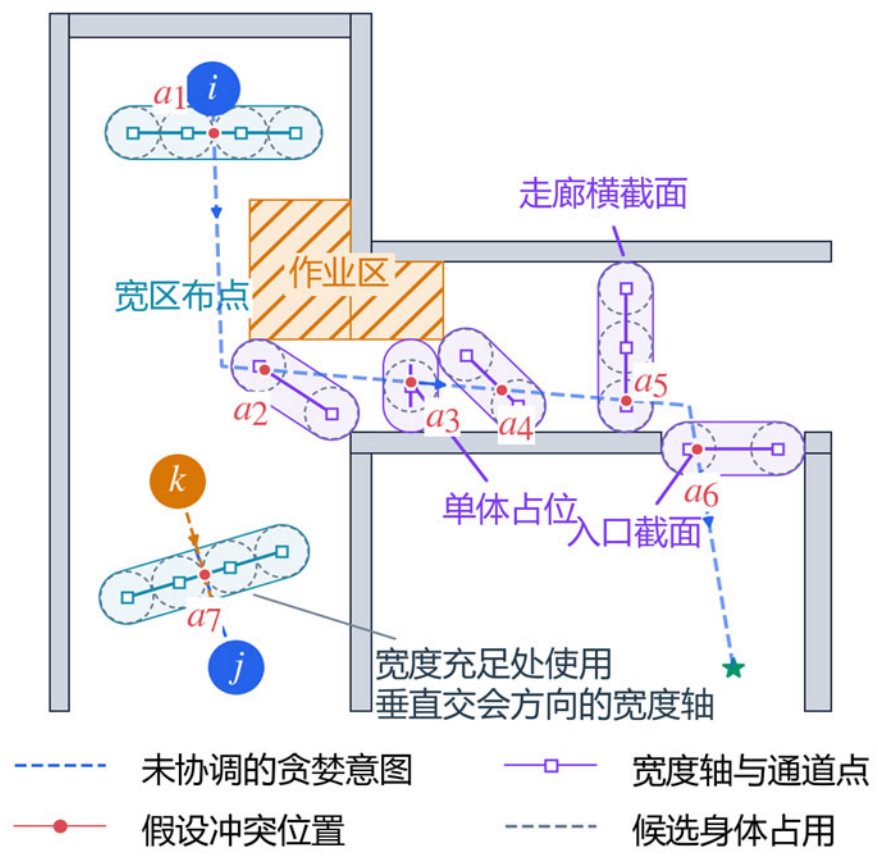

**图 10**　局部宽度与共同通道。$a_1$—$a_6$ 位于示意贪婪意图上，分别表示假设交会的测量原点；实际规划仅在检测到的交会处生成栅格。$a_2$—$a_4$ 由圆盘扫掠测量选择最窄候选方向，$a_3$ 的测量轴为竖直方向；$a_5, a_6$ 为走廊及入口截面。方点表示身体中心，端部半圆和两侧包络共同显示净空范围。$a_7$ 为宽度充足处，采用垂直交会方向的宽度轴，并复核圆盘净空；交会方向由两条来源路线的无向平分方向确定。

### 4.3.3 交会关系下的相容通道分配

通道的使用方式取决于代理如何接近和离开交会处。同向跟随可保持间距先后通过同一点，迎面通过需要分离占位，交叉离开还须协调横向次序。 图 11 以避让前的来源路线展示这些关系。

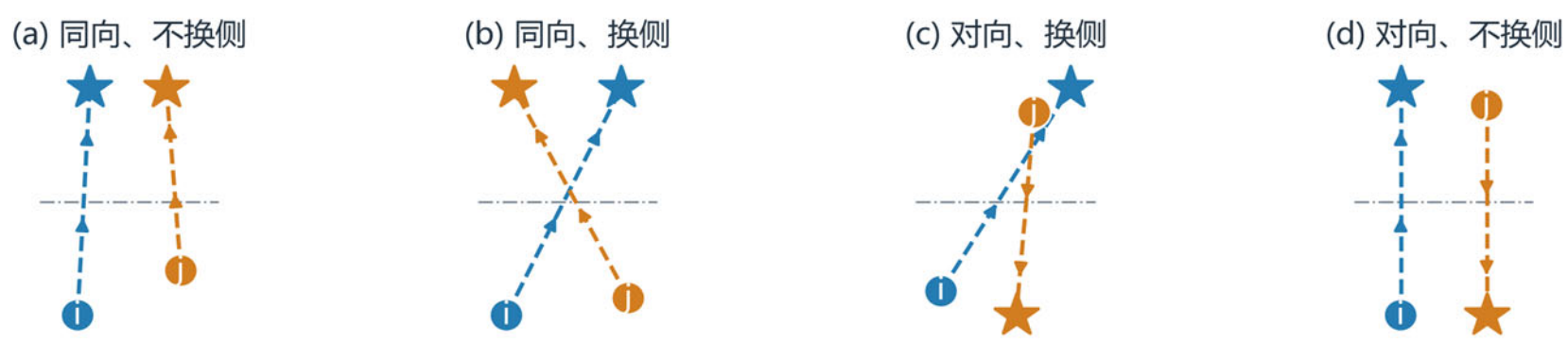

**图 11**　来源路线的四种进出关系。(a) 同向接近并保持横向次序；(b) 同向接近后交叉离开；(c) 相向接近并交叉通过；(d) 相向接近且保持横向次序。圆盘为当前身体，虚线为来源路线，箭头与星形分别表示运动方向和局部离开位置。

表 1　来源路线的四类交会关系及合法通道分配。$S, O$ 分别表示同向、对向，下标 0、1 分别表示不换侧、换侧；第 4.5 节以这些类型约束共同分配。

| 交会类型 | 来源路线的关系 | 通道分配与通过方式 |
|---|---|---|
| 同向不换侧 $S_0$ | 跟随或汇入，横向次序保持 | 共点先后通过；分点保持原侧向方位 |
| 同向换侧 $S_1$ | 同向关系中的横向次序反转 | 共点先后通过；分点可保持或交换方位 |
| 对向不换侧 $O_0$ | 相反侧进入、分别越至对侧，横向次序保持 | 使用两个分离点，保持原侧向方位 |
| 对向换侧 $O_1$ | 对向穿越且横向次序反转 | 使用两个分离点，可保持或交换方位 |

以过冲突点的宽度轴为共同参考直线，比较来源路线在交会前后的代表位置。从直线相反两侧进入并各自通过至对侧，称为**对向**；其余局部关系归入**同向**组，典型情形为跟随和汇入。再沿宽度轴比较两条路线的横向次序，前后反转称为**换侧**，保持称

为不换侧。参考轴反向只改变左右命名，分类保持不变；包含多次转向和往返的运动按相继局部交会处理。测点迁移后仍以原来源路线确定交会关系。

按式(2)保持中心距 $d$ 的不同通道点能够容纳分离的身体。同向代理可在前方身体释放占位后通过同一点；对向和无法形成跟随的大角度交叉采用分离点。四类分配规则见表 1。

换侧可以在通道处完成，也可在保持通道侧序的前提下由接近或离开路线完成；有关连接统一接受路线与同步运动检查。身体临近通道时优先保持当前侧向方位，减少接近段交叉，见图 12。

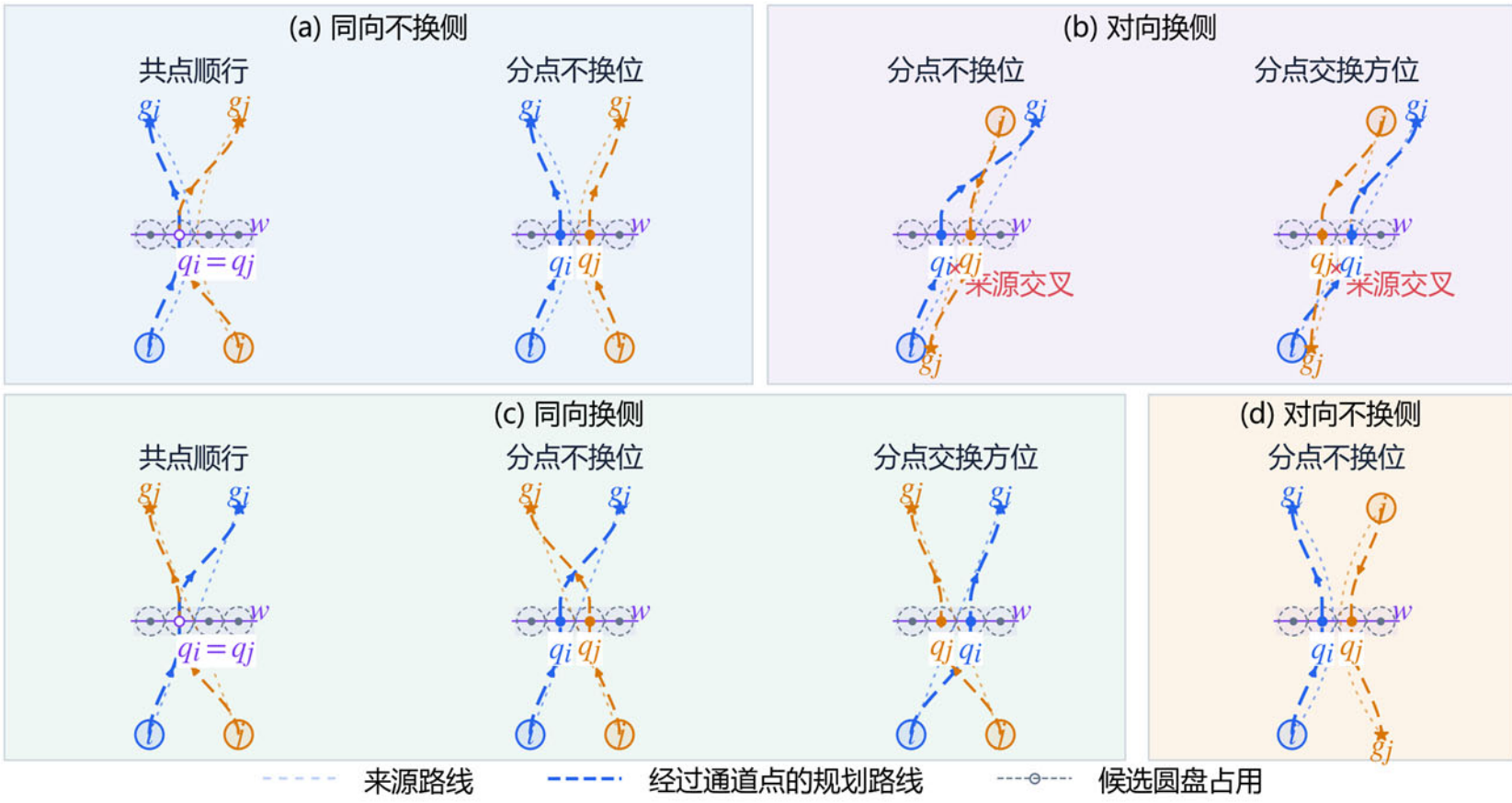

**图 12**　四类交会及合法通道分配。淡色底带标记通道候选区域。浅色虚线为来源路线，深色虚线为通过锚点的路线，星形为目标；$q_i, q_j$ 为通道锚点。(b) 对向换侧关系中，分点选择决定换侧在通道之前还是之后完成。共用点由身体先后通过。

### 4.3.4 宽度不足时寻找替代交会点

定理 1 将单点完整换位与最小中心弦宽联系起来；定理 2 进一步要求组合运动所缺少的方向空间能在可达交会点获得。WP 优先寻找一个可组织通过的替代点，以减少对多基点运动组合的联合搜索。对局部对向通过，双方纵向次序反转时必有纵向位置相同的时刻，此时横向中心距至少为 $d$，因此候选宽度轴上须能放置两个分离身体。

原点无法提供两个合法分离点时，将冲突点投影到相关代理的当前意图，沿意图反向有限探测，并随投影位置更新测量方向。每个新测点重新生成通道，原冲突点仍标识交会来源。双方均能从实际位置接入并继续离开时采用该候选；未找到候选时保留安全安排，随空间和交会变化继续规划（图 13）。

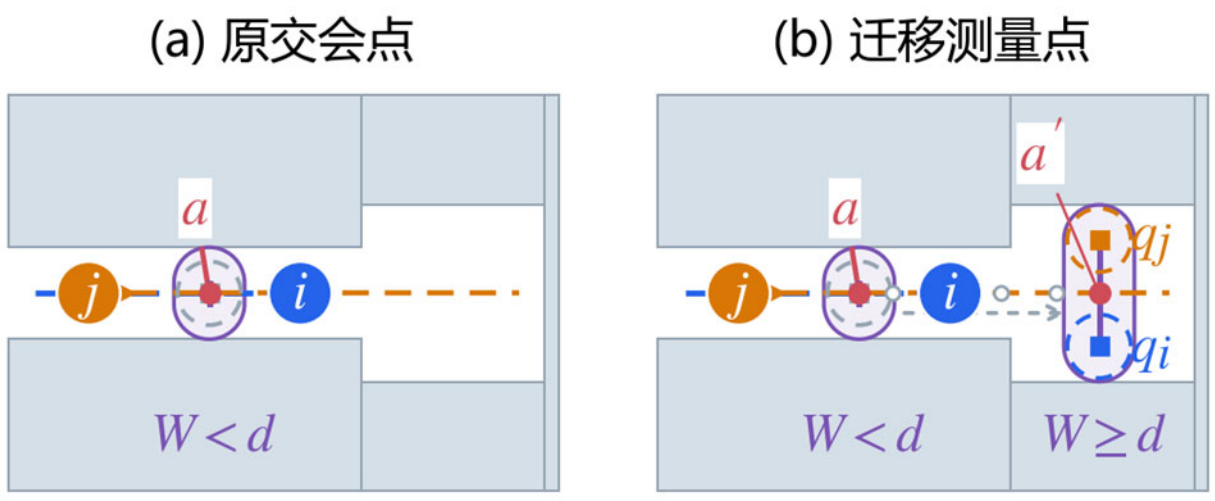

**图 13**　从狭窄交会点探测较宽位置。(a) 原点的中心可用宽度小于 $d$，只能提供一个通道点。(b) 沿代理 $i$ 的意图反向测量，在 $a'$ 获得两个分离候选 $q_i, q_j$。紫色圆头带为规划净空包络，测宽轴上的实线为合法中心区间，虚线圆为候选身体。灰色测点与点线为几何探测，双方身体位置保持；新位置通过路线接入和离开检查后进入规划。

## 4.4 路线与执行：锚点约束下的空间避碰与速度调整

通道分配给出交会的必经位置。WP 依次连接这些位置形成连续路线，先保持交会安排调整路线形状，再以速度偏让或停留错开身体运动；接续仍受阻时，重新分配相关锚点。

### 4.4.1 锚定路线及状态更新

沿用通道点作为锚点，并将获准保留的原路线通过位置同样纳入锚点。代理 $i$ 尚需访问的 $m_i$ 个锚点按次序组成 $A_i = (q_{i,1}, \dots, q_{i,m_i})$；由构造时的当前位置 $x_i$ 出发，经这些点连接目标 $g_i$：

$$x_i \to q_{i,1} \to \cdots \to q_{i,m_i} \to g_i. \tag{27}$$

按第 4.1 节在区域、入口次序和身体净空约束下求取相邻必经点间的最短连接。合并连续重复位置后，得到顶点 $p_0, \dots, p_K$，包括当前位置、锚点、绕障位置和目标；$K$ 为连接段数。末锚点至目标采用该点出发的当前意图；无锚点时采用当前位置的意图。

采用分段三次 Bézier 曲线平滑连接：固定端点保留必经位置，端点附近的控制点调节曲线形状，相邻段共用切向保持方向连续。记第 $k$ 段为 $B_k(u)$，$u \in [0,1]$，其端点为 $B_k(0) = p_{k-1}$、$B_k(1) = p_k$；要求整段位于圆盘中心自由空间 $F_t$，且 $B_k'(u) \neq 0$。将各段依次接续并以弧长 $s$ 参数化，得到实施锚定路线 $\gamma_i(s)$。记前 $k$ 段累计长度为 $S_k$，$S_0 = 0$，总长 $L_i = S_K$，则

$$\gamma_i(s) = B_k(u), \qquad s = S_{k-1} + \int_0^u \| B_k'(v) \| \, \mathrm{d}v, \qquad S_{k-1} \le s \le S_k. \tag{28}$$

$v$ 为积分变量，$B_k'(v)$ 为曲线对段内参数的导数，其范数将参数增量换为弧长。将上述连接和平滑构造记为 $G_t(x_i, A_i, g_i)$，其中切向、控制长度及净空条件按当前连接选定；局部更新沿用保留区间及其边界切向，仅重建第 4.5.2 节确定的改动区间。

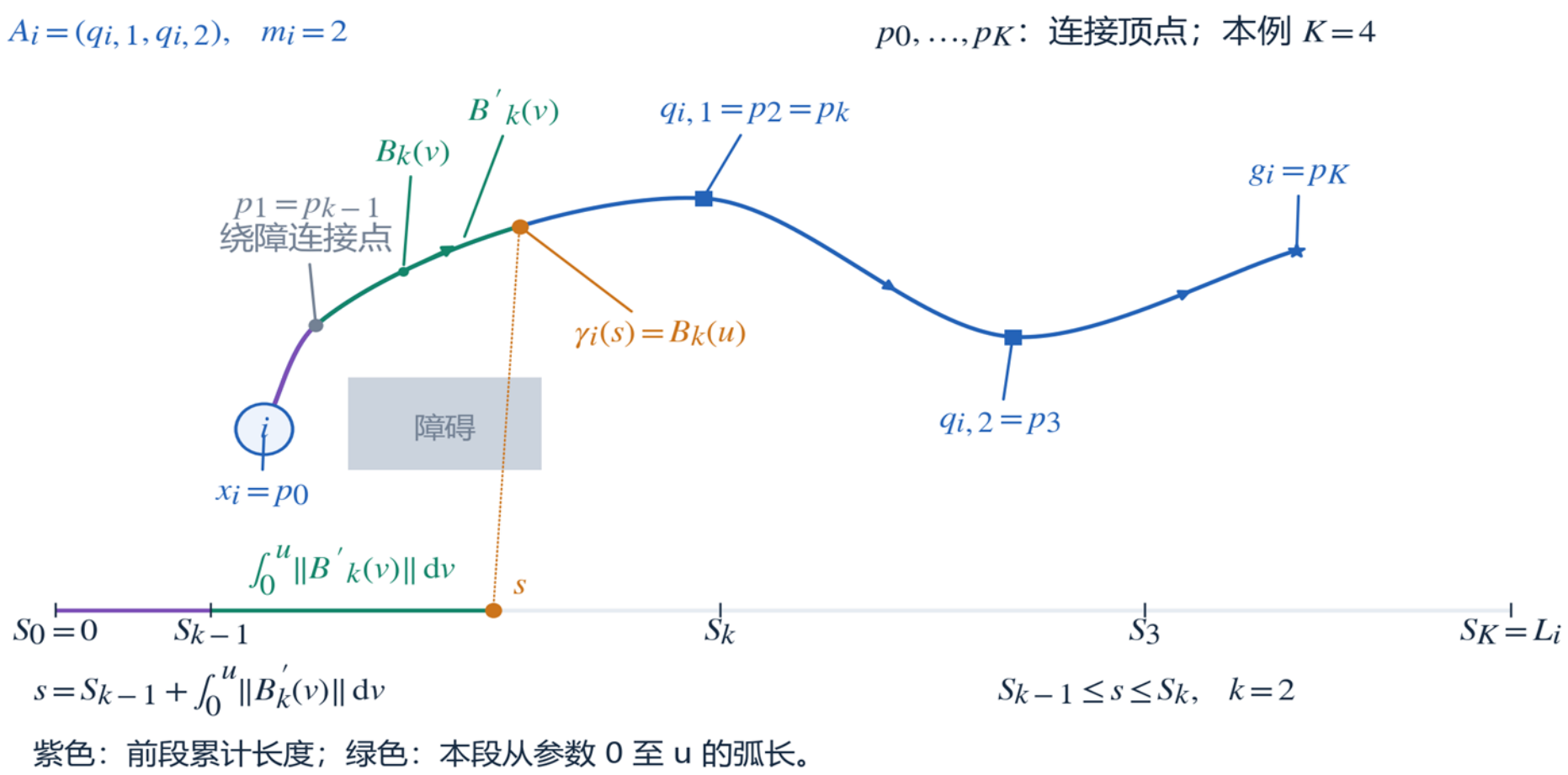

**图 14** 锚点约束与路线弧长。方点、灰点和星形分别表示锚点、绕障连接点和目标。第 $k=2$ 段上的橙点同时表示 $B_k(u)$ 与 $\gamma_i(s)$；紫色前段累计长度 $S_{k-1}$ 与绿色段内弧长相加得到 $s$。

沿路线运动只累积进度 $s_i$，停留时进度不变。速度偏让使身体离开路线，或曲线、锚点、目标、环境变化要求重接时，从实际位置重新构造。用 $\eta_i = 0,1$ 分别表示沿用与重建，上标“+”表示更新后的量：

$$(\gamma_i^+, s_i^+) = \begin{cases} (\gamma_i, s_i + \Delta s_i), & \eta_i = 0, \\ (G_{t^+}(x_i^+, A_i^+, g_i^+), 0), & \eta_i = 1. \end{cases} \tag{29}$$

$\Delta s_i \in [0, L_i - s_i]$ 为本步沿路进度，停留时为零；$A_i^+, g_i^+$ 为更新后的锚点与目标。步内重规划取 $t^+ = t$，身体位置不变；执行后取 $t^+ = t + \Delta t$，$x_i^+$ 为实际终点，$\Delta t$ 为步长。已访问的锚点从序列中移除，身体清出要求继续保留，原路线余段仍可沿用。重建成功后进度归零；失败时舍弃候选并继续调整。预测与执行据此使用一致的路线和进度，锚点插入、替换与移除见第 4.5 节。

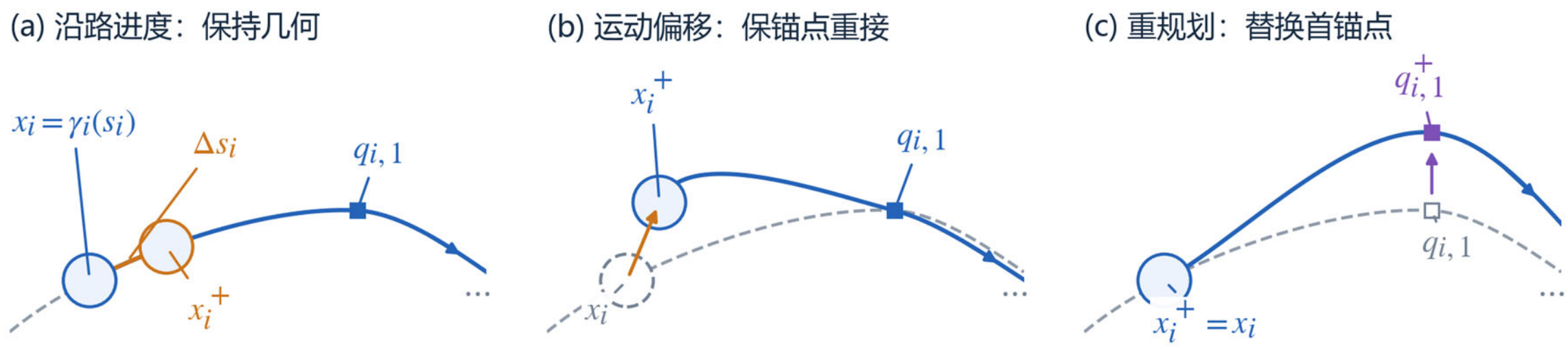

图 15　实施锚定路线的状态更新。(a) 沿原路线前进，几何保持；停留时进度不变。(b) 偏让后，从实际终点 $x_i^+$ 接续原锚点。(c) 步内替换首锚点后，从不变的身体位置接入更新序列。两种重建均使进度归零；右侧箭头与省略号表示后续锚点及目标，三例均保持目标不变。

### 4.4.2 固定锚点下的路线避碰

通道点分离后，接近和离开路线仍可能产生同步接触。记顶点 $p_j$ 的单位切向为 $d_j$；第 $k$ 段两端至相邻控制点的距离为 $a_k, b_k$，相应控制点为 $p_{k-1} + a_k d_{k-1}$ 和 $p_k - b_k d_k$。起点 $p_0 = x_i$ 的参照切向取满足区域和入口约束的首段最短连接方向；首锚点可直达时即指向该点，其余顶点沿用当前锚定路线在该点的切向作为参照。

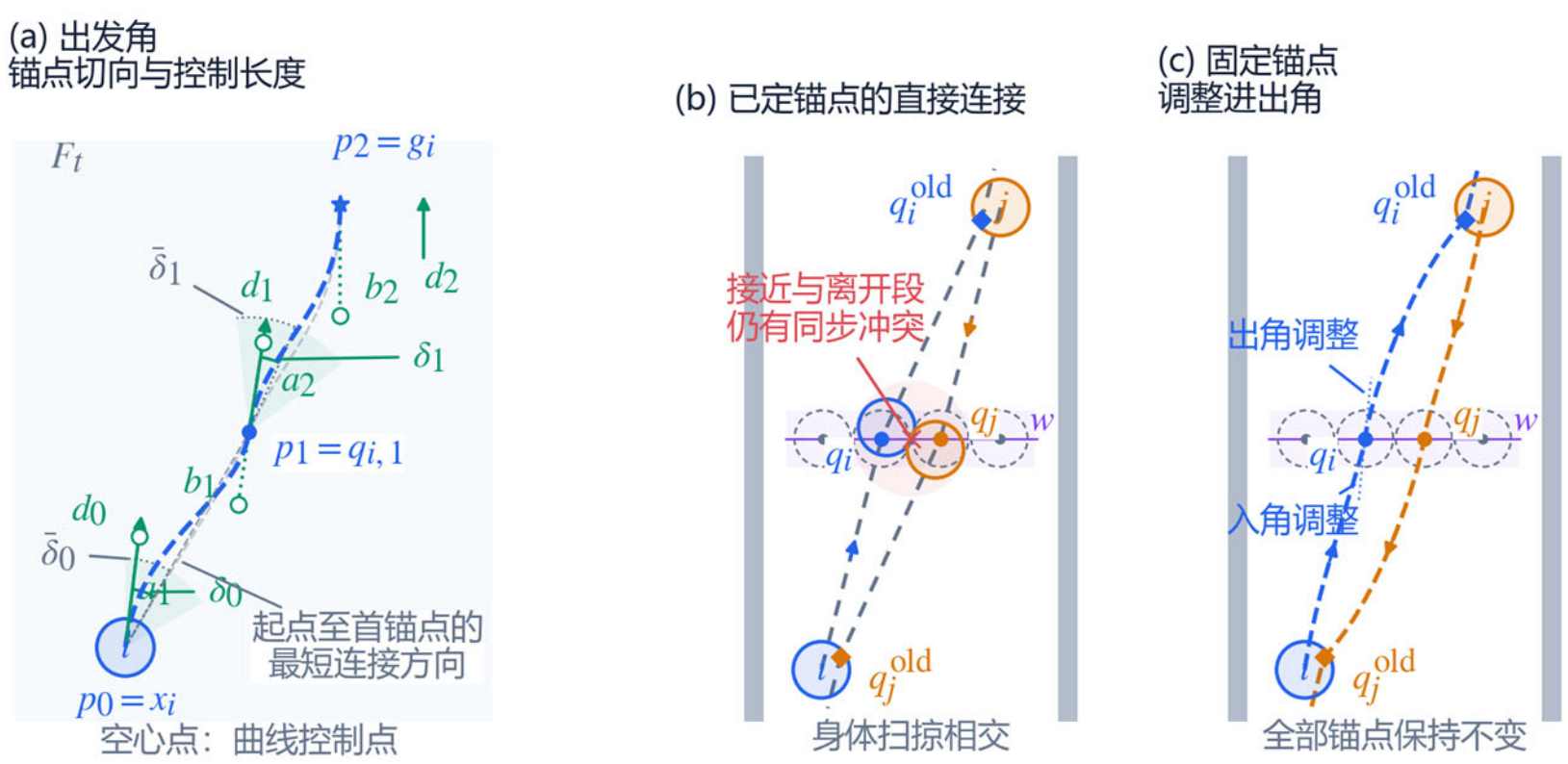

图 16　固定锚点的曲线避让。(a) $p_0 = x_i$、$p_1 = q_{i,1}$、$p_2 = g_i$ 固定，四个空心控制点由 $a_1, b_1, a_2, b_2$ 与 $d_0, d_1, d_2$ 确定。$\delta_0$ 相对起点最短连接方向，$\delta_1$ 相对锚点参照切向；淡色扇区为允许范围。(b、c) 相同起止位置与锚点下，曲线调整改变近端的同步身体间距。

各顶点切向相对参照方向旋转 $\delta_j$，满足

$$\begin{aligned} |\delta_j| \le \bar{\delta}_j \ (j = 0, \dots, K), \\ 0 < a_k, b_k \le \bar{h}_k (k = 1, \dots, K). \end{aligned} \tag{30}$$

$\bar{\delta}_j$ 为方向调整界，固定方向时取零；$\bar{h}_k$ 为邻接长度与净空限定的控制长度上界。旋转后的单位切向仍记为 $d_j$；起点和终点切向分别影响首、末段，内部切向联动相邻两段。有限候选保留锚点、访问次序、既定侧向方位及未受影响的连接，并按式(1)复核障碍净空和同步身体间距。

### 4.4.3 速度偏让

路线调整仍不能提供近端安全运动时，进一步调整速度。WP 采用 RVO 启发的有限偏角候选，处理交会处路线切向近似同向或对向的情形。记切向夹角为 $\theta_{ij} \in [0,\pi]$，当 $\min\{\theta_{ij}, \pi-\theta_{ij}\} \le \bar{\theta}$ 时允许速度偏让，$0 \le \bar{\theta} < \pi/2$ 为角度容差；其余斜交由曲线调整、停留或锚点重规划处理。以当前路线单位切向 $t_i$ 及其左转直角的单位法向 $n_i$ 构造

$$\begin{aligned} \boldsymbol{v}_i^R &= v_i(\cos\alpha_i\, t_i + \sin\alpha_i\, n_i), \qquad |\alpha_i| \le \bar{\alpha}_i, \\ z_i &= x_i + \Delta t\, \boldsymbol{v}_i^R. \end{aligned} \tag{31}$$

$v_i$ 为额定速度，上标 $R$ 标记 RVO 候选，$\alpha_i$ 为偏角，$\bar{\alpha}_i$ 为其上界。按交会侧向关系选取偏向，兼顾朝锚点的进展和身体间距；候选运动 $y_i(\tau) = x_i + \tau \boldsymbol{v}_i^R$ 须满足式(1)，并能从终点 $z_i$ 沿该速度方向重接剩余锚点。将候选与其他代理的当前运动共同预测，在时长 $H \ge \Delta t$ 内检查近端接触；对实际一步取 $H = \Delta t$ 检查全部身体扫掠。执行后按式(29)重接，实际终点等于已检验的 $z_i$ 时复用该接续路线。

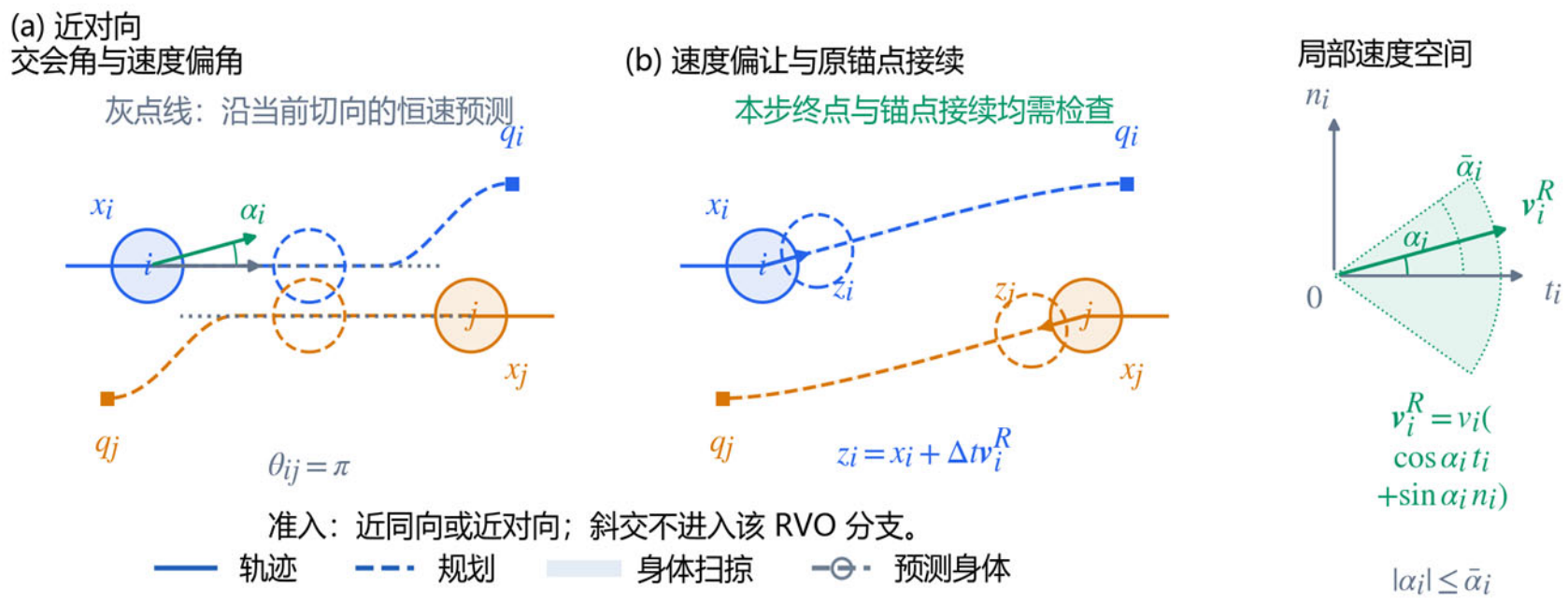

图 17　交会角与速度偏角。(a) 局部近对向，$\theta_{ij} = \pi$，原路线前进将造成接触；灰点线为原切向恒速预测。局部速度图以 $t_i, n_i$ 定位候选 $\boldsymbol{v}_i^R$，淡绿扇区为 $|\alpha_i| \le \bar{\alpha}_i$。(b) 候选通过检查后，从 $z_i = x_i + \Delta t \boldsymbol{v}_i^R$ 接续原锚点，$j$ 同理。

### 4.4.4 停留让行

交会角不满足准入条件、偏角超界，或速度候选无法兼顾安全与接续时，检验双方一停一行能否错开通过时刻。用 $\chi_i = 0,1$ 表示停留与沿路前进：

$$y_i(\tau) = \gamma_i(\min\{s_i + \chi_i v_i \tau, L_i\}), \qquad 0 \le \tau \le \Delta t. \tag{32}$$

式(32)将进度截断于路线终点 $L_i$。停留时位置为 $\gamma_i(s_i)$，路线、锚点和进度均保持。停留身体与其他拟运动身体共同接受式(1)检查；其位置留出对方的通过空间时，原有分点或有序共点安排可继续执行。每步检验，沿路运动恢复安全后解除停留。

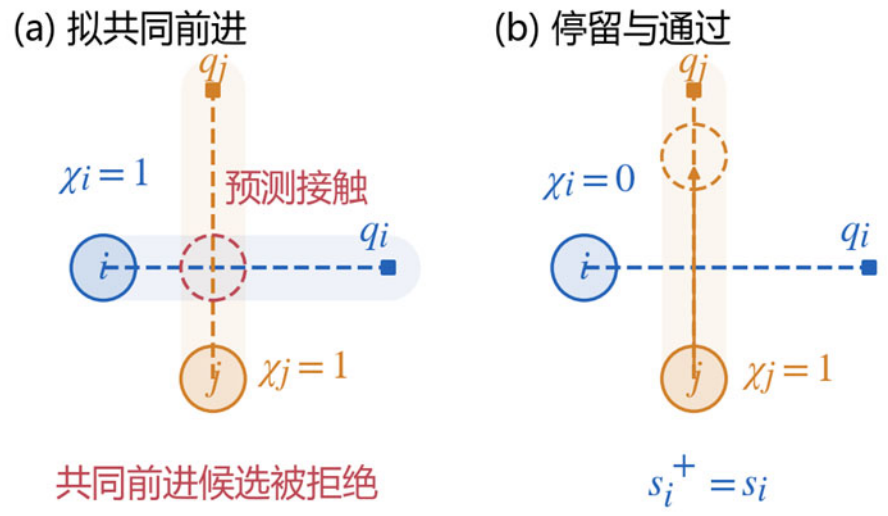

图 18　停留与进度。(a) $\chi_i = \chi_j = 1$ 导致预测身体接触，候选不予执行。(b) $i$ 停留，$\chi_i = 0, s_i^+ = s_i$；$j$ 前进，$\chi_j = 1, s_j^+ = s_j + v_j \Delta t$，到达终点按式(32)截断。两图锚点、路线相同；淡色带为身体扫掠，停留身体位于对方通过范围之外。

停留仍无法恢复接续时，按第 4.5.2 节调整共同安排，并由第 4.5.5 节算法继续协调。

## 4.5 多代理协同：共享通道与交会接续

两代理的避让路线可能影响第三方，已有交会也可能因新成员加入而改变。WP 沿当前实施路线检测这些冲突：能够共用通道的交会在同一栅格上协调，相继发生的交会由锚点序列连接。每次改线后重新预测，优先处理妨碍当前接续的冲突，保留其余有效安排。

### 4.5.1 从新冲突到共享交会栅格

按第 4.2 节从当前路线检出的冲突代理对组成集合 $\mathcal{P}$。优先处理妨碍当前步接续的冲突，其余按双方到预测交会位置的较短剩余行程由近及远选择；并列时按固定代理编号次序选择。目标、空间占用或实际运动改变后，先按式(29)更新路线，再计算 $\mathcal{P}$。没有预测冲突的代理继续沿路运动。

新冲突优先沿用邻近交会栅格的点集，不适用时按第 4.3 节重新测宽。将已有栅格 $h$ 的合法中心线段按净空半径 $R$ 扫掠，定义其覆盖范围

$$D_h = \bigcup_{u_L \le u \le u_R} (a + uw + B_R). \tag{33}$$

$a, w, [u_L, u_R]$ 沿用第 4.3 节测宽结果，$B_R$ 为半径 $R$ 的圆盘；$D_h$ 为带体厚度 $2R$ 的圆头带。新冲突点 $a' \in D_h$ 时，$h$ 成为复用候选。候选点须仍有圆盘净空，能够按表 1 安排交会并接续双方路线。先尝试双方已有栅格，再尝试其他邻近栅格；均不适用时重新测宽，宽度不足则按第 4.3.4 节迁移测点。

新冲突双方沿用同一栅格后，就与该处尚未完成交会的代理共同选择通道。记这些代理为成员集合 $M_h$；当 $|M_h| \ge 3$ 时，两代理交会扩展为共享栅格上的多代理交会。不同方向的栅格仍可分别保留，其接近和离开路线通过后续冲突检测协调，见图 19。

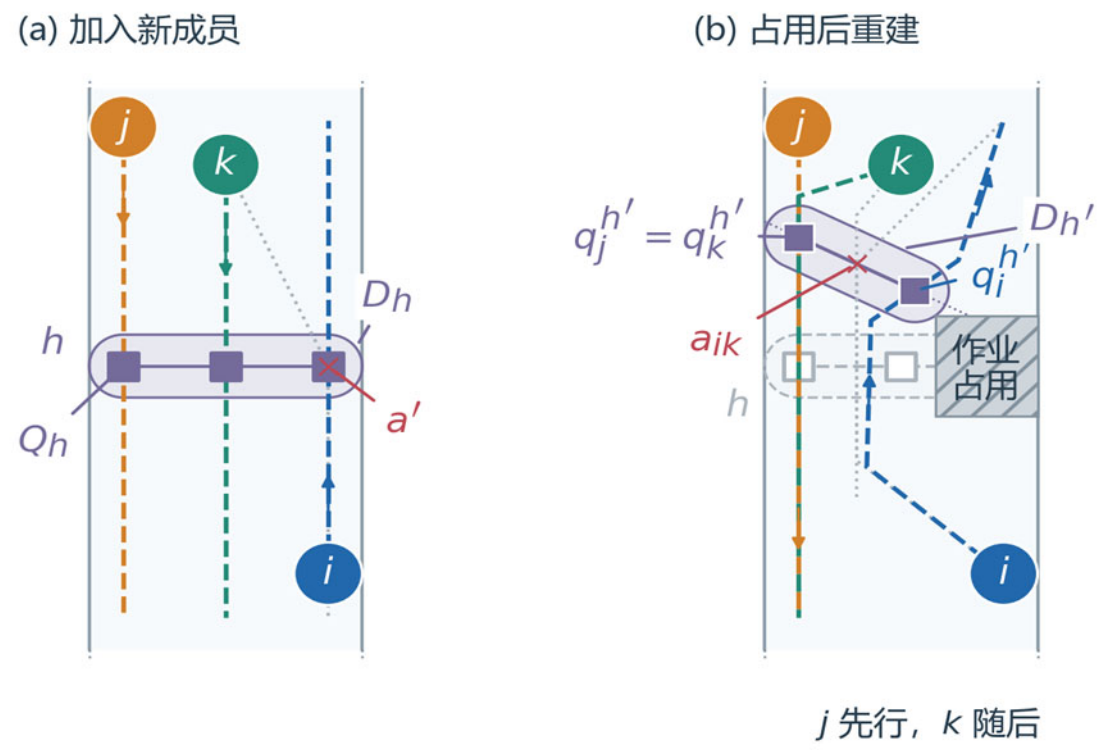

**图 19** 共享通道的形成。(a) 原点集仍可承担新交会，新增成员与已有成员共同分配。(b) 作业占用侵入后，在 $i$ 与 $k$ 的新冲突点 $a_{ik}$ 重新测宽，宽度轴从左侧边界延伸至占用上缘。新栅格包含两个通道点：$j, k$ 沿左侧通道先后通过，$i$ 使用右侧通道。圆头带表示式(33)的覆盖范围，灰色虚线为调整前的路线或栅格。

### 4.5.2 共同分配与锚定路线接续

**同一栅格上的共同分配。** 记成员 $i$ 在栅格 $h$ 上本次通过的位置为 $q_i^h$。新位置取自 $Q_h$；原路线与测量轴的通过位置满足相同净空、间距及表 1 关系时，可继续保留。先为新成员寻找与已有安排相容的通道，无法兼容时再调整相关原位置。

同一成员参与多对交会时，这些交会共同使用其本次通过位置。用 $C_{ij}^h(q_i^h, q_j^h)$ 表示表 1 的合法分配判断：给定来源路线的交会类型，点对满足相应规则时取 1，否则取 0。共同分配要求

$$C_{ij}^h(q_i^h, q_j^h) = 1, \qquad i, j \in M_h, \quad i < j. \tag{34}$$

式中 $i < j$ 使每对成员只检验一次。对向及不能形成跟随的交叉采用分离点；不换侧关系保持原横向次序；允许跟随的同向关系可以共点。不同位置的中心距至少为 $d = 2r + b$，新选栅格点由式(24)满足这一距离。共点时按照前后次序通过，沿路间距和速度按第 4.4 节检验。

需要改点时，在可接续方案中优先选择改动较小、绕行较短者。正在通过的身体继续参加占位检查，后续分配为其保留清出空间。

**从共同位置到个人路线。** $q_i^h$ 按访问次序加入式(27)的锚点序列 $A_i$。同次通过涉及多对交会时只列一个锚点；再次经过同一栅格时另列访问项，并用 $q_{i,k}$ 的序列下标区分。每个锚点关联所属交会和参与者，使局部改线能够确定需复核的共同安排。

新锚点 $q$ 在已有序列 $A_i$ 中的**插入次序**，由其在当前剩余路线上的最近投影弧长确定：

$$s_q = \min \arg\min_{s_i \le s \le L_i} \| \gamma_i(s) - q \|. \tag{35}$$

argmin 给出最近投影对应的弧长，外层 min 选择其中最小者。记已有锚点中访问弧长不大于 $s_q$ 的项数为 $k$，则

$$A_i^+ = (q_{i,1}, \dots, q_{i,k}, q, q_{i,k+1}, \dots, q_{i,m_i}), \qquad 0 \le k \le m_i. \tag{36}$$

$m_i$ 为原剩余锚点数；$k = 0$ 和 $k = m_i$ 分别对应序列首端、末端。同弧长的旧锚点保持原次序，新点排在其后。重选同次交会位置时，先移除原访问项，再按新点的投影弧长插入；其余锚点相对次序保持。双方分别按自己的剩余路线排序。

新锚点排在两个保留端点 $u, v$ 之间时，将连接 $u \to v$ 改为 $u \to q \to v$。替换或删除锚点也只重建相邻连接，按第 4.4.1 节的连接与更新规则保留其余路线及连接边界的切向；局部无法接续时再扩大改动范围。涉及其他交会的锚点或连接时，一并复核相关成员的分配，满足表 1 且能够接续后采用改动。各栅格由此通过同一代理的锚点序列相连。

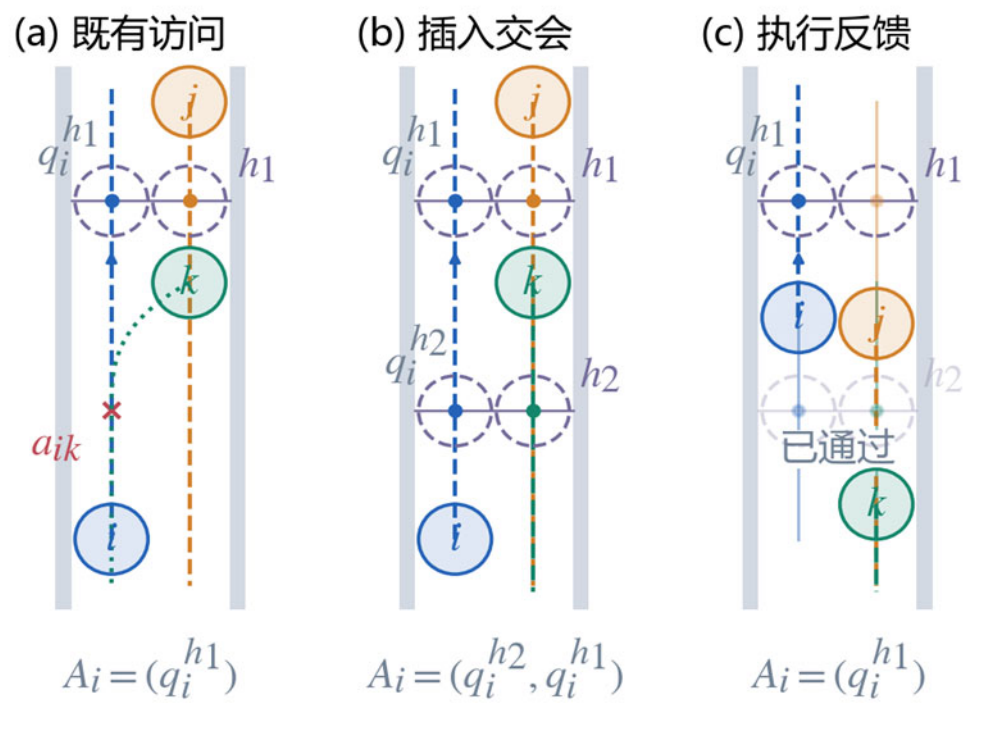

**图 20**　相继交会的路线接续。$q_i^{h_1}, q_i^{h_2}$ 分别为栅格 $h_1, h_2$ 上本次交会的通过位置；新点插在原点之前时，只重建所涉及的连接。通过该点后移除已完成锚点，身体清出后解除相应占位。

### 4.5.3 改线后的冲突回流与实际反馈

**改线后重新预测。** 采用新的通道位置与锚点连接后，按式(19)—(20)重算受影响代理与其他代理之间的接触。前段改线或速度变化引起的到达时间变化同时传递到后段。候选无法接续时，继续尝试其他合法分配或测点。

改线可能引出更早的交会。只要一方将在自己的首锚点前与另一方接触，就继续按第 4.5.1—4.5.2 节安排通道并接入新锚点；双方接触均在各自首锚点之后时，先执行当前接续，随实际推进继续检测。任一方没有剩余锚点时，直接处理其当前预测范围内的冲突。本步运动始终接受式(1)的身体检查。

**执行后保持路线与实际状态一致。** 沿线运动累积进度，停留保持进度；速度偏让造成偏移时，按式(29)从实际位置重接剩余锚点。更新后的路线和到达时序再次用于冲突预测。目标改变或作业占用使原接续失效时，撤去失效的未来安排，重建受影响连接，其余有效安排继续保留。

锚点访问完成后，从序列中移除该必经点；相应占位保留到身体完全清出交会范围。目标改变或锚点重选也保留尚未清出的身体检查。成员完成该处全部通过要求并清出后退出 $M_h$；成员与实际占位均结束时撤销栅格。

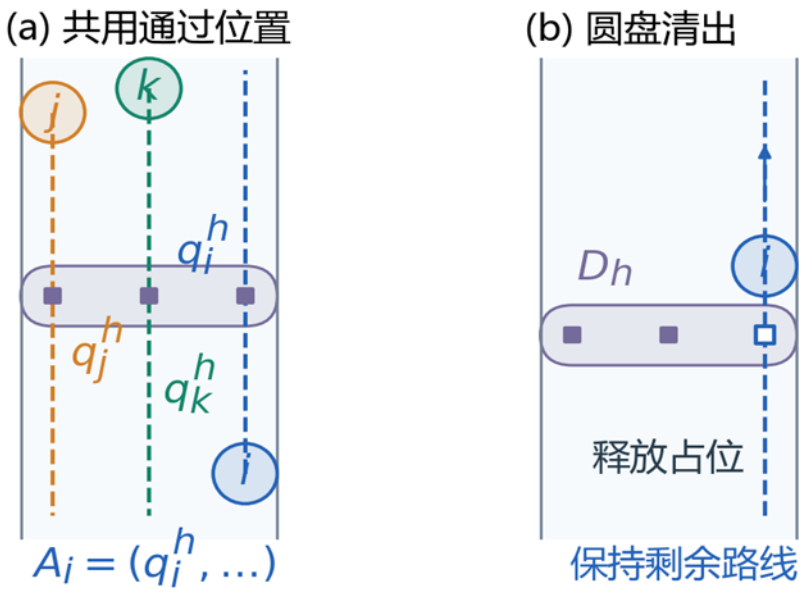

**图 21** 共同安排与实际清出。(a) 共同通过安排；(b) 身体清出后的占位释放。一个通道位置可以同时关联多项成对交会，但在个人路线中只形成一个本次通过锚点。中心经过锚点后继续沿原路线前进，圆盘完全清出后再释放占位。

若交会变化后首锚点已无避让作用，可尝试直接接续下一锚点或目标。短接须保留其他有效安排，在相同预测时域内满足净空与同步身体检查；检查通过后解除该点，并将改线重新纳入预测。锚点为空时恢复式(18)的贪婪意图。

### 4.5.4 接入受阻时的通道疏通

共同通道提供交会处的分离位置，接近各自锚点的身体仍可能相互阻挡。路线调整、速度偏让和停留均无法恢复推进时，沿当前占用通道的规定方向组织顺行，清出后从实际位置接续剩余有效锚点；尚未占用通道的身体按最近合法通道的方向生成接入候选。未清出身体被前方代理阻挡时，将阻挡者也纳入同向疏通；其前方仍有阻挡者，则继续向前扩展，直至连通可接纳身体的出口空间。同一身体同时受到不同通道的疏通请求时，保留已采用的方向，并按第 4.5.1—4.5.2 节重新协调不相容的安排。

取通道点 $q$ 处垂直于宽度轴 $w$、指向规定出口的单位向量 $n$。对纳入该方向疏通的每名代理，在具有圆盘净空的路段内保留建议位移 $u$ 的横向分量，并补足正向分量：

$$u' = u - (u \cdot n)n + \max\{u \cdot n, \beta \parallel u \parallel\} n, \qquad 0 < \beta < 1. \tag{37}$$

$u \cdot n$ 是原位移沿通道的分量，$u - (u \cdot n)n$ 是横向分量；$\beta$ 控制候选向出口推进的程度。原位移为零时，直接生成速度允许的正向候选。候选先按本步速度上限截取，再与其他代理的同一步运动共同检验式(1)，通过后按式(29)接续锚点。

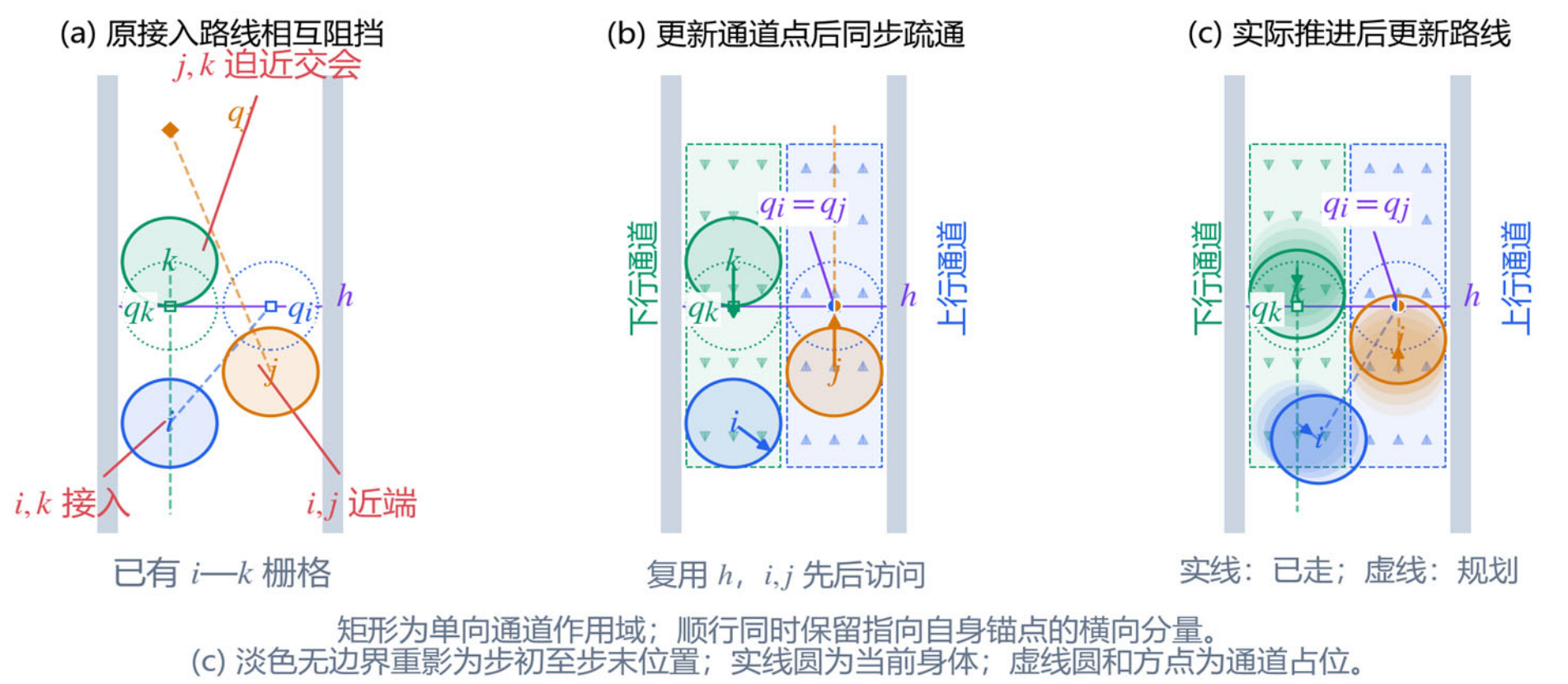

**图 22**　占用通道顺行。受阻身体沿当前通道向出口推进，并在空间允许时接续自身锚点。淡色身体与实色边界分别表示步初和候选步末位置。

定理 1 说明局部换位的空间相容要求，式(26)直接给出分离通道片段的身体间距保证。沿各自通道轴线运动时，不同通道上的身体保持横向分离；同一通道内，后方身体依次跟随前方身体。多代理相互阻挡因而转为相邻双体的间距协调。纳入前方阻挡者后，队首进入可用出口空间，后续身体便依次获得前进空间。

为界定清出时间，固定本轮扩展后的有限成员集合，以各队列前方的可用出口作为清出边界。设通道及出口在清出期间保持可用，成员不后退，且每步至少一名未清出的队首前进 $\min\{\delta, D_i(k)\}$ 或更长距离。其中 $\delta > 0$ 为这一阶段的最小推进尺度，$D_i(k)$ 为第 $k$ 步后成员 $i$ 沿规定方向至身体完全越过该出口边界的剩余纵向距离，已清出时取零。定义

$$\Phi(k) = \sum_i \left\lceil \frac{D_i(k)}{\delta} \right\rceil, \qquad \Phi(k+1) \le \Phi(k) - 1 \quad 当\ \Phi(k) > 0. \tag{38}$$

$\lceil\cdot\rceil$ 表示向上取整，求和遍历本轮待清出的成员；$\Phi$ 将各成员的剩余清出距离换算为推进次数。每步至少一项减少，其余项不增，因此这批成员至多在 $\Phi(0)$ 步内清出。若疏通中又纳入新的前方阻挡者，则对扩展后的成员重新计量剩余清出距离。

递归纳入的身体沿受阻通道规定的方向共同推进；清出后按式(29)从实际位置接续剩余有效锚点。若作业占用截断通道，则重新测宽并调整接续；该类空间改变与前方身体阻挡分别处理。所有候选继续接受式(1)检查，暂不可推进的身体保持停留，其余身体执行可行运动。

### 4.5.5 多代理交会自动规划算法

算法 1 在每步实际运动前，按第 4.5.3 节规定的范围处理当前路线上的紧迫冲突。Predict 依据式(19)—(20)检测交会，Coordinate 按第 4.5.1—4.5.2 节完成栅格选择、共同分配与局部接续；保留锚点即可处理时优先采用第 4.4 节的调整。每步仅进行有限次候选尝试，并跳过同一状态下已失败或重复的方案；发现新的前置冲突时，在剩余尝试范围内继续协调。达到本步尝试上限或无未试候选时，结束内层规划，检验路线跟随、速度偏让和停留；近端冲突仍阻碍接入时，将前方阻挡者纳入通道疏通。尚未解决的冲突随实际状态进入下一步。实际执行只采用满足式(1)的共同运动。

**Algorithm 1. Repeated multi-agent encounter planning**

**Input:** Current positions, goals and free space; routes $\gamma_i$, progress $s_i$, timing $\ell_i$, anchors $A_i$ and active grids.

**Output:** Admissible joint motion and updated routes and encounter arrangements.

```
while the lifelong task continues do
    Update anchors and occupied space from actual motion
    Reconnect changed routes by (29)
    Try first-anchor release while preserving other arrangements // §4.5.3
    P ← Predict current routes and timing by (19)-(20)
    while an eligible conflict in P and a trial remain do
        (i, j) ← Select the most urgent eligible conflict with an untried
  candidate
        proposal ← Coordinate(i, j)                         //
  §§4.5.1-4.5.2
        if proposal is feasible and has not been tried then
            Adopt its common allocation and route connections
            P ← Predict changed routes and timing          // §4.5.3
        end if
        Record the trial
    end while
    Test joint route-following, velocity and yielding motions // §4.4
    If access remains blocked, include forward blockers and test directed
  clearing // §4.5.4
    Execute a safe advancing motion, or remain at rest
    Update progress and actual positions; t ← t + Δt
end while
```

规划内循环保持身体位置不变；实际运动外循环更新路线、占位和交会关系。

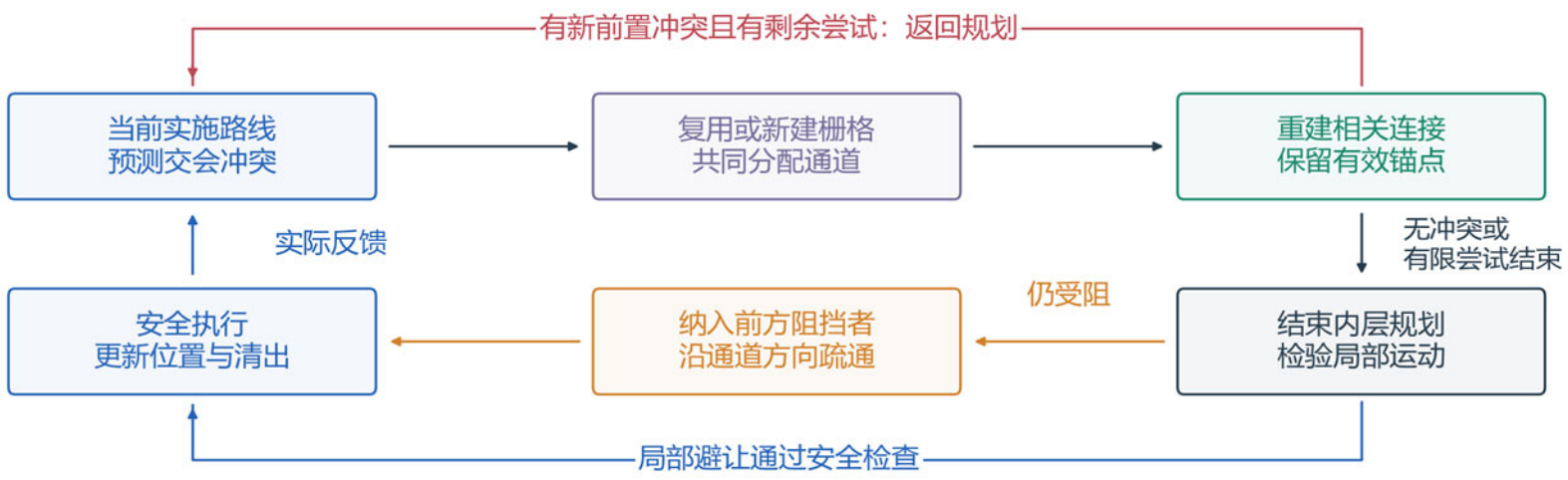

**图 23** 有限重试与执行反馈。新前置冲突仅在本步仍有未试候选和尝试余量时返回共同规划；内层规划结束后检验局部运动，仍受阻则转入通道疏通。所有执行候选均经安全检查，随后依据实际位置和清出状态开始下一步。

# 5. 实验与结果

本章从交会过程、持续到达效率和计算代价三个方面评价 WP。交会回放与运行状态统计用于考察目标更新、成员接入及作业占用变化后的运动接续；与四种代表性方法的比较用于评价完整观察时窗内的目标到达量、行程用时和计算负担。

## 5.1 案例与实验场景

本文以某大厦标准层的楼层布局和施工作业需求构造多代理终身路径规划计算场景。预制装配式与现浇结构混合体系中，垂直构件的分项作业同时产生运动目标与空间占用。不同工种的作业人员往返于材料堆场与作业位置，并在相邻构件之间转移。各人的行程随任务产生，却共同经过入口和连接通道；作业操作空间又压缩邻近通行范围，使交会位置与可用宽度随施工状态改变。

图 24 给出楼层布局与运动对象。墙体（W）、柱子（C）和预制件（PC）构成主要施工对象，围挡及构件之间的开口连接各活动区域。作业人员分为钢筋（R）、模板（F）、预制件接头处理（JC）、预制件灌浆（G）和吊装（HI）五类工种，选择类型相容的任务，并以等半径圆盘表示身体。进行中的作业占据相应操作空间，作业位置与堆场交替成为人员的运动目标。

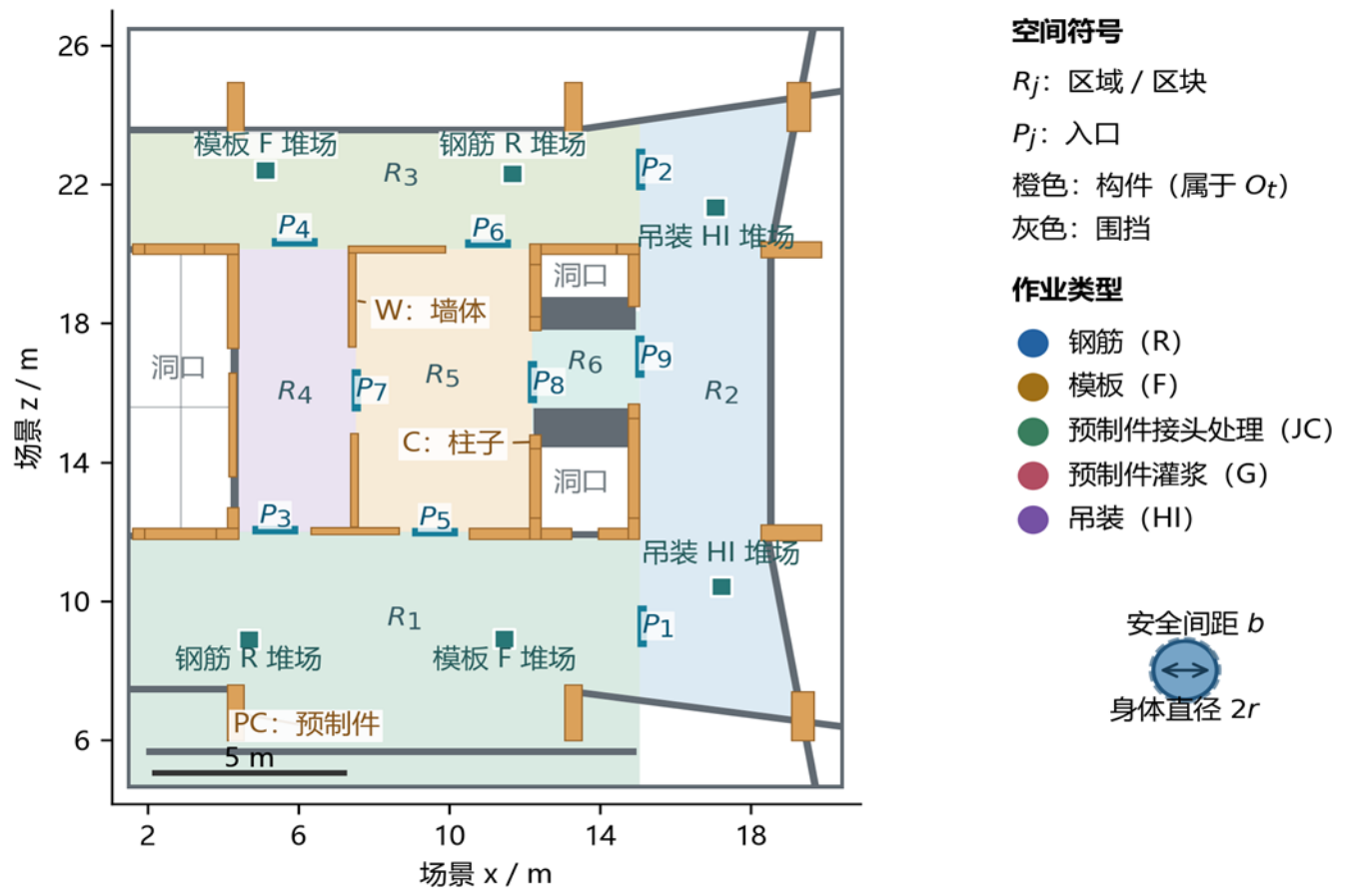

**图 24**　施工布局与圆盘运动模型。区域以 $R_j$ 标记，通过入口 $P_j$ 相连；橙色表示构件，灰色表示围挡。右侧列出作业类型，圆盘外的窄环表示第 3.1 节定义的安全间距 $b$，环的厚度仅作示意。

为在有限时窗内形成反复交会，实验加快作业推进与材料补给，并减小单次携带量，使人员频繁往返于作业位置和堆场。后续目标从类型相容且可达的位置中选择；任务产生的转移需求不足时，以同类型可达位置补充运动需求。该设置保留作业占用对通行空间的影响，并增加持续运动与交会的观测。

实验采用 1000 模拟步（对应 200 s）的统一观察时窗，结合交会回放与多次完整运行统计；对照试验设置 5—10 名代理的六种配置，每种方法在各配置下重复运行。

## 5.2 动态交会的通过结果与运行状态

从保存的规划状态和逐步运动记录中，选取两人通过、目标更新、占用变化及多方接续过程。图 25 — 图 29 各以三个连续阶段呈现通道安排、路线变化与实际通过。

### 5.2.1 两代理通过与目标更新后的接续

图 25 中，两名代理经共同宽度上的不同通道点，沿相反方向通过连接处。末帧中双方已分向离开，呈现了分点安排下的身体分离与实际通过。

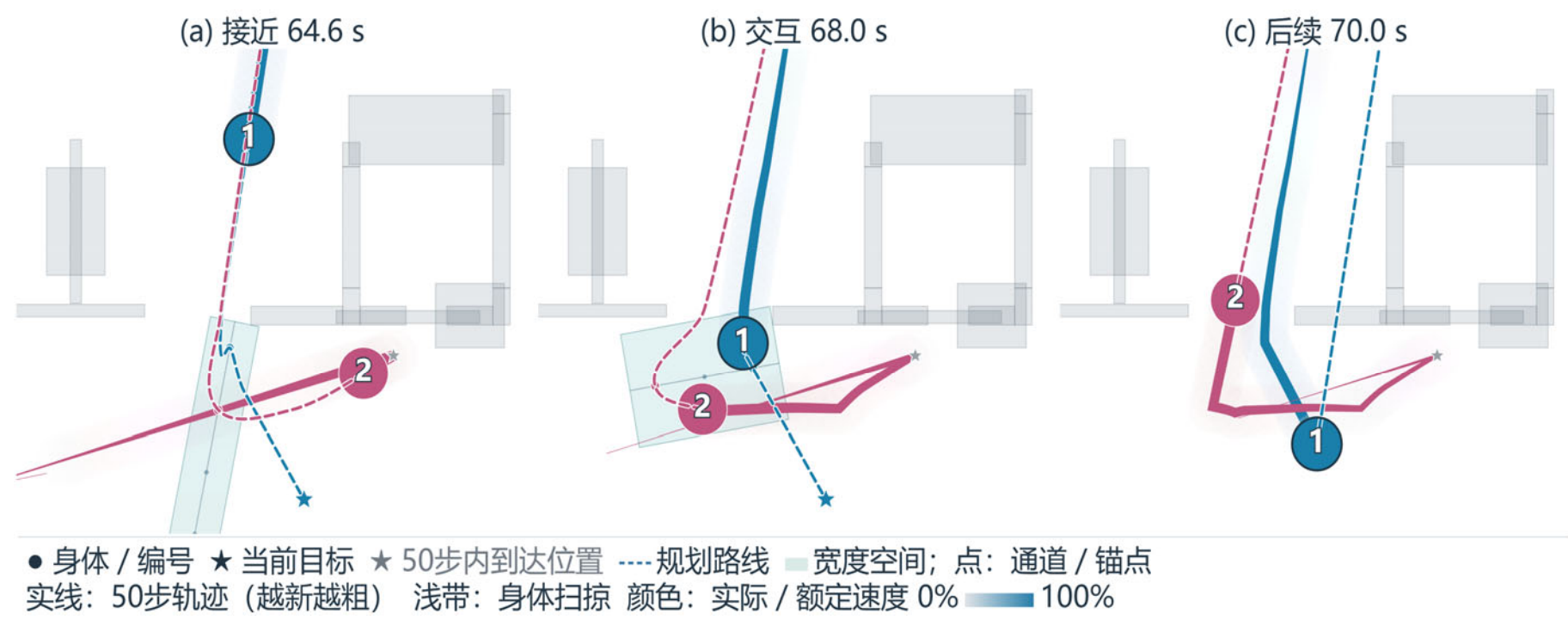

**图 25**　入口附近的两代理交会。双方先接近共同宽度，再从不同通道位置通过。末帧中代理 1 向下、代理 2 向上继续运动；灰色星标标示代理 2 此前实际到达的位置。

图 26 中，代理 2 到达目标后转向下方，与向上接近的代理 1、3 形成新的通过关系。共同通道及接续路线随之更新，三者随后分向离开，显示了目标变化后三方交会的接续过程。

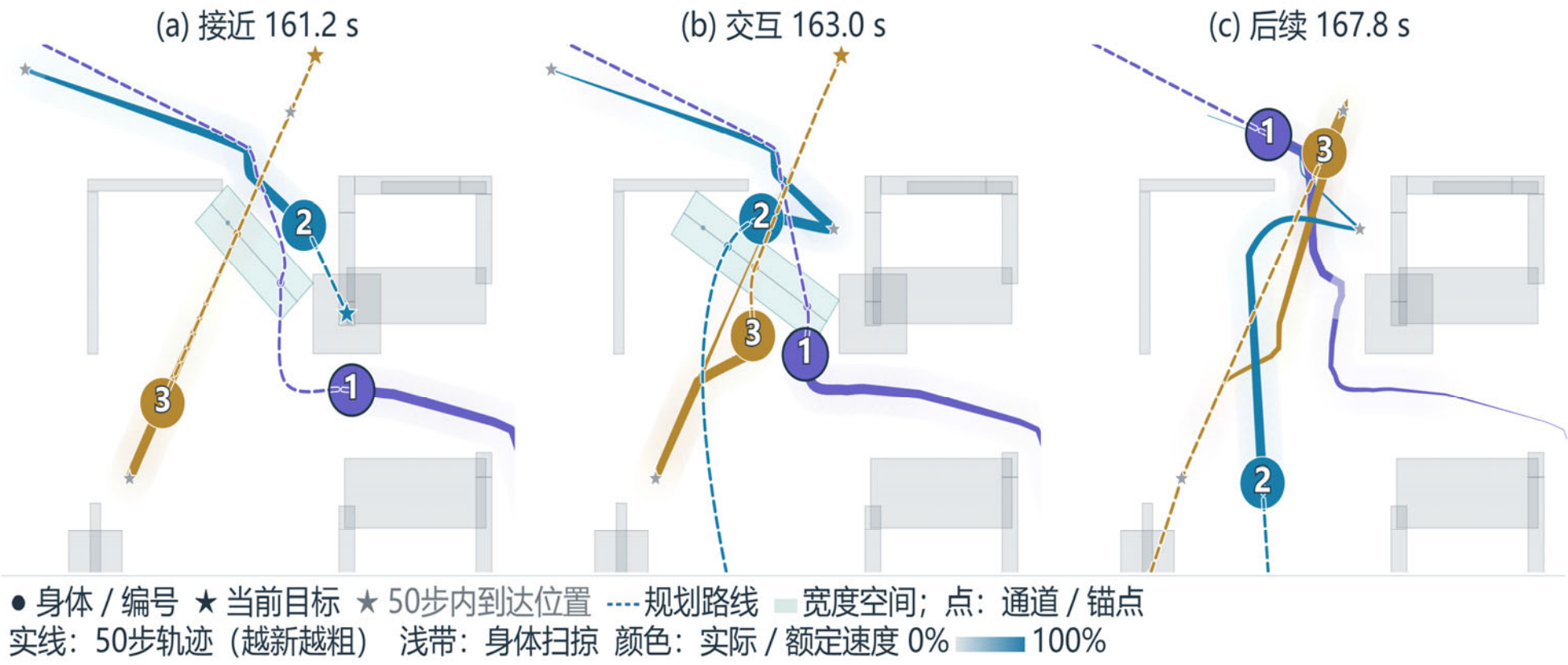

**图 26**　三代理在共同栅格上的接续协调。代理 2 到达目标后转向下方，代理 1、3 从下方接近并向上通过。中帧显示共同宽度及更新后的路线连接，末帧显示三者继续分向运动。

### 5.2.2 新增作业占用后的路线恢复

图 27 中，代理 2 开始作业后，新增操作空间侵入代理 1、3 原路线的身体扫掠范围。两者保持各自目标，从实际位置更新受影响的连接，分别向下方和上方继续运动。该例显示，同一占用变化下，不同方向的行程均可通过局部改线继续执行。

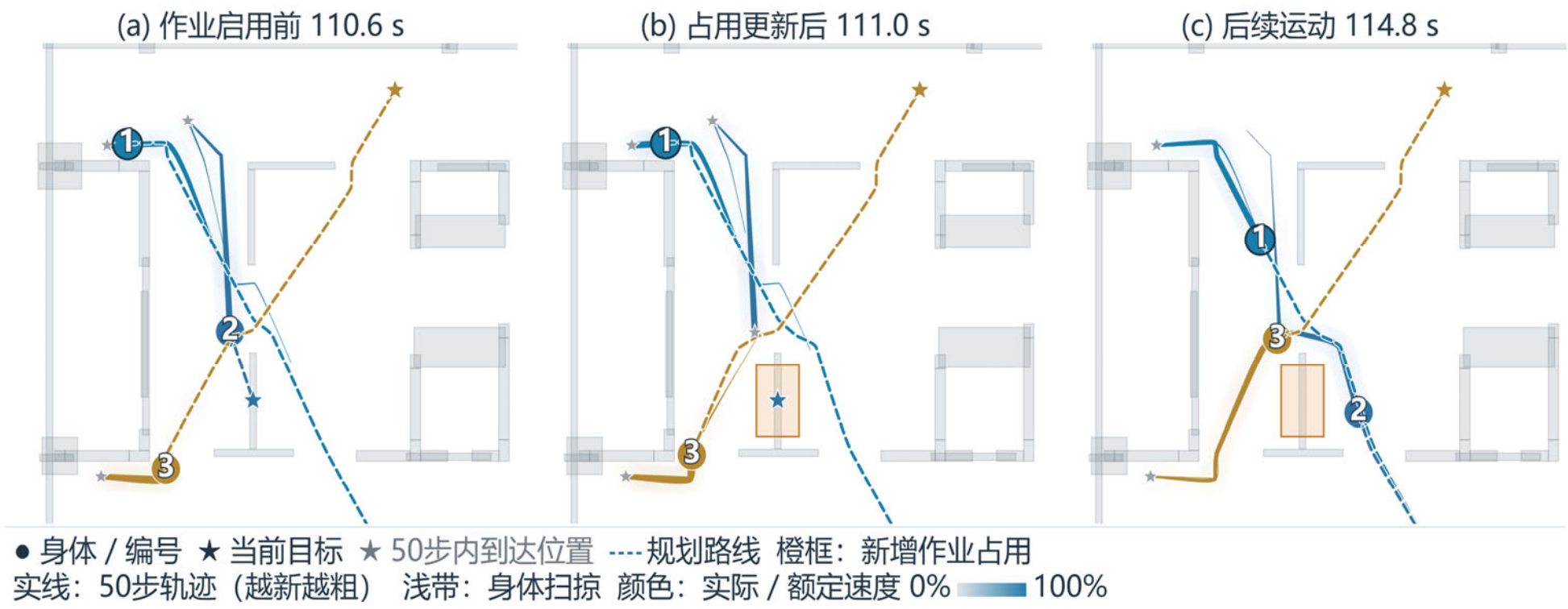

**图 27**　同一新增作业占用对两条路线的影响。(a) 110.6 s，代理 2 接近作业位置，代理 1、3 分别从上方和下方接近；(b) 111.0 s，新增操作空间以橙色边框标示，代理 2 暂离导航平面，代理 1、3 更新连接；(c) 114.8 s，两者沿调整后的路线继续接近原目标，代理 2 已返回导航平面并向下移动。代理 1、3 的目标在所示过程中保持不变。

### 5.2.3 多代理接入与交会接续

图 28 中，两人交会随新成员接近扩展为四人交会。本次交会结束后，代理 1、2 继续离开，代理 3、4 进入后续局部关系。成员随当前冲突加入、随身体清出退出，使局部通过安排沿连续行程建立、解除并接续。

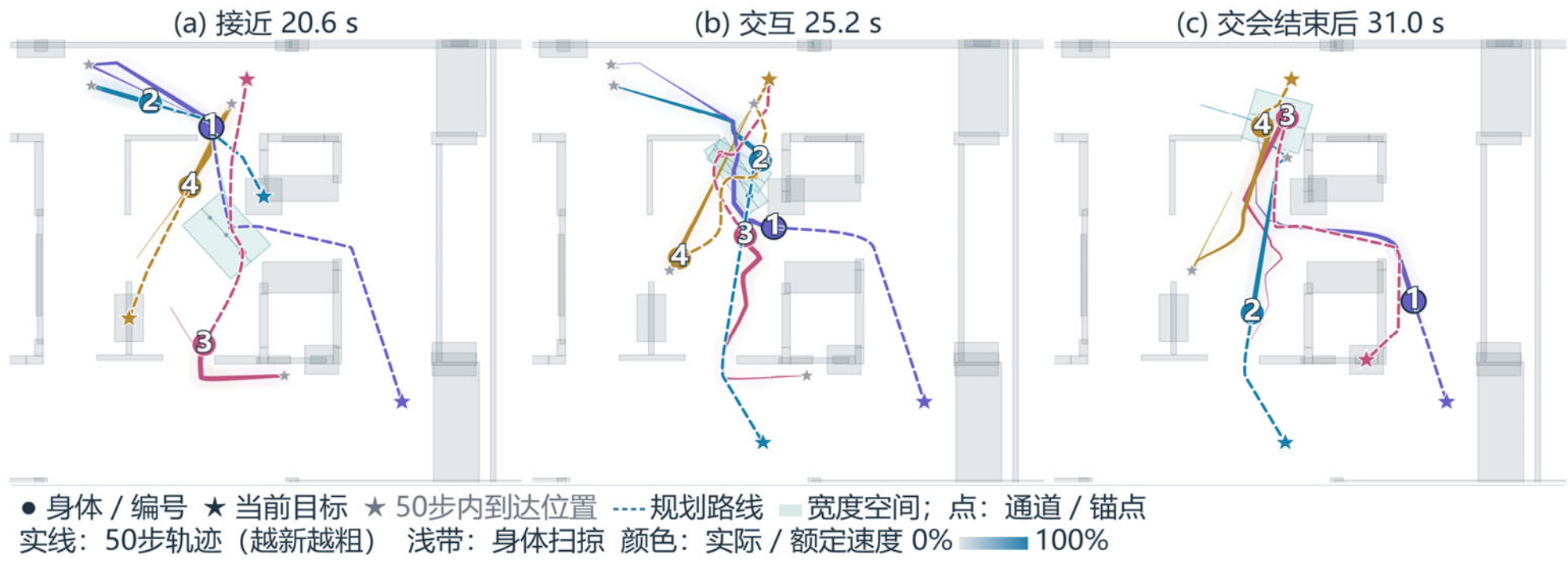

**图 28** 两人关系扩展形成的四人交会。20.6 s 时各成员逐步接近；25.2 s 时在连接区形成多方协调；31.0 s 时本次交会已结束，代理 1、2 继续离开，代理 3、4 开始参与后续局部关系。三个时刻按事件阶段选取。

图 29 中，五名代理于 195.0 s 集中进入协调，本次交会于 201.8 s 结束，随后恢复分散运动。该过程展示了五方安排从集中通过至身体清出的接续。

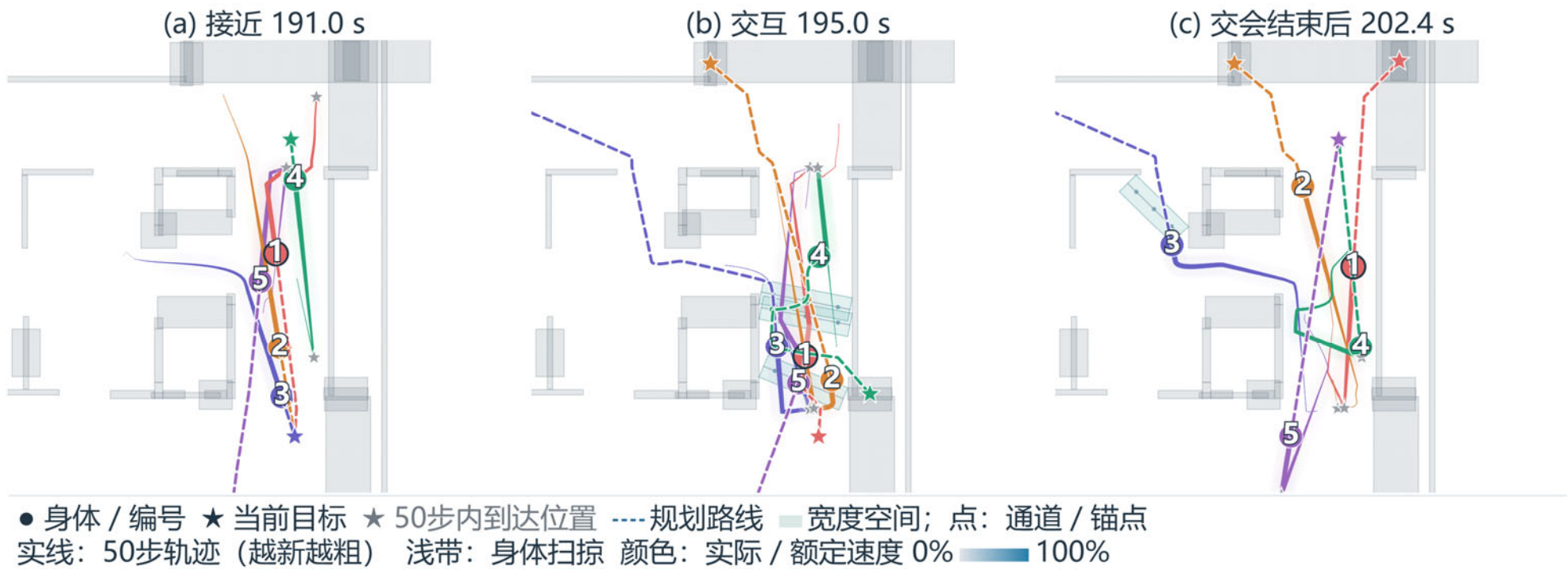

**图 29** 五人交会中的接入、集中通过与分散。三个时刻分别为 191.0、195.0 和 202.4 s。为观察本次交会于 201.8 s 结束后的状态，局部跟踪延续至统一观察时窗之外；效率统计仍采用前 200 s。

### 5.2.4 近停比例与持续时间

局部让行和执行约束可能使代理暂时减速，其影响取决于发生比例与持续时间。将一个代理在一个运动步内的状态计为一个“代理步”，按实际行驶距离与该步额定可行驶距离之比划分导航状态：近停为不超过 1%，低速为高于 1%且低于 20%，其余为达到或超过 20%的导航运动；作业状态单独统计。

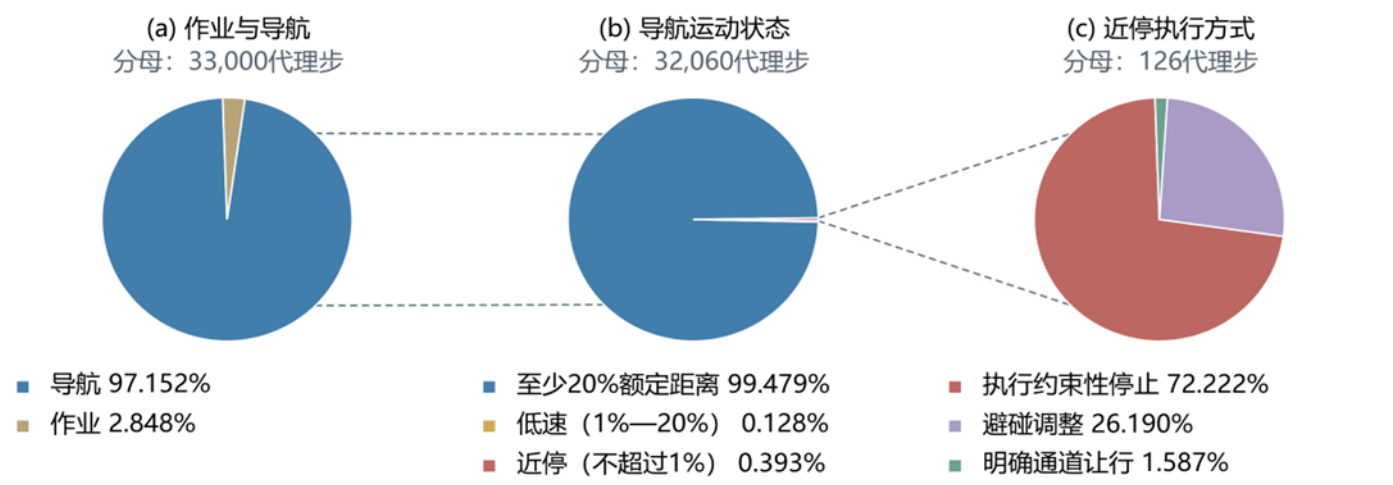

**图 30** 运行状态及短暂停止。三幅图的分母依次为 33,000 个全部代理步、32,060 个导航代理步和 126 个近停代理步。(a) 导航与作业分别为 32,060、940 步；(b) 不低于额定距离 20%的运动、低速与近停分别为 31,893、41、126 步；(c) 执行约束性停止、避碰调整与明确通道让行分别为 91、33、2 步。虚线表示从前一图相应扇区展开的条件分布。

在作业与材料补给加速的实验条件下，多次完整运行中导航占全部代理步的 97.15%。近停共 126 个代理步，占导航代理步的 0.393%；低速占 0.128%（图 30）。按同一代理、同一目标行程统计，最长连续近停为 5 步，即 1.0 s。近停记录主要归于

执行约束和避碰调整，明确标记为通道让行的仅占少部分。所测反复交会与目标更新过程中，近停比例较低、持续时间较短，完整方法总体保持了运动接续。

图 25 — 图 30 展示了通道安排下的身体通过、占用变化后的路线恢复和新成员接入后的多方接续；完整运行统计显示，近停比例较低且持续时间较短。两类证据共同支持 WP 在所测动态交会中的运动接续能力。下节进一步比较完整时窗内的到达效率与计算代价。

## 5.3 对照实验

### 5.3.1 比较目的与实验设计

对照实验选取离散时空搜索、路径引导下的互惠避碰和连续轨迹搜索三类方法。互惠避碰分别设置独立使用与结合局部 MAPF 恢复两种对照，比较受阻恢复安排不同的运行表现。表 2 列出四种对照方法的原理、已有应用及比较维度。

表 2　对照方法的选择依据。

| 方法 | 基本原理与已有应用 | 本实验考察的差异 |
| --- | --- | --- |
| RHCR＋PBS[5,23] | RHCR 将持续目标需求转化为滚动时窗内的 MAPF 问题，PBS 依据冲突搜索代理间的优先关系；原研究在自动化仓储仿真中展示了持续任务处理及规模扩展能力。 | 离散时空协调与局部连续交会组织的差异。 |
| Theta*＋ORCA[4,19,31] | Theta*生成任意角度路径，ORCA 依据互惠责任选择避碰速度；ORCA 已在移动机器人和密集群体仿真中展示高效局部避碰，二者组合提供路径引导下的反应式导航。 | 个体路径与速度避让能否应对受限空间中的持续交会。 |
| ORCA＋局部MAPF[31,32] | 在互惠避碰受阻时，为相关代理构造局部离散规划问题；原研究在狭窄、受限场景中通过局部协同提高目标到达成功率。 | 结合局部重规划的反应式导航在持续交会中的表现。 |
| K-CBS 滚动适配[6] | 冲突树约束与单体采样规划相结合，直接生成连续轨迹；原研究在开阔、狭窄及杂乱环境的车式机器人案例中展示了连续协调能力。 | 连续轨迹搜索与通道锚点决策在运动效率及计算代价上的差异。 |

五种方法在 5—10 名代理的六种配置下分别多次运行，共享楼层布局、身体尺度、额定速度、初始目标和任务更新规则。后续目标按各自的到达时刻更新。网格方法采用安全格点到达，连续方法采用接触到达；比较对应这些条件下的适配方法。

评价指标为**目标到达量、已完成行程时长和平均单步计算耗时**。目标到达量计数代理到达任务区及堆场的次数，反映固定观察时窗内的持续转移能力；行程时长记录完成一次目标行程所用的模拟时间；单步计算耗时记录每个运动步的在线计算时间。对至少完成一次目标行程的完整运行，后两项的均值为

$$\bar{T} = \frac{1}{L}\sum_{\ell=1}^{L} T_\ell, \qquad \bar{w} = \frac{1}{K}\sum_{k=1}^{K} w_k. \tag{39}$$

其中，$L$ 为已完成行程数，$T_\ell$ 为第 $\ell$ 次行程的模拟时长；$K$ 为运行步数，$w_k$ 为第 $k$ 步的计算耗时。$\bar{T}$ 对该次运行的已完成行程取均值，$\bar{w}$ 对全部运行步取均值；各组再对每次完整运行的指标等权汇总。到达量与计算耗时统计完成 200 s 时窗的运行，行程时长统计其中已完成的行程，因此结果描述完整观测条件下的效率。

计算耗时包括每步规划、执行检查和任务更新，在同一 AMD Ryzen 9 9950X3D 主机上测量；计时条件不同的记录单独标注。

### 5.3.2 密集交会的实际轨迹比较

图 31 — 图 35 分别选取五种方法在 8 代理运行中的密集交会区域，各呈现 9 s 内的三个时刻，统一显示实际轨迹。比较关注持续推进、同一目标下的局部回折，以及停留后的运动恢复。灰色星标标示目标到达引起的正常转向；逐步位置、运动距离与目标记录用于核对回折和近停。各方法的后续目标与局部状态已随运行分化，片段用于比较运动形态，整体效率按完整时窗评价。

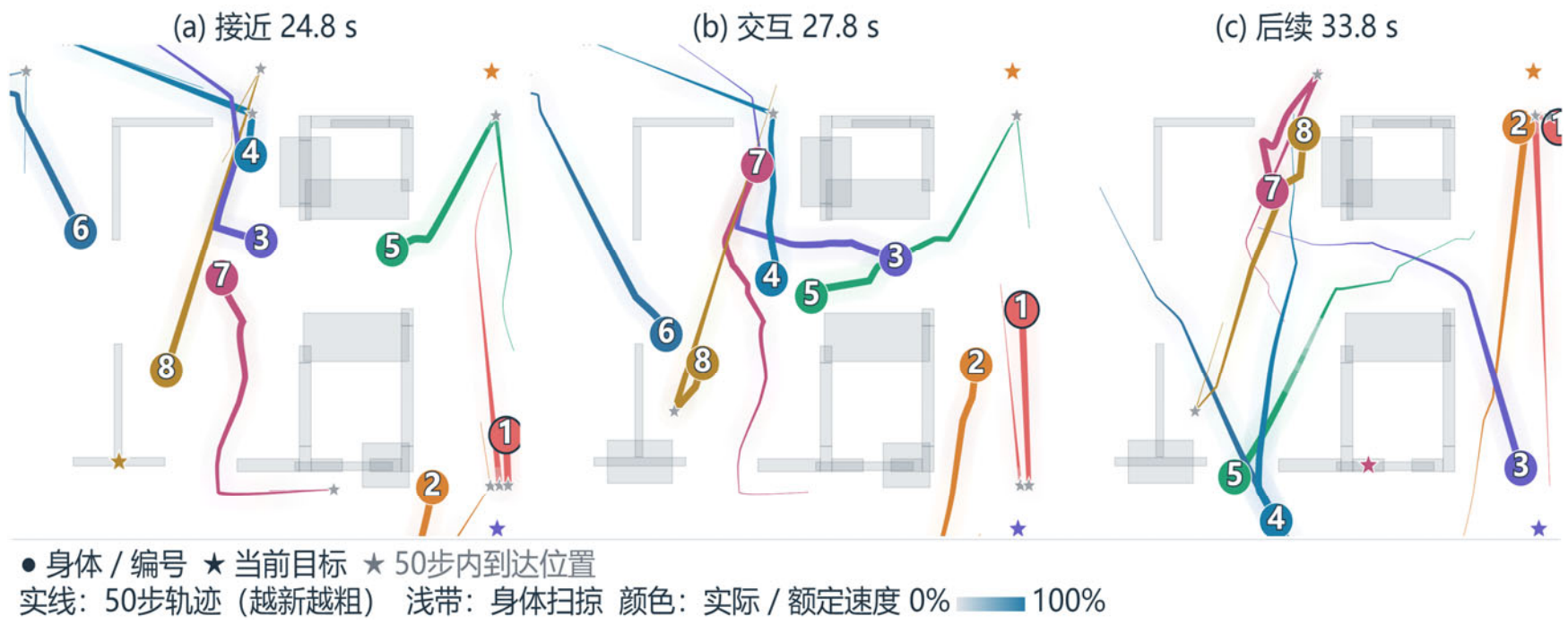

**图 31** WP 在 24.8、27.8 和 33.8 s 的局部运动。代理 3 向右通过，代理 4、5 逐步向下离开；代理 8 在下方到达后转向上行，代理 7 在上方到达后转向下行。灰色星标标示近 50 步内的实际到达位置，图中仅显示实际轨迹。

WP 片段中，多方向运动在中央连接处汇集后继续分散。代理 3、4、5 保持各自目标，经右侧和下方空间离开；代理 8 和 7 分别于 26.0 s 和 30.2 s 到达目标，其折返属于新行程。其他局部方向修正伴随持续净位移，主要成员继续离开当前交会位置，呈现了目标更新与密集交会下的多向通过。

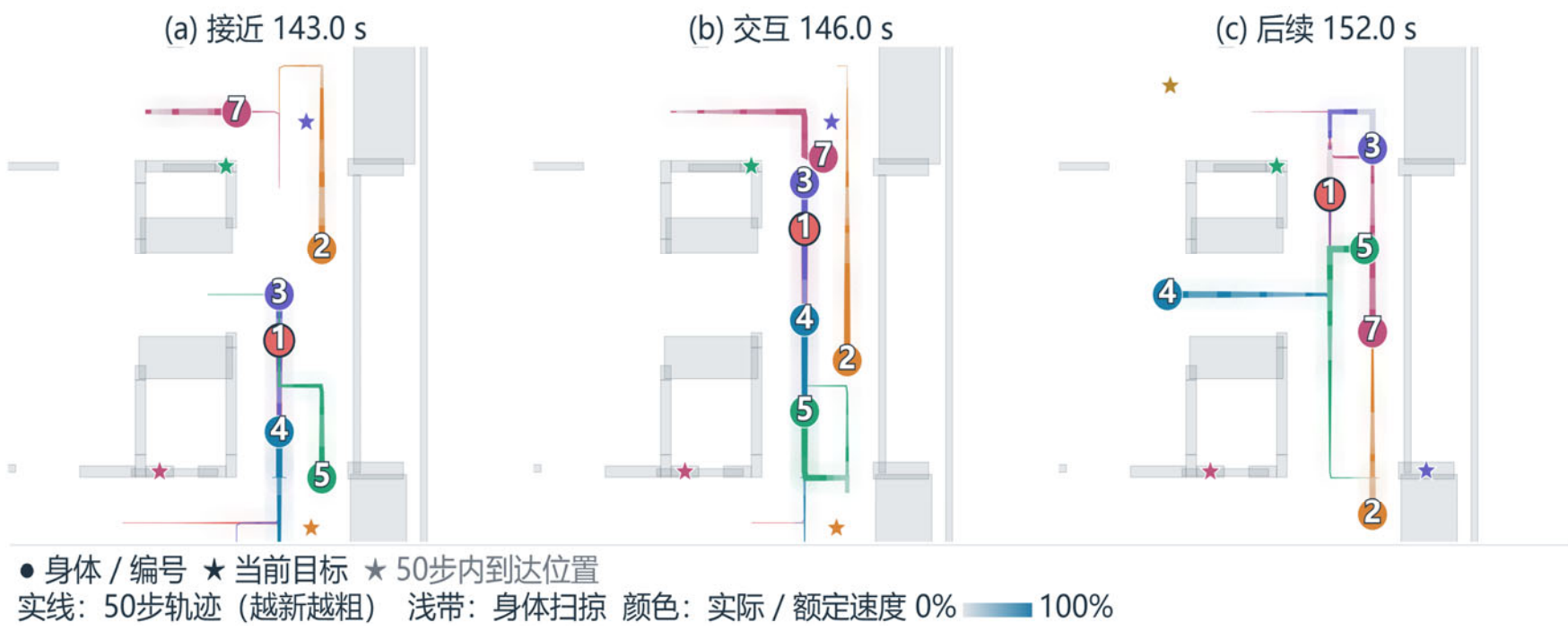

**图 32** RHCR＋PBS 在 143.0、146.0 和 152.0 s 的局部运动。代理 1、3、4、5 沿右侧通道接续，代理 7 由上方转入；轨迹主要由水平与竖直段连接。

RHCR＋PBS 的轨迹沿离散通行方向展开，以转角连接直线路段。 图 32 中，代理 3 先向上接近，再横向进入相邻位置；代理 4 从纵向通道转向左侧，部分代理等待后恢复前进。该片段呈现了离散时空协调下的有序接续，所见直角变化主要对应图结构中的路线转换。

Theta*＋ORCA 沿个体引导路径调整避碰速度。 图 33 中，代理 1、5 在同一目标下反复小幅改向，累计运动仅产生很小的净推进；代理 4 则向下离开相邻群体。个体绕避与局部滞留在该片段中并存：持续速度调整尚未使右侧受阻成员及时清出。

ORCA＋局部 MAPF 以局部协同规划恢复受阻运动。 图 34 中，代理 4 在下方连接处先向外侧运动，再转回通道；代理 5 连续近停，代理 6 向上推进。持续推进、局部回折与连续近停同时出现，表明采用局部恢复后，群体仍有部分成员暂时受阻。

K-CBS 滚动适配在连续空间中协调相容轨迹，允许斜向及多方向连接。图 35 中，多数成员持续移动，中央区域的身体分布随通过过程改变；代理 3、7、8 仍出现同一目标下的局部回折。该片段呈现了连续轨迹协调下的多方接续及执行中的方向修正。图中折线连接逐步位置，可辨认转向，连续曲率则需原轨迹描述。

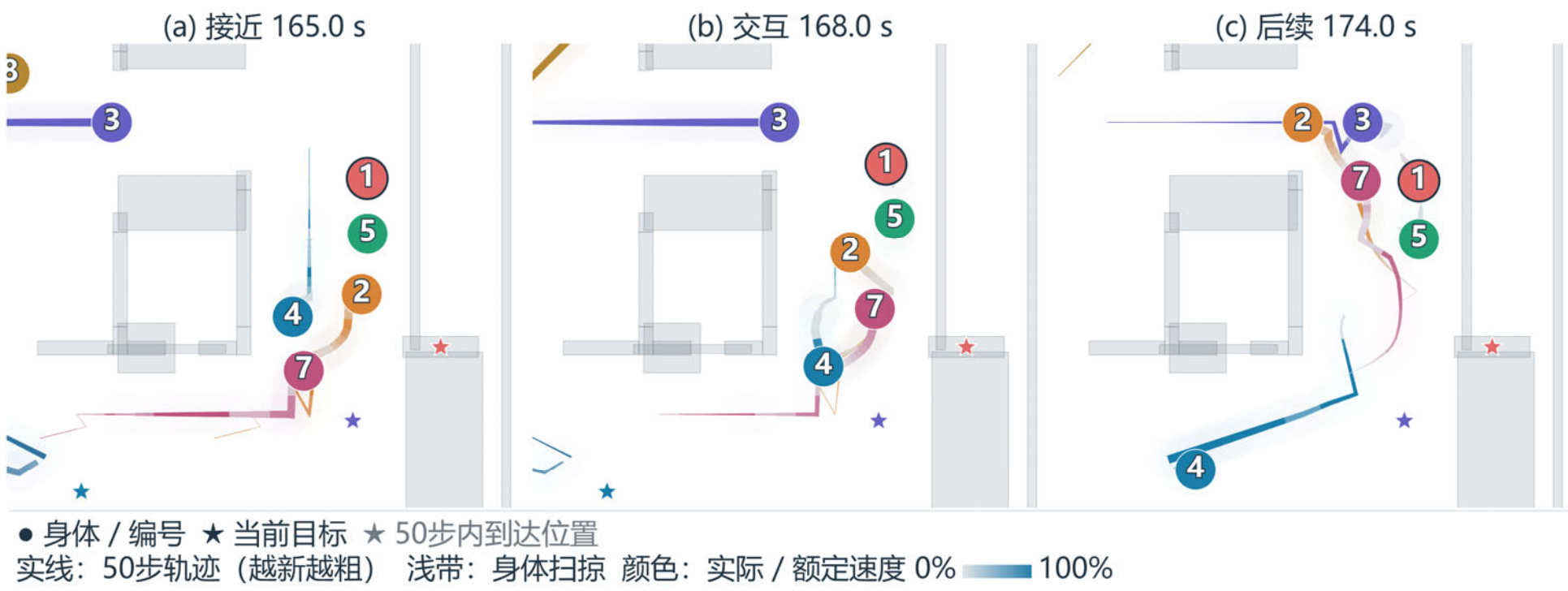

图 33 Theta*＋ORCA 在 165.0、168.0 和 174.0 s 的局部运动。代理 1、5 持续位于右侧相近位置，代理 2、3、7 逐渐汇集，代理 4 向下离开。

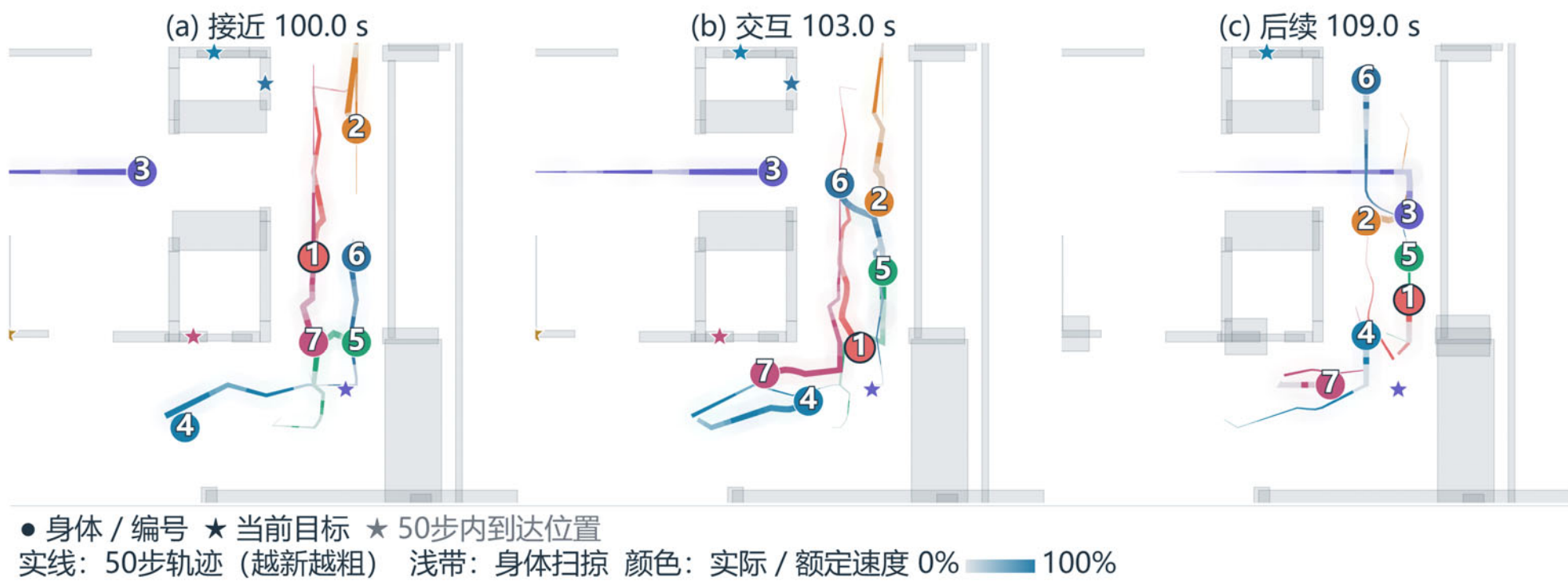

图 34 ORCA＋局部 MAPF 在 100.0、103.0 和 109.0 s 的局部运动。代理 6 向上推进，代理 3 由左侧接入；代理 1、2、5 位于通道中，代理 4、7 在下方连接处调整。

这些片段显示，局部转向既可能对应目标更新，也可能伴随避碰调整或持续受阻，需结合净进展、持续时间与共享空间清出判断。WP 的多向通过、离散协调的有序接续以及两种互惠避碰对照中的局部滞留，为后续完整时窗的效率比较提供了运动过程解释；各方法的整体表现以下节到达量与行程时长为依据。

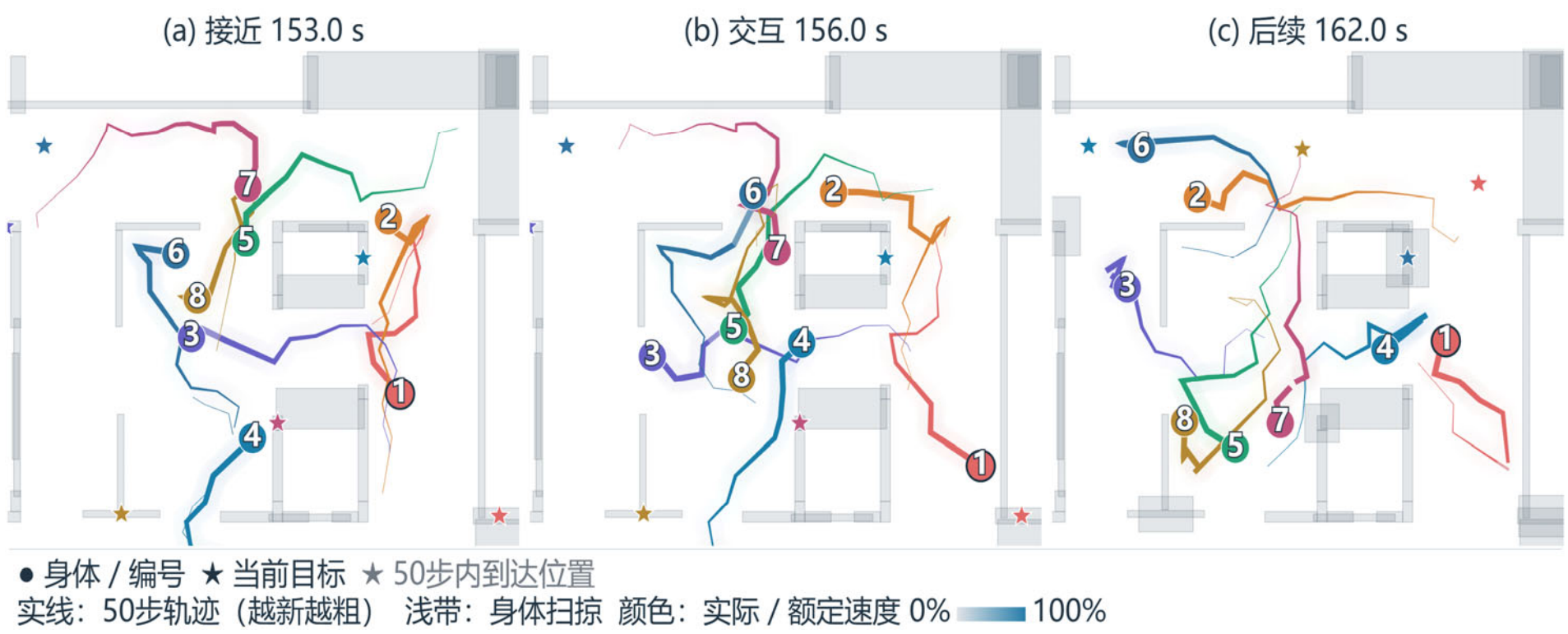

图 35 K-CBS 滚动适配在 153.0、156.0 和 162.0 s 的局部运动。代理 2、6 经上方空间通过，代理 5、7、8 在中央连接处接续；实际位置由逐步记录连接。

### 5.3.3 目标到达量与持续推进比较

表 3 汇总多次完整运行在 200 s 内的目标到达量，报告均值及观测范围； 图 36 给出对应的累计到达过程。

表 3　200 s 内目标到达量：均值 [最小值，最大值]。

| 代理数 | WP | RHCR＋PBS | Theta*＋ORCA | ORCA＋局部 MAPF | K-CBS 滚动适配 |
|---|---|---|---|---|---|
| 5 | 96.7 [95, 98] | 65.7 [65, 66] | 49.3 [41, 55] | 57.3 [54, 61] | 76.0 [70, 80] |
| 6 | 111.0 [106, 115] | 72.7 [72, 73] | 46.7 [40, 59] | 68.0 [66, 69] | 88.7 [82, 94] |
| 7 | 136.0 [126, 143] | 85.3 [84, 86] | 57.3 [52, 62] | 71.0 [66, 76] | 104.3 [102, 107] |
| 8 | 153.0 [148, 160] | 94.3 [91, 98] | 57.3 [51, 63] | 78.0 [71, 88] | 110.7 [110, 111] |
| 9 | 164.3 [162, 166] | 98.7 [97, 100] | 65.7 [62, 70] | 70.7 [68, 74] | 108.0 [73, 126] |
| 10 | 189.0 [185, 194] | 109.7 [108, 112] | 67.3 [65, 71] | 79.0 [76, 83] | 133.7 [132, 137] |

表 3 中，WP 在六种人数配置下均取得最高平均目标到达量，较各配置最佳对照 K-CBS 滚动适配提高 25.2%—52.2%。5—7 代理的增幅为 25.2%—30.4%，8—10 代理为 38.3%—52.2%，较高密度配置的相对优势更大。在初态配对的 8 代理组中，WP 平均到达 153.0 次，比 K-CBS 滚动适配的 110.7 次高 38.3%，为 RHCR＋PBS 的 1.62 倍；各匹配初态下均取得更多到达，差值为 37—50 次。WP 的优势因此体现为完整观察时窗内的持续到达。

图 36 中，WP 持续累积目标到达，与其他方法的平均差距随运行推进总体扩大，并延续至观察末端。5—10 代理配置下，WP 每代理平均到达 18.3—19.4 次，总到达量随人数增加而上升，单位代理到达量则保持相近。该趋势限于当前楼层、任务规则和完整样本；部分配置使用不同初态集合，尚不能单独归因于代理人数变化。

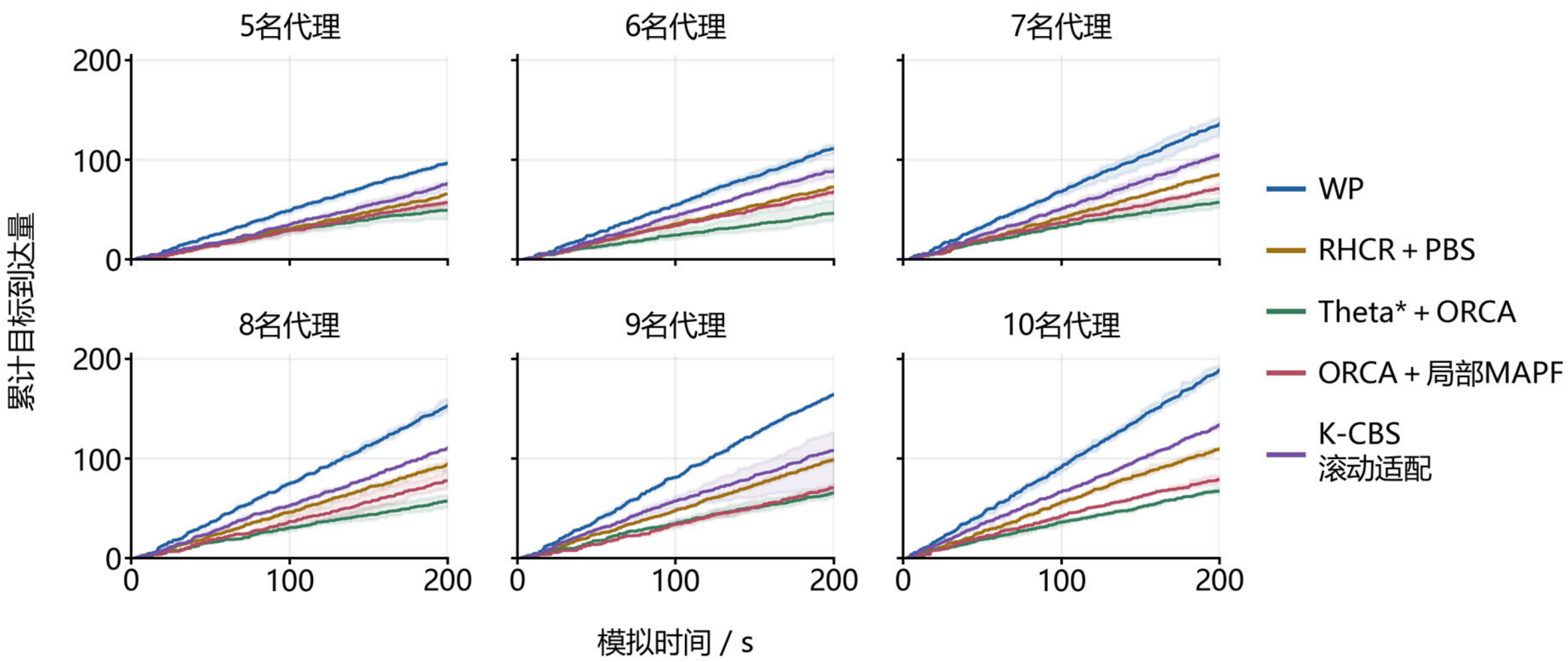

**图 36**　不同代理人数下的累计目标到达量。曲线为完整运行的均值，阴影为观测最小—最大值；各面板采用相同坐标尺度。

### 5.3.4 已完成行程时长与平均计算耗时比较

表 4 列出已完成行程的平均时长和平均单步计算耗时，分别反映模拟行程用时与在线计算负担。各组对单次完整运行的指标等权汇总。

表 4　不同代理人数下的已完成行程时长与平均计算耗时。

| 代理数 | 方法 | 已完成行程平均时长 / s | 平均单步计算耗时 / ms |
|---|---|---|---|
| 5 | **WP**（本文方法） | 9.66 | 19.7 |
| 5 | RHCR＋PBS | 14.44 | 7.5 |

| 代理数 | 方法 | 已完成行程平均时长 / s | 平均单步计算耗时 / ms |
|---|---|---|---|
| 5 | Theta*＋ORCA | 16.20 | 76.0 |
| 5 | ORCA＋局部 MAPF | 15.70 | 89.7 |
| 5 | K-CBS 滚动适配 | 12.51 | 139.9 |
| 8 | **WP**（本文方法） | 9.83 | 65.5 |
| 8 | RHCR＋PBS | 15.81 | 12.1 |
| 8 | Theta*＋ORCA | 18.89 | 87.1 |
| 8 | ORCA＋局部 MAPF | 19.14 | 116.7 |
| 8 | K-CBS 滚动适配 | 13.32 | 227.1 |
| 10 | **WP**（本文方法） | 9.86 | 107.8 |
| 10 | RHCR＋PBS | 16.01 | 14.5 |
| 10 | Theta*＋ORCA | 22.74 | 104.9† |
| 10 | ORCA＋局部 MAPF | 21.60 | 121.2† |
| 10 | K-CBS 滚动适配 | 13.00 | 357.9 |

† 补充批次受同期回放整理影响，计时条件存在差异，均值仅供参考。

8 代理组中，WP 与 K-CBS 滚动适配的已完成行程平均时长分别为 9.83 s 和 13.32 s，平均单步计算耗时分别为 65.5 ms 和 227.1 ms。WP 取得更多到达，同时具有较短的已完成行程时长和较低的平均计算耗时。RHCR＋PBS 以 12.1 ms 的平均单步计算耗时取得 94.3 次到达，计算代价更低。各方法后续完成的行程组成不同，时长比较对应各自实际完成的需求，不能直接换算为相同行程的时间节约。

表 3 与表 4 显示，WP 在当前场景中兼顾了持续到达与计算代价：相对于 RHCR＋PBS，以较高计算代价取得较多到达；相对于 K-CBS 滚动适配，在 8 代理配对组中同时取得较多到达和较低平均计算代价。这一比较支持所述运动需求与适配条件下完整方法的规划效率。

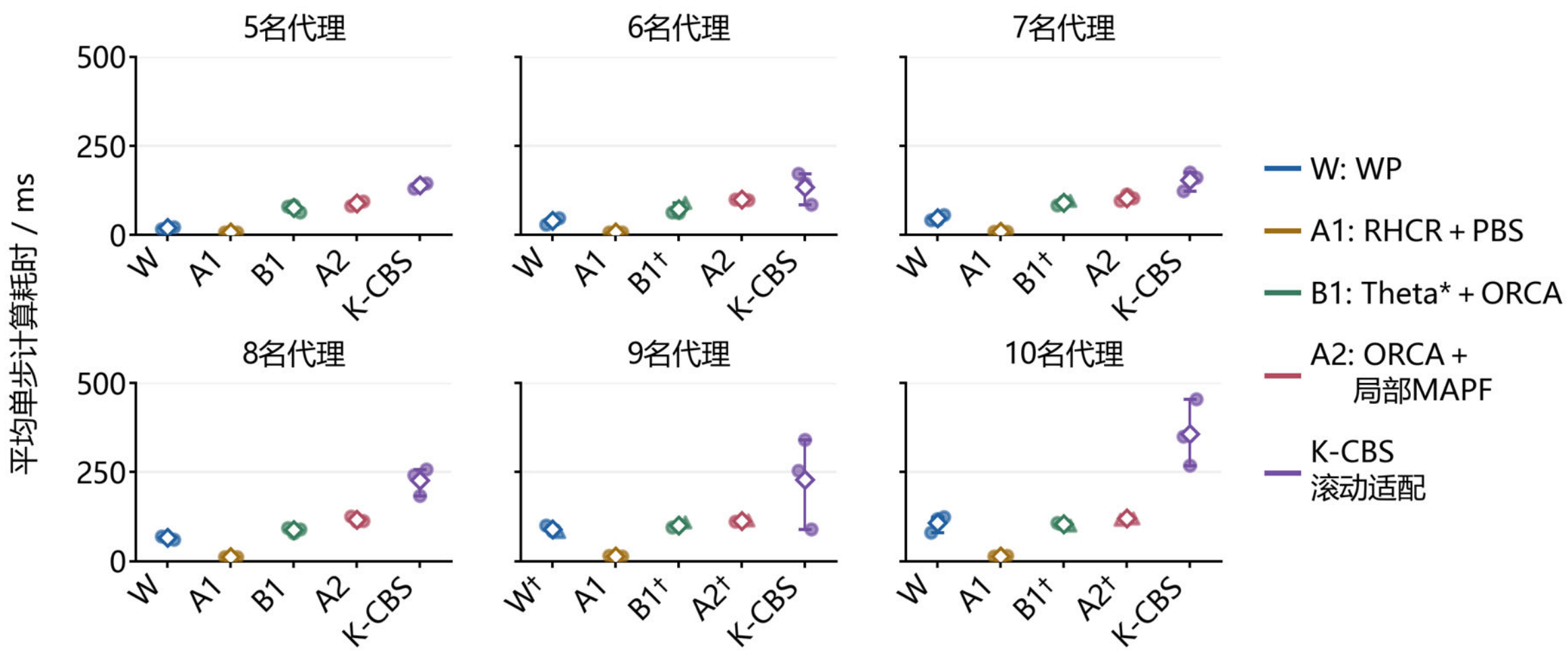

**图 37**　平均单步计算耗时及运行间差异。圆点与三角形表示不同批次的完整运行，空心菱形为组均值，须线为观测范围；各面板采用相同毫秒尺度。†为计时条件存在差异的参考均值。W、A1、B1、A2 和 K-CBS 依次对应 WP、RHCR＋PBS、Theta*＋ORCA、ORCA＋局部 MAPF 和 K-CBS 滚动适配。

图 37 中，WP 的平均单步计算耗时随人数增加而上升；在 5—10 代理配置下，其组均值均低于 K-CBS 滚动适配、高于 RHCR＋PBS。连续通道规划因而仍有随配置扩大而增加的计算负担。平均耗时描述整段运行的计算代价，实时期限是否满足还取决于单步时延分布。

## 6. 讨论

作业启停改变局部净空，改线又影响后续交会，使当前通过与后续接续相互依赖。局部冲突求解、宽度引导、空间分离和依赖维护已有研究基础[7－14,18,25,28,29]。WP 的创新在于以实际交会需要连接这些处理：在交会处依据实时空间建立共同通过关系，在相邻交会之间保留贪婪引导的连续路线，随空间和参与者变化更新受影响安排。相对于以网格位置、分区边界或采样连接组织运动的表达[6,18,19]，这一架构明确了共同约束在何处生成、怎样接续以及何时更新，并可继续利用区域拓扑和几何路径求解。

空间相容判据为这种组织方式提供了几何依据。在所述局部空间条件下，定理 1 给出定点换位的充要宽度条件；定理 2 给出组合换位必须覆盖的分离方向，分别指导通道构造和交会位置迁移。多代理在共享栅格上按式(34)协调通过位置，以式(35)—(36)接入各自锚定路线，改线后重新预测，实际运动统一满足式(1)。因此，局部空间条件不仅用于选择一次交会的位置，也参与后续通过关系的保持与重建。

共同通道同时联系时间让行与受阻疏通：分离位置保留通过空间，让行错开到达时刻，顺行协调将前方阻挡者递推纳入同向运动。通道轴线上的横向分离由式(26)保证，有限步清出依赖式(38)的出口接纳与实际进展。两至五人交会及占用变化后的记录展示了通过和清出的接续；完整 WP 运行中，近停占导航代理步 0.393%，同一代理、同一目标行程内最长连续近停为 1.0 s。这些结果支持完整方法的运动接续表现，各机制的独立收益尚需区分。

5—10 代理的完整运行比较中，WP 平均目标到达量较最佳对照提高 25.2%—52.2%；更多到达伴随已完成行程平均时长的缩短，二者相互印证。其平均计算耗时低于 K-CBS 滚动适配，支持 WP 在所测场景中兼顾持续到达与计算代价。这些系统结果为交会处共同协调与交会间路线自由相结合提供了计算证据。

本文由此形成的贡献，是从实时可用空间构造局部通过关系、并沿连续运动维护这些关系的规划原则。几何判据给出构造依据，共享通道与锚点接续使局部安排参与多方协同，紧迫更新和实际清出决定关系的保持与解除，共同支撑了对作业与运动空间耦合过程的处理。这为其他兼有任务占用与持续转移需求的工程场景提供了方法依据。

## 7. 结论

本文针对作业与运动空间耦合条件下局部交会可通过性的构造与持续保持，提出了 WP 方法。创新在于以预测身体冲突组织共同决策，将实时空间相容分配、连续路线接续与受阻疏通纳入按需生成、持续更新的规划架构。定点换位的充要宽度条件与组合换位的必要方向条件提供了通道构造和位置迁移的几何依据；共享通道与有序锚点连接多方交会，紧迫更新保留有效关系并重建受影响部分。由此形成的贡献是以局部通过关系组织连续群体运动的规划原则，将共同协调与贪婪路线自由相结合，扩展了终身路径规划对作业占用变化与持续运动需求的联合处理。

楼层计算实验记录了作业占用变化后的路线接续及多方交会清出，完整 WP 运行中的近停比例较低、持续时间较短。5—10 代理的 200 s 完整运行比较中，WP 各配置平均目标到达量较最佳对照提高 25.2%—52.2%，相较 K-CBS 滚动适配的平均计算耗时较低。这些结果为该架构在工程生产场景中的运动协调提供了计算证据；局部疏通仍受相应几何与执行条件约束，各机制的独立效应及跨场景适用性有待受控检验。

# 参考文献